\documentclass[reprint, amsmath, amssymb, showpacs, superscriptaddress, aps, prl]{revtex4-2}

\usepackage{microtype}
\usepackage{graphicx,soul}
\usepackage{dcolumn}
\usepackage{bm}

\usepackage{multirow}
\usepackage{xcolor}
\usepackage[export]{adjustbox}

\usepackage{url}
\usepackage{mathtools}

\usepackage{comment}
\usepackage[normalem]{ulem}
\usepackage{nameref}

\usepackage{multirow}
\usepackage{xcolor}
\usepackage[export]{adjustbox}

\usepackage[colorlinks=true,
						linkcolor=blue,
						urlcolor=blue,
						citecolor=blue,
						bookmarks=true,
						pdfborder={0 0 0}]{hyperref}

\graphicspath{{figure/}}

\begin{document}
\title{
Higher-Order Functional Networks from Event Time Series
}

\author{Andrea Civilini}
\email{andrea.civilini@icm-institute.org}
\affiliation{Sorbonne Université, Paris Brain Institute (ICM), CNRS UMR7225, INRIA Paris, INSERM U1127, Hôpital de la Pitié Salpêtrière, AP-HP, Paris 75013, France}
\affiliation{School of Mathematical Sciences, Queen Mary University of London, London E1 4NS, United Kingdom}
\affiliation{Dipartimento di Fisica ed Astronomia, Universit\`a di Catania and INFN, Catania I-95123, Italy}

\author{Fabrizio de Vico Fallani}
\affiliation{Sorbonne Université, Paris Brain Institute (ICM), CNRS UMR7225, INRIA Paris, INSERM U1127, Hôpital de la Pitié Salpêtrière, AP-HP, Paris 75013, France}

\author{Vito Latora}
\affiliation{School of Mathematical Sciences, Queen Mary University of London, London E1 4NS, United Kingdom}
\affiliation{Dipartimento di Fisica ed Astronomia, Universit\`a di Catania and INFN, Catania I-95123, Italy}
\affiliation{Complexity Science Hub Vienna, A-1080 Vienna, Austria}

\begin{abstract}
Inferring relations among the units of a complex system from their activity is a fundamental problem across many fields.
Here, we introduce a method to identify  functional dependencies of arbitrary order from  multivariate event time series. The method maps the data onto a symbolic sequence, and uses a Bayesian framework to extract statistically significant motifs of symbols.
These motifs define the hyperedges of a hypergraph, providing
a compact representation of
higher-order functional dependencies
in their original data and enabling their
analysis with network-based tools. 
Applications to neural and social systems reveal nontrivial higher-order dependencies associated with collective brain activity and social behaviour. 
Our framework thus provides a general approach for uncovering higher-order functional organization directly from event data.
\end{abstract}

\maketitle

\setcounter{secnumdepth}{4}

\paragraph*{\bf Introduction.} 
Complex systems exhibit collective behaviours of extraordinary richness. Notable examples include the emergence of synchronization in biological systems \cite{strogatz1993coupled} and in human cognition \cite{dehaene_cognitive_1998}, as well as cooperation \cite{axelrod_evolution_1981,nowak_supercooperators_2012} and epidemics in social groups \cite{pastor_epidemic_review_2015}. Such behaviours directly stem from the interactions among the system units, such as the individuals in a society, or the neurons in the brain \cite{strogatz_exploring_2001, boccaletti_complex_2006, fallani_graphbrain_2014, chiaron_EEG_2023}. 
A convenient way to represent the pattern of relations in a complex system is through a graph, whose nodes correspond to the  units of the complex system and the edges stand for direct or indirect couplings, 
or other dependencies or correlations between two units \cite{newman_networks_2010, latora_complex_2017, cimini_statistical_2019}.
However, graphs can only describe pairwise relations, while real-world systems often 
exhibit higher-order dependencies and correlations involving three or more units \cite{Martignon1995, LeeWilson2004, Diekman2009, Shimazaki2012, robiglio_synergistic_ho_2025}.
This is the reason why in the last decade we have witnessed an increasing use of higher-order representations, such as simplicial complexes and hypergraphs, in the study of complex systems  \cite{battiston_networks_2020, battiston_physics_2021, iacopini_simplicial_2019, alvarez_pgg_ho_2021, civilini2024explosive, yu_ho_cortical_2011, petri_scaffold_2014, squartinigarlaschelli_book_2017, squartini_reconstruction_2018, chelaru_high_order_2021, santoro_ho_connectonomics_2024, abiad_hypergraph_roadmap_2026}.
A particularly important problem is, thus, how to construct such graphs or hypergraphs of relations from the  empirical measurement of the temporal activity of the system units.
Unlike structural networks, whose links represent underlying physical interactions, functional networks encode statistical dependencies among the dynamics of the system's components, providing a complementary perspective on the system's organization and behaviour. Various methods have been proposed over the years to infer networks from multivariate time-series data, including approaches based on statistical dependencies, dynamical models, information theory, and causal inference \cite{yu_estimating_network_2006, timme_revealing_net_2007, han_reconstruction_prl_2015, brugere_network_extraction_2018, marrelec_partial_2006, granger1969investigating, guo_partial_granger_2008, schreiber2000transfer}. However, these approaches have primarily focused on pairwise relations, while methods specifically designed to identify higher-order relations have only emerged in recent years \cite{musciotto_detecting_2021, young_hypergraph_2021, lizotte_hypergraph_2023, malizia_reconstructing_2024, delabays_hyper_recons_2025,
santoro_higher_order_2023, varley_partialentropy_2023, keogh_dimensionality_2001, lin_sax_2007, 
wang_full_2022}. 
Such methods generally require at least one of the following:
\textit{i)} the assumption of continuous, differentiable time series \cite{malizia_reconstructing_2024, delabays_hyper_recons_2025}, 
\textit{ii)} 
specific transformations of continuous signals together with assumptions about the statistical properties of the resulting representation
\cite{santoro_higher_order_2023, keogh_dimensionality_2001, lin_sax_2007}, 
\textit{iii)} prior knowledge of the system’s underlying dynamical rules and their functional form 
\cite{malizia_reconstructing_2024, wang_full_2022}, 
or \textit{iv)} 
the estimation of high-dimensional probability distributions, whose statistical and computational cost grows rapidly with interaction order
\cite{panzeri2007correcting, williams2010nonnegative, rosas2019o}.
These requirements are often too restrictive in real-world scenarios, where the exact dynamical rules generating the time series are often unknown \cite{delabays_hyper_recons_2025}. Additionally, the activity of individual units is frequently neither continuous nor differentiable; instead, it tends to occur at discrete moments in time, with events that are effectively instantaneous relative to the typical timescale of observation.
Relevant examples are the spiking activity of neurons, seismic events, user activity on social platforms and sell-buy orders in stock markets. 
In each of these situations, the dynamics can be represented as a collection of discrete time series consisting of $0$ and $1$, where entries equal to $1$ mark the occurrence of significant events.

In this Letter, we introduce a general method to identify higher-order functional relations in discrete multivariate time series. The method does not require any assumption on the underlying dynamics and 
is specifically designed for, and naturally applies to, all the systems described above.
The method works in the following way: first, the original discrete multivariate time series is mapped into a sequence of symbols; then,  Bayesian statistics is used to infer the significant patterns of adjacent symbols and their higher-order organization. 
In this way, the approach combines the descriptive power of symbolic dynamics \cite{Beck_Schögl_1993, daw_review_2003} 
with the flexibility of Bayesian methods and the  ability of hypergraphs to model complex interactions.  
While the framework is presented here for event time series, it naturally extends to other discrete-valued sequences, since any finite-state process can be represented through a collection of binary variables.
More generally, it can also be applied to continuous-valued signals after suitable preprocessing procedures that map them into symbolic sequences \cite{lin_sax_2007,Schafer2012SFA,Kedem1986ZeroCrossing,Kennel2003SFNN,Hirata2004Shadowing,Aminikhanghahi2017Survey,Truong2020Selective}.
In addition, it can be applied directly to detect motifs in any ordered sequence of symbols, such as DNA sequences or textual data~\cite{Kandel1996, vanDerHeyden1998, Das2007, sinatra_motifs_2010, Hashim2019}.

\begin{figure}
    \centering
    \includegraphics[width=1\linewidth]{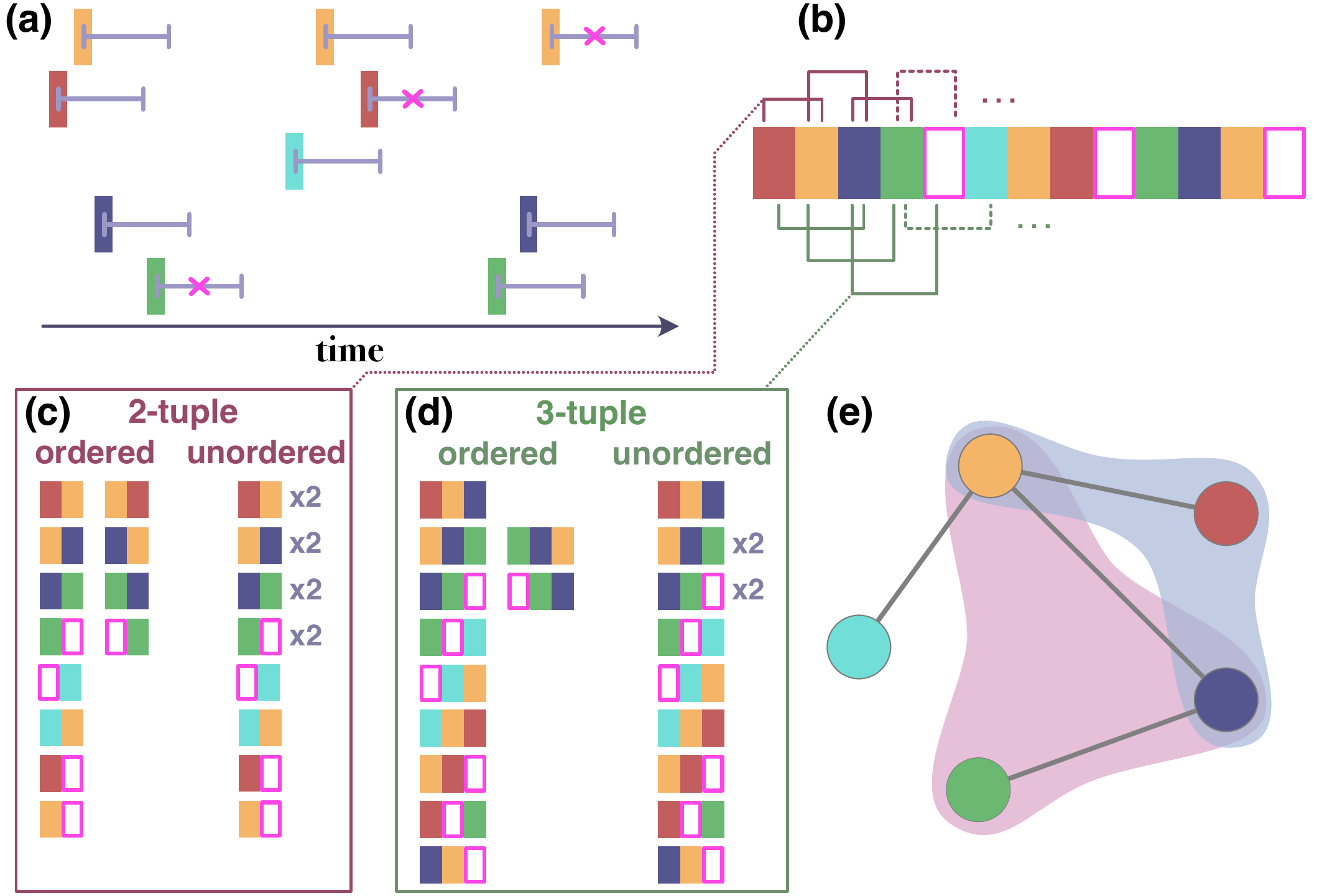}
    \caption{From event time series to higher-order functional networks. \textbf{(a)}~
    Each time series $x_i(t)$, with $i=1,\ldots N$, is associated to one of $N$ different symbols/colours. 
    \textbf{(b)} Then the set of time series is converted into a symbolic sequence: 
    if an event $x_i(t)=1$ is followed by $x_j(t') = 1$ within a time interval $\Delta t$ (grey bar in (a)), their corresponding symbols are placed adjacently in the symbolic sequence. 
    Otherwise (pink cross) a ``space" symbol is inserted in the sequence. \textbf{(c-d)}~Lists of ordered and unordered 
    $2$-tuples and $3$-tuples found in the sequence.   
\textbf{(e)} Statistically significant tuples (motifs) are finally the hyperedges of an hypergraph.
}
    \label{fig:scheme_model}
\end{figure}

\paragraph*{\bf Method.} 
Let us consider a multivariate time series of $N$ binary signals $x_i(t) \in \{0,1\}, \forall i=1,...,N$, with $t \in [0,T]$, where $x_i(t)=1$ represents the occurrence of event of type $i$ at time $t$, and $T$ is the observation time. We want to construct a hypergraph whose $N$ nodes are associated with the different types of events, and the hyperedges represent groups of correlated events. 
In order to do this, we first convert the multivariate time series into a symbolic sequence, Fig.~\ref{fig:scheme_model}(a,b). 
We consider an alphabet $\cal A$ 
of $N + 1$ symbols, with the first $N$ symbols (colours in Figure) 
respectively associated to each one of the $N$ time series, plus a special ``space" (or empty) symbol.  
We then construct an ordered sequence $\cal S$ of such symbols from the original set of $N$ time series in the following way. 
When the event $x_i(t) = 1$ is followed by the event $x_j(t') = 1$ within a given time interval $\Delta t$ (i.e., when $t'-t<\Delta t$), the symbols corresponding to $i$ and $j$ are placed in adjacent positions in the symbolic sequence $\cal S$. 
If instead no activity occurs within $\Delta t$, the space symbol is inserted in the sequence after $i$.
Notice that the parameter $\Delta t$ controls the temporal correlation, e.g. a larger $\Delta t$ allows events far away in time to be potentially adjacent.
Finally, from the symbolic sequence $\cal{S}$, we extract the multiset $\mathcal{M}^{(k)}$ of all overlapping $k$-tuples 
(allowing repetitions) 
with cardinality $|\mathcal{M}^{(k)}| = M^{(k)}$, 
and the set $\mathcal{U}^{(k)}$ of $k$-tuples (without repetitions), with cardinality $|\mathcal{U}^{(k)}| = U^{(k)}$, see Fig.~\ref{fig:scheme_model}(c,d).
A $k$-tuple $\boldsymbol{s} = (s_1, s_2, \dots, s_{k}) \in \mathcal{U}^{(k)}$  is an ordered (or unordered) sequence of $k$ adjacent symbols with $s_i \in \mathcal{A} \text{, for }  i=1,2,\ldots k$. 
In this way, from a sequence 
$\mathcal{S}$ of length $L$, we get $M^{(k)}= L - k + 1$ tuples of length $k$.

This approach allows to build a hypergraph $\mathcal{H}$ 
of the higher-order correlations in the original multivariate time-series. The $N$
nodes of the hypergraph correspond to the $N$ different types of events, while 
the $E$ hyperedges of sizes $k$ (with $k \ge 2$), representing groups of $k$ 
units, are defined by the
$k$-motifs in $\cal S$, i.e. by the statistically significant $k$-tuples of adjacent symbols. 
This is shown in  Fig.~\ref{fig:scheme_model}(e) for the case of $k=2$ and $3$.
As for any functional network, these hyperedges encode statistical dependencies, which may also reflect indirect effects or the temporal overlap of events, rather than direct physical interactions among the units.
To assess the statistical significance of a $k$-tuple $\boldsymbol{s}$, we test the null hypothesis ($H_0$) that the sequence is generated by a stationary Markov process of order $k-2$. Under this hypothesis, any correlation of length $k$ is simply a consequence of the correlations present at length $k-1$ and shorter.
For $k>2$, this means the probability of a symbol $s_{k}$ depends only on the preceding $k-2$ symbols, such that $p(s_{k} | s_1, \dots, s_{k -1}) = p(s_{k} | s_2, \dots, s_{k -1})$. The expected probability,  under $H_0$, is then:
\begin{align}
    p_{\text{exp}}(s_1, \dots, s_{k}) &= p_{\text{obs}}(s_1, \dots, s_{k -1}) p_{\text{obs}}(s_{k} | s_2, \dots, s_{k -1}) \nonumber \\
    &= \frac{p_{\text{obs}}(s_1, \dots, s_{k -1}) p_{\text{obs}}(s_2, \dots, s_{k})}{p_{\text{obs}}(s_2, \dots, s_{k -1})}
    \label{eq:exp_lengthk}
\end{align}
where the probabilities on the right-hand side are replaced by their maximum likelihood estimates from the data:
\begin{align}
    p_{\text{obs}}(\boldsymbol{s}) = \frac{n_{\text{obs}}(\boldsymbol{s})}{L - k + 1}
\end{align}
where $n_{\text{obs}}(\boldsymbol{s})$ is the number of times the tuple $\boldsymbol{s}$ appears in the sequence of length $L$.
For the case of $k=2$, the null hypothesis simplifies to a Markov process of order $0$, where symbols are independent:
\begin{align}
    p_{\text{exp}}(s_1, s_2) &= p_{\text{obs}}(s_1)p_{\text{obs}}(s_2) \label{eq:exp_length2}
\end{align}
A significant deviation of $p_{\text{obs}}(\boldsymbol{s})$ from $p_{\text{exp}}(\boldsymbol{s})$ provides evidence against the null hypothesis, indicating the presence of irreducible higher-order correlations~\cite{sinatra_motifs_2010}. 

If we are interested in the case of unordered tuples, we can simply sum over $S_p(\boldsymbol{s})$, the set of  unique permutations $\boldsymbol{\sigma}$ of the ordered tuple $\boldsymbol{s}=(s_1, \dots, s_{k})$, obtaining: $ p_{\text{exp}}^{\text{un}}(\boldsymbol{s}) = \sum_{\boldsymbol{\sigma} \in S_p(\boldsymbol{s})}p_{\text{exp}}(\boldsymbol{\sigma})$
and $n_{\text{obs}}^{\text{un}}(\boldsymbol{s}) = \sum_{\boldsymbol{\sigma} \in S_p(\boldsymbol{s})}n_{\text{obs}}(\boldsymbol{\sigma})$,
where $p_{\text{exp}}(\boldsymbol{\sigma})$ and $n_{\text{obs}}(\boldsymbol{\sigma})$ are respectively the expected probability and the observed occurrences of the permutation $\boldsymbol{\sigma}$ for the ordered case.
We denote with 
$|\text{supp}(p_{\text{exp}}(\boldsymbol{s}))| = |\mathcal{Q}^{(k)}|=Q^{(k)}$ the cardinality of the support $\mathcal{Q}^{(k)}$ of the expected probability given by Eqs.~\eqref{eq:exp_lengthk},\eqref{eq:exp_length2}. It is worth noticing that $U^{(k)} \leq Q^{(k)}$. 

We then use a Bayesian approach to reject or accept the null hypothesis that each $k$-tuple $\boldsymbol{s}_i \in \mathcal{Q}^{(k)}$ where $i \in 1,\dots, Q^{(k)}$ appears with probability $ p_{\text{exp}}(\boldsymbol{s}_i)$.
Let us define the vector of the observed count of $k$-tuples in $\mathcal{S}$: $\mathbf{n}^{(k)}_{\text{obs}} =  \left[n_{\text{obs}}(\boldsymbol{s}_1), \dots, n_{\text{obs}}(\boldsymbol{s}_{Q^{(k)}}) \right]$, 
where $n_{\text{obs}}(\boldsymbol{s}_i)$ is the number of times each tuple $\boldsymbol{s}_i \in \mathcal{Q}^{(k)}$ is repeated in $\mathcal{M}^{(k)}$.
We assume that each $k$-tuple represents the outcome of a random experiment over $L - k + 1$ independent trials.
That is, the outcome of each trial can be tuple $\boldsymbol{s}_i \in \mathcal{Q}^{(k)}$ with probability $p(\boldsymbol{s}_i)$, such that $\sum_i p(\boldsymbol{s}_i)=1$.
Under this assumption, the likelihood of 
$\mathbf{n}^{(k)}_{\text{obs}} = \left[ n_{\text{obs}}(\boldsymbol{s}_1), ..., n_{\text{obs}}(\boldsymbol{s}_{Q^{(k)}})\right]$ is a multinomial distribution:
\begin{align}
\mathcal{L}(\mathbf{n}^{(k)}_{\text{obs}} \mid \mathbf{p}^{(k)}) 
= \frac{(L - k + 1)!}{\prod_{i=1}^{Q^{(k)}} n_{\text{obs}}(\boldsymbol{s}_i)!}
\prod_{i=1}^{Q^{(k)}} p(\boldsymbol{s}_i)^{n_{\text{obs}}(\boldsymbol{s}_i)}
\end{align}
where $\mathbf{p}^{(k)} = \left[ p(\boldsymbol{s}_1),\dots,p(\boldsymbol{s}_{Q^{(k)}})\right]$ is the vector of the occurrence probability of each tuple. 
We note that the multinomial likelihood neglects the covariance between counts of overlapping tuples, although such dependencies are naturally present under the Markov-process null hypothesis used to derive $p_{\text{exp}}$. 
This approximation remains accurate when temporal correlations decay rapidly relative to the sequence length, as supported by the synthetic benchmarks on higher-order Markov processes in SM Sec.~VIII, in which correlations decay within a few symbols.

For the prior $\Pi(\mathbf{p}^{(k)})$, as commonly done,  we take the conjugate distribution of the multinomial, namely the Dirichlet distribution:
\begin{align}
    \Pi(\mathbf{p}^{(k)}) =& \text{Dirichlet}(\alpha'_1, \alpha'_2, \dots, \alpha'_{Q^{(k)}}) \nonumber
    \\
     =& \frac{\Gamma\left(\sum_{i=1}^{Q^{(k)}} \alpha'_i\right)}
{\prod_{i=1}^{Q^{(k)}} \Gamma(\alpha'_i)}
\prod_{i=1}^{Q^{(k)}} p(\boldsymbol{s}_i)^{\alpha'_i - 1}
\end{align}
where $\alpha'_i$ is the concentration parameter of the tuple $\boldsymbol{s}_{i}$, which can be interpreted as prior counts or pseudo-counts, and $\Gamma(\cdot)$ is the Gamma function.
We adopt an empirical Bayesian approach, where part of the information of the data is used to compute the prior.
Namely, to compute the concentration parameter of a $k$-tuple, we use the expected probability in Eqs.~\eqref{eq:exp_length2},\eqref{eq:exp_lengthk}: $\alpha'_i = n_{\text{exp}}(\boldsymbol{s}_i)$.
Here, $n_{\text{exp}}(\boldsymbol{s}_i) = p_{\text{exp}}(\boldsymbol{s}_i) (L - k + 1)$ is the expected number of occurrences of the $k$-tuple $\boldsymbol{s}_i$ in the sequence.
We then use the likelihood to update the prior, obtaining the posterior $P(\mathbf{p}^{(k)}\mid\text{Data})$ which 
is also a Dirichlet distribution:
\begin{align}
    P(\mathbf{p}^{(k)}\mid\text{Data}) = \text{Dirichlet}(&\alpha'_1 + n_{\text{obs}}(\boldsymbol{s}_1), \alpha'_2 + n_{\text{obs}}(\boldsymbol{s}_2), \nonumber
    \\
    &\dots, \alpha'_{Q^{(k)}} + n_{\text{obs}}(\boldsymbol{s}_{Q^{(k)}}))
\end{align}
To assess the significance of the $k$-tuple $\boldsymbol{s}_i$,  
we compute the Jensen-Shannon distance $ d_{\text{JS}}(\boldsymbol{s}_i) $
between the marginal distribution of the prior and that of the posterior\cite{lin_jensen_shannon_1991, abbott_LIGO_prior_posterior, pratten_binaries_2020}.
Notice that the marginal of the Dirichlet distribution 
$[p(\boldsymbol{s}_1),\dots,p(\boldsymbol{s}_{N})] \sim \text{Dirichlet}(\alpha_1, \alpha_2, \dots, \alpha_{N})$ is the Beta distribution:
\begin{equation}
    p(\boldsymbol{s}_i) \sim \text{Beta}(\alpha_i, \beta_i = \alpha_0 - \alpha_i) \label{eq:beta_general}
\end{equation}
where $\alpha_0=\sum_i \alpha_i$.
We recall that for the prior: 
\begin{equation}
    \alpha_i = n_{\text{exp}}(\boldsymbol{s}_i)=p_{\text{exp}}(\boldsymbol{s}_i)(L-k+1)
    \label{eq:alpha_prior}
\end{equation} 
and since $\sum_{i=1}^{Q^{(k)}}p_{\text{exp}}(\boldsymbol{s}_i)=1$, we have:
\begin{equation}
    \beta_i=\alpha_0-\alpha_i=(1-p_{\text{exp}}(\boldsymbol{s}_i))(L-k+1)
    \label{eq:beta_prior}
\end{equation}
For the posterior, instead: 
\begin{equation}
    \alpha_i = n_{\text{exp}}(\boldsymbol{s}_i) + n_{\text{obs}}(\boldsymbol{s}_i)
    \label{eq:alpha_posterior}
\end{equation} and given that $\sum_i n_{\text{obs}}(\boldsymbol{s}_i) = L-k+1$, we have:
\begin{equation}
    \beta_i = \alpha_0-\alpha_i=(2-p_{\text{exp}}(\boldsymbol{s}_i))(L-k+1)-n_{\text{obs}}(\boldsymbol{s}_i)
    \label{eq:beta_posterior}
\end{equation}
If $d_{\text{JS}}(\boldsymbol{s}_i) $ is large, the null  hypothesis that the probability of observing $\boldsymbol{s}_i$ can be estimated from the probabilities of the shorter tuples in $\boldsymbol{s}_i$ must be rejected. Hence, the $k$-tuple is considered a $k$-motif and, as such, is a $k$-hyperedge (a hyperedge with $k$ nodes) of the hypergraph $\mathcal{H}$. 
Our method naturally allows us to associate to each hyperedge a 
weight ranging from $0$ to $\sqrt{\ln(2)}$, the maximum value   of the 
Jensen--Shannon distance.

Note that, when $L - k + 1$ is small relative to the number of possible tuples $Q(k)$, or for heterogeneous symbol frequencies, the expected count $n_{\text{exp}} = p_{\text{exp}}(L - k + 1)$ can become very small. In such cases ($\alpha_i \ll 1$, $\beta_i \gg 1$), the marginal Beta distribution in Eq.~\eqref{eq:beta_general} diverges causing computational overflows and spurious significance. To prevent this, we apply an additive (Laplace) smoothing \cite{manning2008introduction} by adding $\varepsilon = 1$ to both parameters $\alpha_i$ and $\beta_i$ in Eqs.~\eqref{eq:alpha_prior}-\eqref{eq:beta_posterior}, effectively adding one pseudo-count to successes and failures.

Being $d_{\text{JS}}$ a metric, it does not take into account for the relative position of the prior and posterior distributions. We therefore associate to the hyperedge  
$\boldsymbol{s}_i$ a weight equal to
$d_{\text{JS}}(\boldsymbol{s}_i)$ only when the posterior 
is concentrated around larger values than the prior. 
That is, we consider as potentially significant only those tuples for which the mean of the prior is smaller than the mean of the posterior, but a similar analysis could be done also for under-represented tuples. An unweighted version of the hypergraph $\mathcal{H}$ can then be obtained by retaining only $k$-hyperedges whose weights are larger than a given threshold $d_{\text{JS}}^{\text{thr}}(k)$. 
In this way, the Jensen-Shannon distance between the prior and posterior distributions of each tuple becomes a score of statistical significance. 
Hereinafter we will indicate this score as the BJS-score (Bayesian-Jensen-Shannon) and we will refer to our method as SYMBA (SYMbolic BAyesian).
An algorithmic complexity analysis shows that SYMBA is asymptotically optimal in computational time and memory usage, as both scale 
linearly with the sequence length $L$ (see SM Sec.I).

\paragraph*{\bf Benchmarking on synthetic datasets.} 

We tested SYMBA on artificial symbolic sequences of various lengths, alphabet sizes,
tunable numbers of $2$-, $3$- and $4$-motifs, 
and different types and levels of noise (see SM Secs.~II-VII). 
The symbols in the noise 
are sampled with an arbitrary rank--frequency distribution $f(r) \sim r^{-\gamma}$, where $r$ is the rank of each symbol ordered according to its frequency $f$. 
For $\gamma = 1$, this reduces to the Zipf's law, describing, among other quantities, the abundance of letters and words in a text 
\cite{newman_zipf_2005, latora_complex_2017}, while for $\gamma =0 $ and $\gamma = -1$  we get the constant and the linear (degenerate) distributions,  respectively. 
For $\gamma>0$ we have different levels of heterogeneity, the larger $\gamma$ the more heterogenous being the frequency distribution.
We generated sequences of  length $L = L_{\text{s}} + L_{\text{n}}$, where $L_{\text{s}}$ and $L_{\text{n}}$ denote, respectively, the number of symbols in the signal (motifs) and in the noise, see SM Sec.~II. We notice that $L$ depends linearly on the noise-to-signal ratio $r_{\text{ns}} = L_{\text{n}} / L_{\text{s}}$. In particular, $L = (1 + r_{\text{ns}})L_{\text{s}}$, so varying $r_{\text{ns}}$ at fixed $L_{\text{s}}$ also benchmarks our method across different sequence lengths $L$. 
In SM Sec.~IV we identify analytically, in the case $\gamma = 0$, two relevant regimes as a function of $r_{\text{ns}}$, namely the \emph{data sparsity} and \emph{noise dominated} one. When data are sparse, increasing $r_{\text{ns}}$ can have a positive impact on the performance, as it increases the available statistics. 
However, as $r_{\text{ns}}$ increases further, the system enters the noise-dominated regime, where the significant tuples (the signal) are completely obscured by the noise.

\begin{figure}[htp!]
    \centering
      \includegraphics[width=1.\linewidth]{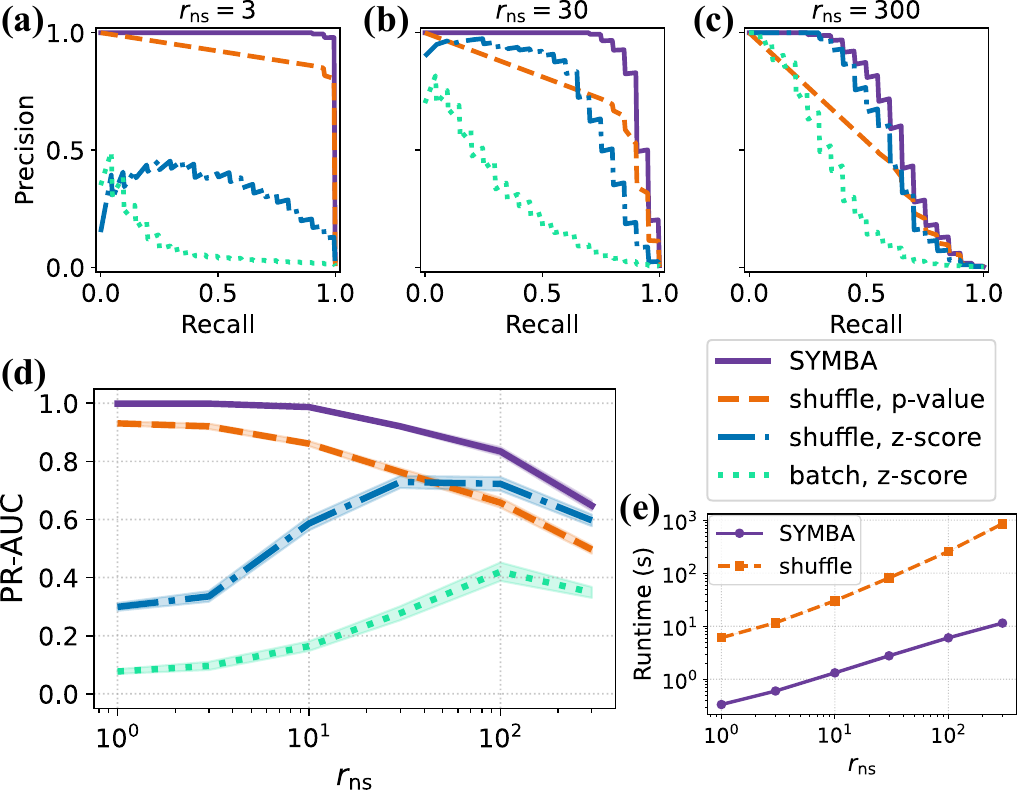}
    \caption{Performance of SYMBA compared to shuffle and batch methods for the identification of $3$-motifs. 
\textbf{(a--c)} Precision-Recall (PR) curves for three values of the noise-to-signal ratio ($r_{\text{ns}}=3, 30, 300$). 
\textbf{(d)}  Area Under the PR Curve (PR-AUC) as a function of $r_{\text{ns}}$. 
\textbf{(e)} Execution time as a function of $r_{\text{ns}}$ for SYMBA and the shuffle method with $1000$ reshuffles.
Artificial sequences were generated using an alphabet of $30$ symbols and contain $20$ distinct ordered $2$-, $3$-, and $4$-motifs, each repeated $10$ times. The rank--frequency distribution of the noise symbols follows a Zipf distribution ($\gamma=1$). Results are averaged over $20$ independent runs.}
    \label{fig:pr2and3motifs}
\end{figure}

In Fig.~\ref{fig:pr2and3motifs}, the performance of SYMBA in the identification of higher-order $3$-motifs as a function of $r_{\text{ns}}$ is compared to two alternative methods. 
In the first method, the expected probabilities in Eqs.~\eqref{eq:exp_lengthk},\eqref{eq:exp_length2} are used to compute a $z$-score by dividing the original sequence in $R$ sequential non-overlapping subsequences (batches).
Instead, in the second method, a canonical random shuffle preserving the number of tuples of order $k$, both a $z$-score and a $p$-value for the empirical probability of each tuple are computed using the sequence reshuffled $B$ times (see SM Secs.~X, XI) \cite{Kandel1996, propp1998how, jiang2008ushuffle}.
We generated artificial sequences from an alphabet of $30$ symbols. The signal consists of $20$ ordered $2$-, $3$- and $4$-motifs, each repeated $10$ times and embedded in a noise with a rank-frequency distribution with $\gamma=1$.
Figures~\ref{fig:pr2and3motifs}(a)-(c) report the Precision-Recall (PR) curves for the detection of ordered $3$-motifs for $r_{\text{ns}}=3, 30,$ and $300$. 
Moving from left to right along the PR curves corresponds to decreasing values of the inference threshold \cite{powers_evaluation_2011}. 
An ideal classifier maintains high precision across all thresholds, successfully identifying true motifs with minimal false positives even at high recall, thereby yielding an Area Under the PR Curve (PR-AUC) approaching $1$.
As shown by the PR curves, and by the PR-AUC  reported 
in Fig.~\ref{fig:pr2and3motifs}(d) against the noise level $r_{\text{ns}}$,  SYMBA consistently matches or outperforms the alternative methods.
Moreover, SYMBA provides a computational speedup of more than one order of magnitude, as shown in Fig.~\ref{fig:pr2and3motifs}(e).
Crucially, this computational advantage grows with the number of unique tuples in the sequence (e.g., for longer tuples or larger alphabets), since obtaining reliable empirical significance estimates 
in surrogate-data methods requires an increasingly large number of surrogate sequences.
The PR curves for other values of $\gamma$, as well as the extended analysis of ordered and unordered $2$-, $3$-, and $4$-motifs, can be found in SM Secs.~V and~VI, together with the corresponding Receiver Operating Characteristic (ROC) curves.
Additionally, benchmarks on synthetic sequences generated by higher-order Markov processes, testing SYMBA under the null-hypothesis used to derive $p_{\text{exp}}$ in Eqs.~\eqref{eq:exp_lengthk},\eqref{eq:exp_length2}, are reported in SM Sec.~VIII.

\paragraph*{\bf Applications to real-world datasets.}

As three study cases, we considered multivariate time series from neuronal activities \cite{allen_whitepaper}, fluctuations of stock prices in financial markets \cite{yahoofinance, yfinance, Bonannostocks2001} and email exchanges \cite{klimt_enron_2004}.
For these applications, we consider unordered tuples.
\begin{table}
\begin{tabular}{ |p{4.5cm}||p{1cm}|p{1cm}|p{1cm}| }
\hline
 \bf{Dataset}  & $N$  & $E_2$ & $E_3$  \\
 \hline
 \hline
  Brain activity (micro-scale)       &  346   & 4523   &   1441 \\
  Brain activity (macro-scale)       &  15   & 42   &   130 \\
  Stock price changes (unsigned)        &   24   & 16   &   4 \\
  Stock price changes (signed)        &   48   & 110   &   4 \\
  Email exchanges         &  143   & 131   &   5 \\
  \hline
\end{tabular} 
\caption{Basic properties of the unweighted hypergraphs obtained from multivariate time-series, respectively of  single-cell neuronal activity \cite{allen_whitepaper}, financial stocks \cite{Bonannostocks2001}, and email exchanges \cite{klimt_enron_2004}. %
}  \label{tab:datasets}
\end{table}
In Table~\ref{tab:datasets} we report the number of nodes and the number of undirected $2$- and $3$-hyperedges of the (unweighted) hypergraphs obtained using a threshold 
$\text{BJS}^{\text{thr}} = 0.6$ (see SM Secs.~IX,XII).

For the application to the brain, we studied the spiking activity at the level of individual neurons in the mouse during visual tasks. The multivariate time-series were 
obtained from the electrophysiological recordings in the Neuropixel datasets of the Allen Brain Observatory \cite{allen_whitepaper}. 
We set $x_i(t)=1$ to indicate that an action potential occurred in the neuron $i$ at time $t$. 
We considered two different spatial scales, one that involves individual neurons ($N=346$), and the other that aggregates neurons in the same functional area ($N=15$), see SM Sec.~XII. 
Higher-order motifs were present in both situations. However, only at the macro-scale of functional brain areas the number $E_3$ of  $3$-motifs exceeds the number $E_2$ of pairwise interactions (more than $70\%$ of the total number of hyperedges).
These results are in agreement with a recent study of large-scale electroencephalographic (EEG) brain signals \cite{delabays_hyper_recons_2025}, where a similar percentage of brain dynamics was considered higher-order. 
The significant increase of 3-motifs when moving from the microscale of individual neurons to the macroscale of functional brain areas reveals a non-negligible spatial effect that should be considered in future studies.

As a second application, we considered time series of closure prices of stocks from NASDAQ Stock Market and New York Stock Exchange over $30$ years (1994-2024) \cite{yahoofinance, yfinance}. In particular, we selected $24$ stocks, namely $3$ stocks from each of $8$ different categories: technology, chemicals, banking, entertainment, food and beverage, pharmaceutical, energy, consumer goods (see SM Sec.~XII).
We first converted the data into a binary time series by setting  $x_i(t) = 1$ for stock $i$ at day $t$ if the variation (in absolute value) in closure price of $i$  exceeds $2\%$ of the previous closure price, otherwise $x_i(t)=0$. 
We also considered the signed case, by assigning to each stock two different symbols, one associated to a positive variation and one to a negative variation.
For the case of unsigned price changes 
we found that $\sim 85\%$ of the pairwise motifs involve stocks in the same category. 
Regarding the $3$-motifs, our method pointed out (BAC, C, JPM), with the $3$ banking stocks, and (COP, CVX, XOM), with the $3$ energy stocks, 
as the most significant  motifs, both with a BJS-score of $\sim 0.75$. 
Interestingly, in the case of signed changes, all the 2-motifs are among stock price changes of the same sign ($54$ among symbols with positive and $56$ with negative variations).

Finally, we constructed binary time-series from Enron's dataset of emails internally sent by the company's employees  \cite{klimt_enron_2004}. 
We set $x_i(t) = 1$ at time $t$ when employee $i$ sends an email. Although in this case we only found five 
3-motifs, a centrality analysis of the resulting unweighted hypergraph consistently highlights 
key figures in the company as the nodes with the highest centrality (see SM Sec.~XII).  
Our findings, based on correlations in employee activity, complement those from \cite{tang_enron_2013}, which define the network through direct email sender–receiver interactions.  

\paragraph*{\bf Conclusions.} 
In this Letter we introduced a method to identify higher-order functional dependencies among the components of a complex system from empirical time series of their activities.
The method is based on comparing the empirical data to a semi-analytical null model. Our approach offers a significant advantage over alternative surrogate data testing methods, such as random shuffling or computationally expensive algorithms used to generate null models from the data \cite{theiler_surrogate_1992}. 
Because our method evaluates statistics only for tuples that actually occur in the data sequence, it scales optimally with system size, with computational demands rising linearly with sequence length.
In this way our method enables the higher-order analysis of data from large systems, that is currently not possible with existing approaches.

\paragraph*{Code availability.} The code implementing SYMBA, together with the notebooks that reproduce the analyses of this Letter, will be made publicly available at \url{https://github.com/Ancivi/symba} upon publication.

\paragraph*{Acknowledgements.} A.C. and V.L. acknowledge support from the European Union - NextGenerationEU, GRINS project (Grant E63-C22-0021-20006). A.C and F.d.V.F. acknowledge support from the Agence
Nationale de la Recherche under the France 2030 program
(Grant ANR-23-IACL-0008).

\bibliographystyle{apsrev4-2} 
\bibliography{bibliography}

\begin{thebibliography}{79}%
\makeatletter
\providecommand \@ifxundefined [1]{%
 \@ifx{#1\undefined}
}%
\providecommand \@ifnum [1]{%
 \ifnum #1\expandafter \@firstoftwo
 \else \expandafter \@secondoftwo
 \fi
}%
\providecommand \@ifx [1]{%
 \ifx #1\expandafter \@firstoftwo
 \else \expandafter \@secondoftwo
 \fi
}%
\providecommand \natexlab [1]{#1}%
\providecommand \enquote  [1]{``#1''}%
\providecommand \bibnamefont  [1]{#1}%
\providecommand \bibfnamefont [1]{#1}%
\providecommand \citenamefont [1]{#1}%
\providecommand \href@noop [0]{\@secondoftwo}%
\providecommand \href [0]{\begingroup \@sanitize@url \@href}%
\providecommand \@href[1]{\@@startlink{#1}\@@href}%
\providecommand \@@href[1]{\endgroup#1\@@endlink}%
\providecommand \@sanitize@url [0]{\catcode `\\12\catcode `\$12\catcode `\&12\catcode `\#12\catcode `\^12\catcode `\_12\catcode `\%12\relax}%
\providecommand \@@startlink[1]{}%
\providecommand \@@endlink[0]{}%
\providecommand \url  [0]{\begingroup\@sanitize@url \@url }%
\providecommand \@url [1]{\endgroup\@href {#1}{\urlprefix }}%
\providecommand \urlprefix  [0]{URL }%
\providecommand \Eprint [0]{\href }%
\providecommand \doibase [0]{https://doi.org/}%
\providecommand \selectlanguage [0]{\@gobble}%
\providecommand \bibinfo  [0]{\@secondoftwo}%
\providecommand \bibfield  [0]{\@secondoftwo}%
\providecommand \translation [1]{[#1]}%
\providecommand \BibitemOpen [0]{}%
\providecommand \bibitemStop [0]{}%
\providecommand \bibitemNoStop [0]{.\EOS\space}%
\providecommand \EOS [0]{\spacefactor3000\relax}%
\providecommand \BibitemShut  [1]{\csname bibitem#1\endcsname}%
\let\auto@bib@innerbib\@empty
\bibitem [{\citenamefont {Strogatz}\ and\ \citenamefont {Stewart}(1993)}]{strogatz1993coupled}%
  \BibitemOpen
  \bibfield  {author} {\bibinfo {author} {\bibfnamefont {S.}~\bibnamefont {Strogatz}}\ and\ \bibinfo {author} {\bibfnamefont {I.}~\bibnamefont {Stewart}},\ }\href {https://doi.org/10.1038/scientificamerican1293-102} {\bibfield  {journal} {\bibinfo  {journal} {Scientific American}\ }\textbf {\bibinfo {volume} {269}},\ \bibinfo {pages} {102} (\bibinfo {year} {1993})}\BibitemShut {NoStop}%
\bibitem [{\citenamefont {Dehaene}\ \emph {et~al.}(1998)\citenamefont {Dehaene}, \citenamefont {Kerszberg},\ and\ \citenamefont {Changeux}}]{dehaene_cognitive_1998}%
  \BibitemOpen
  \bibfield  {author} {\bibinfo {author} {\bibfnamefont {S.}~\bibnamefont {Dehaene}}, \bibinfo {author} {\bibfnamefont {M.}~\bibnamefont {Kerszberg}},\ and\ \bibinfo {author} {\bibfnamefont {J.-P.}\ \bibnamefont {Changeux}},\ }\href {https://doi.org/10.1073/pnas.95.24.14529} {\bibfield  {journal} {\bibinfo  {journal} {Proceedings of the National Academy of Sciences}\ }\textbf {\bibinfo {volume} {95}},\ \bibinfo {pages} {14529} (\bibinfo {year} {1998})}\BibitemShut {NoStop}%
\bibitem [{\citenamefont {Axelrod}\ and\ \citenamefont {Hamilton}(1981)}]{axelrod_evolution_1981}%
  \BibitemOpen
  \bibfield  {author} {\bibinfo {author} {\bibfnamefont {R.}~\bibnamefont {Axelrod}}\ and\ \bibinfo {author} {\bibfnamefont {W.~D.}\ \bibnamefont {Hamilton}},\ }\href {https://doi.org/10.1126/science.7466396} {\bibfield  {journal} {\bibinfo  {journal} {Science}\ }\textbf {\bibinfo {volume} {211}},\ \bibinfo {pages} {1390} (\bibinfo {year} {1981})}\BibitemShut {NoStop}%
\bibitem [{\citenamefont {Nowak}\ and\ \citenamefont {Highfield}(2012)}]{nowak_supercooperators_2012}%
  \BibitemOpen
  \bibfield  {author} {\bibinfo {author} {\bibfnamefont {M.~A.}\ \bibnamefont {Nowak}}\ and\ \bibinfo {author} {\bibfnamefont {R.}~\bibnamefont {Highfield}},\ }\href@noop {} {\emph {\bibinfo {title} {{SuperCooperators}: altruism, evolution, and why we need each other to succeed}}},\ \bibinfo {edition} {1st}\ ed.\ (\bibinfo  {publisher} {Free Press},\ \bibinfo {address} {New York, NY},\ \bibinfo {year} {2012})\BibitemShut {NoStop}%
\bibitem [{\citenamefont {Pastor-Satorras}\ \emph {et~al.}(2015)\citenamefont {Pastor-Satorras}, \citenamefont {Castellano}, \citenamefont {Van~Mieghem},\ and\ \citenamefont {Vespignani}}]{pastor_epidemic_review_2015}%
  \BibitemOpen
  \bibfield  {author} {\bibinfo {author} {\bibfnamefont {R.}~\bibnamefont {Pastor-Satorras}}, \bibinfo {author} {\bibfnamefont {C.}~\bibnamefont {Castellano}}, \bibinfo {author} {\bibfnamefont {P.}~\bibnamefont {Van~Mieghem}},\ and\ \bibinfo {author} {\bibfnamefont {A.}~\bibnamefont {Vespignani}},\ }\href {https://doi.org/10.1103/RevModPhys.87.925} {\bibfield  {journal} {\bibinfo  {journal} {Rev. Mod. Phys.}\ }\textbf {\bibinfo {volume} {87}},\ \bibinfo {pages} {925} (\bibinfo {year} {2015})}\BibitemShut {NoStop}%
\bibitem [{\citenamefont {Strogatz}(2001)}]{strogatz_exploring_2001}%
  \BibitemOpen
  \bibfield  {author} {\bibinfo {author} {\bibfnamefont {S.~H.}\ \bibnamefont {Strogatz}},\ }\href {https://doi.org/10.1038/35065725} {\bibfield  {journal} {\bibinfo  {journal} {Nature}\ }\textbf {\bibinfo {volume} {410}},\ \bibinfo {pages} {268} (\bibinfo {year} {2001})}\BibitemShut {NoStop}%
\bibitem [{\citenamefont {Boccaletti}\ \emph {et~al.}(2006)\citenamefont {Boccaletti}, \citenamefont {Latora}, \citenamefont {Moreno}, \citenamefont {Chavez},\ and\ \citenamefont {Hwang}}]{boccaletti_complex_2006}%
  \BibitemOpen
  \bibfield  {author} {\bibinfo {author} {\bibfnamefont {S.}~\bibnamefont {Boccaletti}}, \bibinfo {author} {\bibfnamefont {V.}~\bibnamefont {Latora}}, \bibinfo {author} {\bibfnamefont {Y.}~\bibnamefont {Moreno}}, \bibinfo {author} {\bibfnamefont {M.}~\bibnamefont {Chavez}},\ and\ \bibinfo {author} {\bibfnamefont {D.}~\bibnamefont {Hwang}},\ }\href {https://doi.org/10.1016/j.physrep.2005.10.009} {\bibfield  {journal} {\bibinfo  {journal} {Phys. Rep.}\ }\textbf {\bibinfo {volume} {424}},\ \bibinfo {pages} {175} (\bibinfo {year} {2006})}\BibitemShut {NoStop}%
\bibitem [{\citenamefont {De~Vico~Fallani}\ \emph {et~al.}(2014)\citenamefont {De~Vico~Fallani}, \citenamefont {Richiardi}, \citenamefont {Chavez},\ and\ \citenamefont {Achard}}]{fallani_graphbrain_2014}%
  \BibitemOpen
  \bibfield  {author} {\bibinfo {author} {\bibfnamefont {F.}~\bibnamefont {De~Vico~Fallani}}, \bibinfo {author} {\bibfnamefont {J.}~\bibnamefont {Richiardi}}, \bibinfo {author} {\bibfnamefont {M.}~\bibnamefont {Chavez}},\ and\ \bibinfo {author} {\bibfnamefont {S.}~\bibnamefont {Achard}},\ }\href {https://doi.org/10.1098/rstb.2013.0521} {\bibfield  {journal} {\bibinfo  {journal} {Philosophical Transactions of the Royal Society B: Biological Sciences}\ }\textbf {\bibinfo {volume} {369}},\ \bibinfo {pages} {20130521} (\bibinfo {year} {2014})}\BibitemShut {NoStop}%
\bibitem [{\citenamefont {Chiarion}\ \emph {et~al.}(2023)\citenamefont {Chiarion}, \citenamefont {Sparacino}, \citenamefont {Antonacci}, \citenamefont {Faes},\ and\ \citenamefont {Mesin}}]{chiaron_EEG_2023}%
  \BibitemOpen
  \bibfield  {author} {\bibinfo {author} {\bibfnamefont {G.}~\bibnamefont {Chiarion}}, \bibinfo {author} {\bibfnamefont {L.}~\bibnamefont {Sparacino}}, \bibinfo {author} {\bibfnamefont {Y.}~\bibnamefont {Antonacci}}, \bibinfo {author} {\bibfnamefont {L.}~\bibnamefont {Faes}},\ and\ \bibinfo {author} {\bibfnamefont {L.}~\bibnamefont {Mesin}},\ }\bibfield  {journal} {\bibinfo  {journal} {Bioengineering}\ }\textbf {\bibinfo {volume} {10}},\ \href {https://doi.org/10.3390/bioengineering10030372} {10.3390/bioengineering10030372} (\bibinfo {year} {2023})\BibitemShut {NoStop}%
\bibitem [{\citenamefont {Newman}(2010)}]{newman_networks_2010}%
  \BibitemOpen
  \bibfield  {author} {\bibinfo {author} {\bibfnamefont {M.}~\bibnamefont {Newman}},\ }\href {https://doi.org/10.1093/acprof:oso/9780199206650.001.0001} {\emph {\bibinfo {title} {Networks}}}\ (\bibinfo  {publisher} {Oxford University Press},\ \bibinfo {year} {2010})\BibitemShut {NoStop}%
\bibitem [{\citenamefont {Latora}\ \emph {et~al.}(2017)\citenamefont {Latora}, \citenamefont {Nicosia},\ and\ \citenamefont {Russo}}]{latora_complex_2017}%
  \BibitemOpen
  \bibfield  {author} {\bibinfo {author} {\bibfnamefont {V.}~\bibnamefont {Latora}}, \bibinfo {author} {\bibfnamefont {V.}~\bibnamefont {Nicosia}},\ and\ \bibinfo {author} {\bibfnamefont {G.}~\bibnamefont {Russo}},\ }\href@noop {} {\emph {\bibinfo {title} {Complex networks: principles, methods and applications}}},\ \bibinfo {edition} {1st}\ ed.\ (\bibinfo  {publisher} {Cambridge University Press},\ \bibinfo {year} {2017})\BibitemShut {NoStop}%
\bibitem [{\citenamefont {Cimini}\ \emph {et~al.}(2019)\citenamefont {Cimini}, \citenamefont {Squartini}, \citenamefont {Saracco}, \citenamefont {Garlaschelli}, \citenamefont {Gabrielli},\ and\ \citenamefont {Caldarelli}}]{cimini_statistical_2019}%
  \BibitemOpen
  \bibfield  {author} {\bibinfo {author} {\bibfnamefont {G.}~\bibnamefont {Cimini}}, \bibinfo {author} {\bibfnamefont {T.}~\bibnamefont {Squartini}}, \bibinfo {author} {\bibfnamefont {F.}~\bibnamefont {Saracco}}, \bibinfo {author} {\bibfnamefont {D.}~\bibnamefont {Garlaschelli}}, \bibinfo {author} {\bibfnamefont {A.}~\bibnamefont {Gabrielli}},\ and\ \bibinfo {author} {\bibfnamefont {G.}~\bibnamefont {Caldarelli}},\ }\href {https://doi.org/10.1038/s42254-018-0002-6} {\bibfield  {journal} {\bibinfo  {journal} {Nature Reviews Physics}\ }\textbf {\bibinfo {volume} {1}},\ \bibinfo {pages} {58} (\bibinfo {year} {2019})}\BibitemShut {NoStop}%
\bibitem [{\citenamefont {Martignon}\ \emph {et~al.}(1995)\citenamefont {Martignon}, \citenamefont {von Hasseln}, \citenamefont {Gr\"un}, \citenamefont {Aertsen},\ and\ \citenamefont {Palm}}]{Martignon1995}%
  \BibitemOpen
  \bibfield  {author} {\bibinfo {author} {\bibfnamefont {L.}~\bibnamefont {Martignon}}, \bibinfo {author} {\bibfnamefont {H.}~\bibnamefont {von Hasseln}}, \bibinfo {author} {\bibfnamefont {S.}~\bibnamefont {Gr\"un}}, \bibinfo {author} {\bibfnamefont {A.}~\bibnamefont {Aertsen}},\ and\ \bibinfo {author} {\bibfnamefont {G.}~\bibnamefont {Palm}},\ }\href {https://doi.org/10.1007/BF00199057} {\bibfield  {journal} {\bibinfo  {journal} {Biological Cybernetics}\ }\textbf {\bibinfo {volume} {73}},\ \bibinfo {pages} {69} (\bibinfo {year} {1995})}\BibitemShut {NoStop}%
\bibitem [{\citenamefont {Lee}\ and\ \citenamefont {Wilson}(2004)}]{LeeWilson2004}%
  \BibitemOpen
  \bibfield  {author} {\bibinfo {author} {\bibfnamefont {A.~K.}\ \bibnamefont {Lee}}\ and\ \bibinfo {author} {\bibfnamefont {M.~A.}\ \bibnamefont {Wilson}},\ }\href {https://doi.org/10.1152/jn.01030.2003} {\bibfield  {journal} {\bibinfo  {journal} {Journal of Neurophysiology}\ }\textbf {\bibinfo {volume} {92}},\ \bibinfo {pages} {2555} (\bibinfo {year} {2004})}\BibitemShut {NoStop}%
\bibitem [{\citenamefont {Diekman}\ \emph {et~al.}(2009)\citenamefont {Diekman}, \citenamefont {Sastry},\ and\ \citenamefont {Unnikrishnan}}]{Diekman2009}%
  \BibitemOpen
  \bibfield  {author} {\bibinfo {author} {\bibfnamefont {C.~O.}\ \bibnamefont {Diekman}}, \bibinfo {author} {\bibfnamefont {P.~S.}\ \bibnamefont {Sastry}},\ and\ \bibinfo {author} {\bibfnamefont {K.~P.}\ \bibnamefont {Unnikrishnan}},\ }\href {https://doi.org/10.1016/j.jneumeth.2009.06.018} {\bibfield  {journal} {\bibinfo  {journal} {Journal of Neuroscience Methods}\ }\textbf {\bibinfo {volume} {182}},\ \bibinfo {pages} {279} (\bibinfo {year} {2009})}\BibitemShut {NoStop}%
\bibitem [{\citenamefont {Shimazaki}\ \emph {et~al.}(2012)\citenamefont {Shimazaki}, \citenamefont {Amari}, \citenamefont {Brown},\ and\ \citenamefont {Gr\"un}}]{Shimazaki2012}%
  \BibitemOpen
  \bibfield  {author} {\bibinfo {author} {\bibfnamefont {H.}~\bibnamefont {Shimazaki}}, \bibinfo {author} {\bibfnamefont {S.}~\bibnamefont {Amari}}, \bibinfo {author} {\bibfnamefont {E.~N.}\ \bibnamefont {Brown}},\ and\ \bibinfo {author} {\bibfnamefont {S.}~\bibnamefont {Gr\"un}},\ }\href {https://doi.org/10.1371/journal.pcbi.1002385} {\bibfield  {journal} {\bibinfo  {journal} {PLoS Computational Biology}\ }\textbf {\bibinfo {volume} {8}},\ \bibinfo {pages} {e1002385} (\bibinfo {year} {2012})}\BibitemShut {NoStop}%
\bibitem [{\citenamefont {Robiglio}\ \emph {et~al.}(2025)\citenamefont {Robiglio}, \citenamefont {Neri}, \citenamefont {Coppes}, \citenamefont {Agostinelli}, \citenamefont {Battiston}, \citenamefont {Lucas},\ and\ \citenamefont {Petri}}]{robiglio_synergistic_ho_2025}%
  \BibitemOpen
  \bibfield  {author} {\bibinfo {author} {\bibfnamefont {T.}~\bibnamefont {Robiglio}}, \bibinfo {author} {\bibfnamefont {M.}~\bibnamefont {Neri}}, \bibinfo {author} {\bibfnamefont {D.}~\bibnamefont {Coppes}}, \bibinfo {author} {\bibfnamefont {C.}~\bibnamefont {Agostinelli}}, \bibinfo {author} {\bibfnamefont {F.}~\bibnamefont {Battiston}}, \bibinfo {author} {\bibfnamefont {M.}~\bibnamefont {Lucas}},\ and\ \bibinfo {author} {\bibfnamefont {G.}~\bibnamefont {Petri}},\ }\href {https://doi.org/10.1103/PhysRevLett.134.137401} {\bibfield  {journal} {\bibinfo  {journal} {Phys. Rev. Lett.}\ }\textbf {\bibinfo {volume} {134}},\ \bibinfo {pages} {137401} (\bibinfo {year} {2025})}\BibitemShut {NoStop}%
\bibitem [{\citenamefont {Battiston}\ \emph {et~al.}(2020)\citenamefont {Battiston}, \citenamefont {Cencetti}, \citenamefont {Iacopini}, \citenamefont {Latora}, \citenamefont {Lucas}, \citenamefont {Patania}, \citenamefont {Young},\ and\ \citenamefont {Petri}}]{battiston_networks_2020}%
  \BibitemOpen
  \bibfield  {author} {\bibinfo {author} {\bibfnamefont {F.}~\bibnamefont {Battiston}}, \bibinfo {author} {\bibfnamefont {G.}~\bibnamefont {Cencetti}}, \bibinfo {author} {\bibfnamefont {I.}~\bibnamefont {Iacopini}}, \bibinfo {author} {\bibfnamefont {V.}~\bibnamefont {Latora}}, \bibinfo {author} {\bibfnamefont {M.}~\bibnamefont {Lucas}}, \bibinfo {author} {\bibfnamefont {A.}~\bibnamefont {Patania}}, \bibinfo {author} {\bibfnamefont {J.-G.}\ \bibnamefont {Young}},\ and\ \bibinfo {author} {\bibfnamefont {G.}~\bibnamefont {Petri}},\ }\href {https://doi.org/10.1016/j.physrep.2020.05.004} {\bibfield  {journal} {\bibinfo  {journal} {Phys. Rep.}\ }\textbf {\bibinfo {volume} {874}},\ \bibinfo {pages} {1} (\bibinfo {year} {2020})}\BibitemShut {NoStop}%
\bibitem [{\citenamefont {Battiston}\ \emph {et~al.}(2021)\citenamefont {Battiston}, \citenamefont {Amico}, \citenamefont {Barrat}, \citenamefont {Bianconi}, \citenamefont {Ferraz~de Arruda}, \citenamefont {Franceschiello}, \citenamefont {Iacopini}, \citenamefont {Kéfi}, \citenamefont {Latora}, \citenamefont {Moreno}, \citenamefont {Murray}, \citenamefont {Peixoto}, \citenamefont {Vaccarino},\ and\ \citenamefont {Petri}}]{battiston_physics_2021}%
  \BibitemOpen
  \bibfield  {author} {\bibinfo {author} {\bibfnamefont {F.}~\bibnamefont {Battiston}}, \bibinfo {author} {\bibfnamefont {E.}~\bibnamefont {Amico}}, \bibinfo {author} {\bibfnamefont {A.}~\bibnamefont {Barrat}}, \bibinfo {author} {\bibfnamefont {G.}~\bibnamefont {Bianconi}}, \bibinfo {author} {\bibfnamefont {G.}~\bibnamefont {Ferraz~de Arruda}}, \bibinfo {author} {\bibfnamefont {B.}~\bibnamefont {Franceschiello}}, \bibinfo {author} {\bibfnamefont {I.}~\bibnamefont {Iacopini}}, \bibinfo {author} {\bibfnamefont {S.}~\bibnamefont {Kéfi}}, \bibinfo {author} {\bibfnamefont {V.}~\bibnamefont {Latora}}, \bibinfo {author} {\bibfnamefont {Y.}~\bibnamefont {Moreno}}, \bibinfo {author} {\bibfnamefont {M.~M.}\ \bibnamefont {Murray}}, \bibinfo {author} {\bibfnamefont {T.~P.}\ \bibnamefont {Peixoto}}, \bibinfo {author} {\bibfnamefont {F.}~\bibnamefont {Vaccarino}},\ and\ \bibinfo {author} {\bibfnamefont {G.}~\bibnamefont {Petri}},\ }\href {https://doi.org/10.1038/s41567-021-01371-4} {\bibfield  {journal} {\bibinfo
  {journal} {Nat. Phys.}\ }\textbf {\bibinfo {volume} {17}},\ \bibinfo {pages} {1093} (\bibinfo {year} {2021})}\BibitemShut {NoStop}%
\bibitem [{\citenamefont {Iacopini}\ \emph {et~al.}(2019)\citenamefont {Iacopini}, \citenamefont {Petri}, \citenamefont {Barrat},\ and\ \citenamefont {Latora}}]{iacopini_simplicial_2019}%
  \BibitemOpen
  \bibfield  {author} {\bibinfo {author} {\bibfnamefont {I.}~\bibnamefont {Iacopini}}, \bibinfo {author} {\bibfnamefont {G.}~\bibnamefont {Petri}}, \bibinfo {author} {\bibfnamefont {A.}~\bibnamefont {Barrat}},\ and\ \bibinfo {author} {\bibfnamefont {V.}~\bibnamefont {Latora}},\ }\href {https://doi.org/10.1038/s41467-019-10431-6} {\bibfield  {journal} {\bibinfo  {journal} {Nat. Commun.}\ }\textbf {\bibinfo {volume} {10}},\ \bibinfo {pages} {2485} (\bibinfo {year} {2019})}\BibitemShut {NoStop}%
\bibitem [{\citenamefont {Alvarez-Rodriguez}\ \emph {et~al.}(2021)\citenamefont {Alvarez-Rodriguez}, \citenamefont {Battiston}, \citenamefont {de~Arruda}, \citenamefont {Moreno}, \citenamefont {Perc},\ and\ \citenamefont {Latora}}]{alvarez_pgg_ho_2021}%
  \BibitemOpen
  \bibfield  {author} {\bibinfo {author} {\bibfnamefont {U.}~\bibnamefont {Alvarez-Rodriguez}}, \bibinfo {author} {\bibfnamefont {F.}~\bibnamefont {Battiston}}, \bibinfo {author} {\bibfnamefont {G.~F.}\ \bibnamefont {de~Arruda}}, \bibinfo {author} {\bibfnamefont {Y.}~\bibnamefont {Moreno}}, \bibinfo {author} {\bibfnamefont {M.}~\bibnamefont {Perc}},\ and\ \bibinfo {author} {\bibfnamefont {V.}~\bibnamefont {Latora}},\ }\href {https://doi.org/10.1038/s41562-020-01024-1} {\bibfield  {journal} {\bibinfo  {journal} {Nat. Hum. Behav}\ }\textbf {\bibinfo {volume} {5}},\ \bibinfo {pages} {586} (\bibinfo {year} {2021})}\BibitemShut {NoStop}%
\bibitem [{\citenamefont {Civilini}\ \emph {et~al.}(2024)\citenamefont {Civilini}, \citenamefont {Sadekar}, \citenamefont {Battiston}, \citenamefont {G\'omez-Garde\~nes},\ and\ \citenamefont {Latora}}]{civilini2024explosive}%
  \BibitemOpen
  \bibfield  {author} {\bibinfo {author} {\bibfnamefont {A.}~\bibnamefont {Civilini}}, \bibinfo {author} {\bibfnamefont {O.}~\bibnamefont {Sadekar}}, \bibinfo {author} {\bibfnamefont {F.}~\bibnamefont {Battiston}}, \bibinfo {author} {\bibfnamefont {J.}~\bibnamefont {G\'omez-Garde\~nes}},\ and\ \bibinfo {author} {\bibfnamefont {V.}~\bibnamefont {Latora}},\ }\href {https://doi.org/10.1103/PhysRevLett.132.167401} {\bibfield  {journal} {\bibinfo  {journal} {Phys. Rev. Lett.}\ }\textbf {\bibinfo {volume} {132}},\ \bibinfo {pages} {167401} (\bibinfo {year} {2024})}\BibitemShut {NoStop}%
\bibitem [{\citenamefont {Yu}\ \emph {et~al.}(2011)\citenamefont {Yu}, \citenamefont {Yang}, \citenamefont {Nakahara}, \citenamefont {Santos}, \citenamefont {Nikoli{\'c}},\ and\ \citenamefont {Plenz}}]{yu_ho_cortical_2011}%
  \BibitemOpen
  \bibfield  {author} {\bibinfo {author} {\bibfnamefont {S.}~\bibnamefont {Yu}}, \bibinfo {author} {\bibfnamefont {H.}~\bibnamefont {Yang}}, \bibinfo {author} {\bibfnamefont {H.}~\bibnamefont {Nakahara}}, \bibinfo {author} {\bibfnamefont {G.~S.}\ \bibnamefont {Santos}}, \bibinfo {author} {\bibfnamefont {D.}~\bibnamefont {Nikoli{\'c}}},\ and\ \bibinfo {author} {\bibfnamefont {D.}~\bibnamefont {Plenz}},\ }\href {https://doi.org/10.1523/JNEUROSCI.3127-11.2011} {\bibfield  {journal} {\bibinfo  {journal} {Journal of Neuroscience}\ }\textbf {\bibinfo {volume} {31}},\ \bibinfo {pages} {17514} (\bibinfo {year} {2011})}\BibitemShut {NoStop}%
\bibitem [{\citenamefont {Petri}\ \emph {et~al.}(2014)\citenamefont {Petri}, \citenamefont {Expert}, \citenamefont {Turkheimer}, \citenamefont {Carhart-Harris}, \citenamefont {Nutt}, \citenamefont {Hellyer},\ and\ \citenamefont {Vaccarino}}]{petri_scaffold_2014}%
  \BibitemOpen
  \bibfield  {author} {\bibinfo {author} {\bibfnamefont {G.}~\bibnamefont {Petri}}, \bibinfo {author} {\bibfnamefont {P.}~\bibnamefont {Expert}}, \bibinfo {author} {\bibfnamefont {F.}~\bibnamefont {Turkheimer}}, \bibinfo {author} {\bibfnamefont {R.}~\bibnamefont {Carhart-Harris}}, \bibinfo {author} {\bibfnamefont {D.}~\bibnamefont {Nutt}}, \bibinfo {author} {\bibfnamefont {P.~J.}\ \bibnamefont {Hellyer}},\ and\ \bibinfo {author} {\bibfnamefont {F.}~\bibnamefont {Vaccarino}},\ }\href {https://doi.org/10.1098/rsif.2014.0873} {\bibfield  {journal} {\bibinfo  {journal} {J. R. Soc. Interface}\ }\textbf {\bibinfo {volume} {11}},\ \bibinfo {pages} {20140873} (\bibinfo {year} {2014})}\BibitemShut {NoStop}%
\bibitem [{\citenamefont {Squartini}\ and\ \citenamefont {Garlaschelli}(2017)}]{squartinigarlaschelli_book_2017}%
  \BibitemOpen
  \bibfield  {author} {\bibinfo {author} {\bibfnamefont {T.}~\bibnamefont {Squartini}}\ and\ \bibinfo {author} {\bibfnamefont {D.}~\bibnamefont {Garlaschelli}},\ }\href {https://doi.org/10.1007/978-3-319-69438-2} {\emph {\bibinfo {title} {Maximum-Entropy Networks: Pattern Detection, Network Reconstruction and Graph Combinatorics}}},\ SpringerBriefs in Complexity\ (\bibinfo  {publisher} {Springer},\ \bibinfo {address} {Cham},\ \bibinfo {year} {2017})\BibitemShut {NoStop}%
\bibitem [{\citenamefont {Squartini}\ \emph {et~al.}(2018)\citenamefont {Squartini}, \citenamefont {Caldarelli}, \citenamefont {Cimini}, \citenamefont {Gabrielli},\ and\ \citenamefont {Garlaschelli}}]{squartini_reconstruction_2018}%
  \BibitemOpen
  \bibfield  {author} {\bibinfo {author} {\bibfnamefont {T.}~\bibnamefont {Squartini}}, \bibinfo {author} {\bibfnamefont {G.}~\bibnamefont {Caldarelli}}, \bibinfo {author} {\bibfnamefont {G.}~\bibnamefont {Cimini}}, \bibinfo {author} {\bibfnamefont {A.}~\bibnamefont {Gabrielli}},\ and\ \bibinfo {author} {\bibfnamefont {D.}~\bibnamefont {Garlaschelli}},\ }\href {https://doi.org/https://doi.org/10.1016/j.physrep.2018.06.008} {\bibfield  {journal} {\bibinfo  {journal} {Physics Reports}\ }\textbf {\bibinfo {volume} {757}},\ \bibinfo {pages} {1} (\bibinfo {year} {2018})}\BibitemShut {NoStop}%
\bibitem [{\citenamefont {Chelaru}\ \emph {et~al.}(2021)\citenamefont {Chelaru}, \citenamefont {Eagleman}, \citenamefont {Andrei}, \citenamefont {Milton}, \citenamefont {Kharas},\ and\ \citenamefont {Dragoi}}]{chelaru_high_order_2021}%
  \BibitemOpen
  \bibfield  {author} {\bibinfo {author} {\bibfnamefont {M.~I.}\ \bibnamefont {Chelaru}}, \bibinfo {author} {\bibfnamefont {S.}~\bibnamefont {Eagleman}}, \bibinfo {author} {\bibfnamefont {A.~R.}\ \bibnamefont {Andrei}}, \bibinfo {author} {\bibfnamefont {R.}~\bibnamefont {Milton}}, \bibinfo {author} {\bibfnamefont {N.}~\bibnamefont {Kharas}},\ and\ \bibinfo {author} {\bibfnamefont {V.}~\bibnamefont {Dragoi}},\ }\href {https://doi.org/10.1016/j.neuron.2021.09.042} {\bibfield  {journal} {\bibinfo  {journal} {Neuron}\ }\textbf {\bibinfo {volume} {109}},\ \bibinfo {pages} {3954} (\bibinfo {year} {2021})},\ \bibinfo {note} {publisher: Elsevier}\BibitemShut {NoStop}%
\bibitem [{\citenamefont {Santoro}\ \emph {et~al.}(2024)\citenamefont {Santoro}, \citenamefont {Battiston}, \citenamefont {Lucas}, \citenamefont {Petri},\ and\ \citenamefont {Amico}}]{santoro_ho_connectonomics_2024}%
  \BibitemOpen
  \bibfield  {author} {\bibinfo {author} {\bibfnamefont {A.}~\bibnamefont {Santoro}}, \bibinfo {author} {\bibfnamefont {F.}~\bibnamefont {Battiston}}, \bibinfo {author} {\bibfnamefont {M.}~\bibnamefont {Lucas}}, \bibinfo {author} {\bibfnamefont {G.}~\bibnamefont {Petri}},\ and\ \bibinfo {author} {\bibfnamefont {E.}~\bibnamefont {Amico}},\ }\href {https://doi.org/10.1038/s41467-024-54472-y} {\bibfield  {journal} {\bibinfo  {journal} {Nat. Commun.}\ }\textbf {\bibinfo {volume} {15}},\ \bibinfo {pages} {10244} (\bibinfo {year} {2024})}\BibitemShut {NoStop}%
\bibitem [{\citenamefont {Abiad}\ \emph {et~al.}(2026)\citenamefont {Abiad}, \citenamefont {Arenas}, \citenamefont {Backhausz}, \citenamefont {Balogh}, \citenamefont {Banerji}, \citenamefont {Barbarossa}, \citenamefont {Bianconi}, \citenamefont {Bick}, \citenamefont {Botnan}, \citenamefont {Carletti}, \citenamefont {Cavallaro}, \citenamefont {Civilini}, \citenamefont {Eliassi-Rad}, \citenamefont {Gong}, \citenamefont {Guo}, \citenamefont {Harrington}, \citenamefont {Jost}, \citenamefont {Krapivsky}, \citenamefont {Liò}, \citenamefont {MacArthur}, \citenamefont {Mattsson}, \citenamefont {Mediano}, \citenamefont {Millan}, \citenamefont {Mulas}, \citenamefont {Patania}, \citenamefont {Petri}, \citenamefont {Rathilal}, \citenamefont {J~Sánchez-García}, \citenamefont {Scolamiero}, \citenamefont {Schaub}, \citenamefont {Sun}, \citenamefont {Tian}, \citenamefont {Vaccarino},\ and\ \citenamefont {Xia}}]{abiad_hypergraph_roadmap_2026}%
  \BibitemOpen
  \bibfield  {author} {\bibinfo {author} {\bibfnamefont {A.}~\bibnamefont {Abiad}}, \bibinfo {author} {\bibfnamefont {A.}~\bibnamefont {Arenas}}, \bibinfo {author} {\bibfnamefont {A.}~\bibnamefont {Backhausz}}, \bibinfo {author} {\bibfnamefont {J.}~\bibnamefont {Balogh}}, \bibinfo {author} {\bibfnamefont {C.~R.~S.}\ \bibnamefont {Banerji}}, \bibinfo {author} {\bibfnamefont {S.}~\bibnamefont {Barbarossa}}, \bibinfo {author} {\bibfnamefont {G.}~\bibnamefont {Bianconi}}, \bibinfo {author} {\bibfnamefont {C.}~\bibnamefont {Bick}}, \bibinfo {author} {\bibfnamefont {M.}~\bibnamefont {Botnan}}, \bibinfo {author} {\bibfnamefont {T.}~\bibnamefont {Carletti}}, \bibinfo {author} {\bibfnamefont {L.}~\bibnamefont {Cavallaro}}, \bibinfo {author} {\bibfnamefont {A.}~\bibnamefont {Civilini}}, \bibinfo {author} {\bibfnamefont {T.}~\bibnamefont {Eliassi-Rad}}, \bibinfo {author} {\bibfnamefont {X.}~\bibnamefont {Gong}}, \bibinfo {author} {\bibfnamefont {K.}~\bibnamefont {Guo}}, \bibinfo {author} {\bibfnamefont {H.}~\bibnamefont
  {Harrington}}, \bibinfo {author} {\bibfnamefont {J.}~\bibnamefont {Jost}}, \bibinfo {author} {\bibfnamefont {P.~L.}\ \bibnamefont {Krapivsky}}, \bibinfo {author} {\bibfnamefont {P.}~\bibnamefont {Liò}}, \bibinfo {author} {\bibfnamefont {B.~D.}\ \bibnamefont {MacArthur}}, \bibinfo {author} {\bibfnamefont {C.}~\bibnamefont {Mattsson}}, \bibinfo {author} {\bibfnamefont {P.~A.~M.}\ \bibnamefont {Mediano}}, \bibinfo {author} {\bibfnamefont {A.~P.}\ \bibnamefont {Millan}}, \bibinfo {author} {\bibfnamefont {R.}~\bibnamefont {Mulas}}, \bibinfo {author} {\bibfnamefont {A.}~\bibnamefont {Patania}}, \bibinfo {author} {\bibfnamefont {G.}~\bibnamefont {Petri}}, \bibinfo {author} {\bibfnamefont {C.}~\bibnamefont {Rathilal}}, \bibinfo {author} {\bibfnamefont {R.}~\bibnamefont {J~Sánchez-García}}, \bibinfo {author} {\bibfnamefont {M.}~\bibnamefont {Scolamiero}}, \bibinfo {author} {\bibfnamefont {M.~T.}\ \bibnamefont {Schaub}}, \bibinfo {author} {\bibfnamefont {H.}~\bibnamefont {Sun}}, \bibinfo {author} {\bibfnamefont
  {Y.}~\bibnamefont {Tian}}, \bibinfo {author} {\bibfnamefont {F.}~\bibnamefont {Vaccarino}},\ and\ \bibinfo {author} {\bibfnamefont {K.}~\bibnamefont {Xia}},\ }\href {https://doi.org/10.1088/2632-072X/ae3c4e} {\bibfield  {journal} {\bibinfo  {journal} {Journal of Physics: Complexity}\ }\textbf {\bibinfo {volume} {7}},\ \bibinfo {pages} {022501} (\bibinfo {year} {2026})}\BibitemShut {NoStop}%
\bibitem [{\citenamefont {Yu}\ \emph {et~al.}(2006)\citenamefont {Yu}, \citenamefont {Righero},\ and\ \citenamefont {Kocarev}}]{yu_estimating_network_2006}%
  \BibitemOpen
  \bibfield  {author} {\bibinfo {author} {\bibfnamefont {D.}~\bibnamefont {Yu}}, \bibinfo {author} {\bibfnamefont {M.}~\bibnamefont {Righero}},\ and\ \bibinfo {author} {\bibfnamefont {L.}~\bibnamefont {Kocarev}},\ }\href {https://doi.org/10.1103/PhysRevLett.97.188701} {\bibfield  {journal} {\bibinfo  {journal} {Phys. Rev. Lett.}\ }\textbf {\bibinfo {volume} {97}},\ \bibinfo {pages} {188701} (\bibinfo {year} {2006})}\BibitemShut {NoStop}%
\bibitem [{\citenamefont {Timme}(2007)}]{timme_revealing_net_2007}%
  \BibitemOpen
  \bibfield  {author} {\bibinfo {author} {\bibfnamefont {M.}~\bibnamefont {Timme}},\ }\href {https://doi.org/10.1103/PhysRevLett.98.224101} {\bibfield  {journal} {\bibinfo  {journal} {Phys. Rev. Lett.}\ }\textbf {\bibinfo {volume} {98}},\ \bibinfo {pages} {224101} (\bibinfo {year} {2007})}\BibitemShut {NoStop}%
\bibitem [{\citenamefont {Han}\ \emph {et~al.}(2015)\citenamefont {Han}, \citenamefont {Shen}, \citenamefont {Wang},\ and\ \citenamefont {Di}}]{han_reconstruction_prl_2015}%
  \BibitemOpen
  \bibfield  {author} {\bibinfo {author} {\bibfnamefont {X.}~\bibnamefont {Han}}, \bibinfo {author} {\bibfnamefont {Z.}~\bibnamefont {Shen}}, \bibinfo {author} {\bibfnamefont {W.-X.}\ \bibnamefont {Wang}},\ and\ \bibinfo {author} {\bibfnamefont {Z.}~\bibnamefont {Di}},\ }\href {https://doi.org/10.1103/PhysRevLett.114.028701} {\bibfield  {journal} {\bibinfo  {journal} {Phys. Rev. Lett.}\ }\textbf {\bibinfo {volume} {114}},\ \bibinfo {pages} {028701} (\bibinfo {year} {2015})}\BibitemShut {NoStop}%
\bibitem [{\citenamefont {Brugere}\ \emph {et~al.}(2018)\citenamefont {Brugere}, \citenamefont {Gallagher},\ and\ \citenamefont {Berger-Wolf}}]{brugere_network_extraction_2018}%
  \BibitemOpen
  \bibfield  {author} {\bibinfo {author} {\bibfnamefont {I.}~\bibnamefont {Brugere}}, \bibinfo {author} {\bibfnamefont {B.}~\bibnamefont {Gallagher}},\ and\ \bibinfo {author} {\bibfnamefont {T.~Y.}\ \bibnamefont {Berger-Wolf}},\ }\bibfield  {journal} {\bibinfo  {journal} {ACM Comput. Surv.}\ }\textbf {\bibinfo {volume} {51}},\ \href {https://doi.org/10.1145/3154524} {10.1145/3154524} (\bibinfo {year} {2018})\BibitemShut {NoStop}%
\bibitem [{\citenamefont {Marrelec}\ \emph {et~al.}(2006)\citenamefont {Marrelec}, \citenamefont {Krainik}, \citenamefont {Duffau}, \citenamefont {P{\'e}l{\'e}grini-Issac}, \citenamefont {Leh{\'e}ricy}, \citenamefont {Doyon},\ and\ \citenamefont {Benali}}]{marrelec_partial_2006}%
  \BibitemOpen
  \bibfield  {author} {\bibinfo {author} {\bibfnamefont {G.}~\bibnamefont {Marrelec}}, \bibinfo {author} {\bibfnamefont {A.}~\bibnamefont {Krainik}}, \bibinfo {author} {\bibfnamefont {H.}~\bibnamefont {Duffau}}, \bibinfo {author} {\bibfnamefont {M.}~\bibnamefont {P{\'e}l{\'e}grini-Issac}}, \bibinfo {author} {\bibfnamefont {S.}~\bibnamefont {Leh{\'e}ricy}}, \bibinfo {author} {\bibfnamefont {J.}~\bibnamefont {Doyon}},\ and\ \bibinfo {author} {\bibfnamefont {H.}~\bibnamefont {Benali}},\ }\href {https://doi.org/10.1016/j.neuroimage.2005.12.057} {\bibfield  {journal} {\bibinfo  {journal} {NeuroImage}\ }\textbf {\bibinfo {volume} {32}},\ \bibinfo {pages} {228} (\bibinfo {year} {2006})}\BibitemShut {NoStop}%
\bibitem [{\citenamefont {Granger}(1969)}]{granger1969investigating}%
  \BibitemOpen
  \bibfield  {author} {\bibinfo {author} {\bibfnamefont {C.~W.~J.}\ \bibnamefont {Granger}},\ }\href {https://doi.org/10.2307/1912791} {\bibfield  {journal} {\bibinfo  {journal} {Econometrica}\ }\textbf {\bibinfo {volume} {37}},\ \bibinfo {pages} {424} (\bibinfo {year} {1969})}\BibitemShut {NoStop}%
\bibitem [{\citenamefont {Guo}\ \emph {et~al.}(2008)\citenamefont {Guo}, \citenamefont {Seth}, \citenamefont {Kendrick}, \citenamefont {Zhou},\ and\ \citenamefont {Feng}}]{guo_partial_granger_2008}%
  \BibitemOpen
  \bibfield  {author} {\bibinfo {author} {\bibfnamefont {S.}~\bibnamefont {Guo}}, \bibinfo {author} {\bibfnamefont {A.~K.}\ \bibnamefont {Seth}}, \bibinfo {author} {\bibfnamefont {K.~M.}\ \bibnamefont {Kendrick}}, \bibinfo {author} {\bibfnamefont {C.}~\bibnamefont {Zhou}},\ and\ \bibinfo {author} {\bibfnamefont {J.}~\bibnamefont {Feng}},\ }\href {https://doi.org/10.1016/j.jneumeth.2008.04.011} {\bibfield  {journal} {\bibinfo  {journal} {Journal of Neuroscience Methods}\ }\textbf {\bibinfo {volume} {172}},\ \bibinfo {pages} {79} (\bibinfo {year} {2008})}\BibitemShut {NoStop}%
\bibitem [{\citenamefont {Schreiber}(2000)}]{schreiber2000transfer}%
  \BibitemOpen
  \bibfield  {author} {\bibinfo {author} {\bibfnamefont {T.}~\bibnamefont {Schreiber}},\ }\href {https://doi.org/10.1103/PhysRevLett.85.461} {\bibfield  {journal} {\bibinfo  {journal} {Physical Review Letters}\ }\textbf {\bibinfo {volume} {85}},\ \bibinfo {pages} {461} (\bibinfo {year} {2000})}\BibitemShut {NoStop}%
\bibitem [{\citenamefont {Musciotto}\ \emph {et~al.}(2021)\citenamefont {Musciotto}, \citenamefont {Battiston},\ and\ \citenamefont {Mantegna}}]{musciotto_detecting_2021}%
  \BibitemOpen
  \bibfield  {author} {\bibinfo {author} {\bibfnamefont {F.}~\bibnamefont {Musciotto}}, \bibinfo {author} {\bibfnamefont {F.}~\bibnamefont {Battiston}},\ and\ \bibinfo {author} {\bibfnamefont {R.~N.}\ \bibnamefont {Mantegna}},\ }\href {https://doi.org/10.1038/s42005-021-00710-4} {\bibfield  {journal} {\bibinfo  {journal} {Communications Physics}\ }\textbf {\bibinfo {volume} {4}},\ \bibinfo {pages} {218} (\bibinfo {year} {2021})}\BibitemShut {NoStop}%
\bibitem [{\citenamefont {Young}\ \emph {et~al.}(2021)\citenamefont {Young}, \citenamefont {Petri},\ and\ \citenamefont {Peixoto}}]{young_hypergraph_2021}%
  \BibitemOpen
  \bibfield  {author} {\bibinfo {author} {\bibfnamefont {J.-G.}\ \bibnamefont {Young}}, \bibinfo {author} {\bibfnamefont {G.}~\bibnamefont {Petri}},\ and\ \bibinfo {author} {\bibfnamefont {T.~P.}\ \bibnamefont {Peixoto}},\ }\href {https://doi.org/10.1038/s42005-021-00637-w} {\bibfield  {journal} {\bibinfo  {journal} {Communications Physics}\ }\textbf {\bibinfo {volume} {4}},\ \bibinfo {pages} {135} (\bibinfo {year} {2021})}\BibitemShut {NoStop}%
\bibitem [{\citenamefont {Lizotte}\ \emph {et~al.}(2023)\citenamefont {Lizotte}, \citenamefont {Young},\ and\ \citenamefont {Allard}}]{lizotte_hypergraph_2023}%
  \BibitemOpen
  \bibfield  {author} {\bibinfo {author} {\bibfnamefont {S.}~\bibnamefont {Lizotte}}, \bibinfo {author} {\bibfnamefont {J.-G.}\ \bibnamefont {Young}},\ and\ \bibinfo {author} {\bibfnamefont {A.}~\bibnamefont {Allard}},\ }\href {https://doi.org/10.1038/s41598-023-48081-w} {\bibfield  {journal} {\bibinfo  {journal} {Scientific Reports}\ }\textbf {\bibinfo {volume} {13}},\ \bibinfo {pages} {21364} (\bibinfo {year} {2023})}\BibitemShut {NoStop}%
\bibitem [{\citenamefont {Malizia}\ \emph {et~al.}(2024)\citenamefont {Malizia}, \citenamefont {Corso}, \citenamefont {Gambuzza}, \citenamefont {Russo}, \citenamefont {Latora},\ and\ \citenamefont {Frasca}}]{malizia_reconstructing_2024}%
  \BibitemOpen
  \bibfield  {author} {\bibinfo {author} {\bibfnamefont {F.}~\bibnamefont {Malizia}}, \bibinfo {author} {\bibfnamefont {A.}~\bibnamefont {Corso}}, \bibinfo {author} {\bibfnamefont {L.~V.}\ \bibnamefont {Gambuzza}}, \bibinfo {author} {\bibfnamefont {G.}~\bibnamefont {Russo}}, \bibinfo {author} {\bibfnamefont {V.}~\bibnamefont {Latora}},\ and\ \bibinfo {author} {\bibfnamefont {M.}~\bibnamefont {Frasca}},\ }\href {https://doi.org/10.1038/s41467-024-49278-x} {\bibfield  {journal} {\bibinfo  {journal} {Nat. Commun.}\ }\textbf {\bibinfo {volume} {15}},\ \bibinfo {pages} {5184} (\bibinfo {year} {2024})}\BibitemShut {NoStop}%
\bibitem [{\citenamefont {Delabays}\ \emph {et~al.}(2025)\citenamefont {Delabays}, \citenamefont {De~Pasquale}, \citenamefont {Dörfler},\ and\ \citenamefont {Zhang}}]{delabays_hyper_recons_2025}%
  \BibitemOpen
  \bibfield  {author} {\bibinfo {author} {\bibfnamefont {R.}~\bibnamefont {Delabays}}, \bibinfo {author} {\bibfnamefont {G.}~\bibnamefont {De~Pasquale}}, \bibinfo {author} {\bibfnamefont {F.}~\bibnamefont {Dörfler}},\ and\ \bibinfo {author} {\bibfnamefont {Y.}~\bibnamefont {Zhang}},\ }\href {https://doi.org/10.1038/s41467-025-57664-2} {\bibfield  {journal} {\bibinfo  {journal} {Nat. Commun.}\ }\textbf {\bibinfo {volume} {16}},\ \bibinfo {pages} {2691} (\bibinfo {year} {2025})}\BibitemShut {NoStop}%
\bibitem [{\citenamefont {Santoro}\ \emph {et~al.}(2023)\citenamefont {Santoro}, \citenamefont {Battiston}, \citenamefont {Petri},\ and\ \citenamefont {Amico}}]{santoro_higher_order_2023}%
  \BibitemOpen
  \bibfield  {author} {\bibinfo {author} {\bibfnamefont {A.}~\bibnamefont {Santoro}}, \bibinfo {author} {\bibfnamefont {F.}~\bibnamefont {Battiston}}, \bibinfo {author} {\bibfnamefont {G.}~\bibnamefont {Petri}},\ and\ \bibinfo {author} {\bibfnamefont {E.}~\bibnamefont {Amico}},\ }\href {https://doi.org/10.1038/s41567-022-01852-0} {\bibfield  {journal} {\bibinfo  {journal} {Nature Physics}\ }\textbf {\bibinfo {volume} {19}},\ \bibinfo {pages} {221} (\bibinfo {year} {2023})}\BibitemShut {NoStop}%
\bibitem [{\citenamefont {Varley}\ \emph {et~al.}(2023)\citenamefont {Varley}, \citenamefont {Pope}, \citenamefont {Puxeddu}, \citenamefont {Faskowitz},\ and\ \citenamefont {Sporns}}]{varley_partialentropy_2023}%
  \BibitemOpen
  \bibfield  {author} {\bibinfo {author} {\bibfnamefont {T.~F.}\ \bibnamefont {Varley}}, \bibinfo {author} {\bibfnamefont {M.}~\bibnamefont {Pope}}, \bibinfo {author} {\bibfnamefont {M.~G.}\ \bibnamefont {Puxeddu}}, \bibinfo {author} {\bibfnamefont {J.}~\bibnamefont {Faskowitz}},\ and\ \bibinfo {author} {\bibfnamefont {O.}~\bibnamefont {Sporns}},\ }\href {https://doi.org/10.1073/pnas.2300888120} {\bibfield  {journal} {\bibinfo  {journal} {Proc. Natl. Acad. Sci. U.S.A.}\ }\textbf {\bibinfo {volume} {120}},\ \bibinfo {pages} {e2300888120} (\bibinfo {year} {2023})}\BibitemShut {NoStop}%
\bibitem [{\citenamefont {Keogh}\ \emph {et~al.}(2001)\citenamefont {Keogh}, \citenamefont {Chakrabarti}, \citenamefont {Pazzani},\ and\ \citenamefont {Mehrotra}}]{keogh_dimensionality_2001}%
  \BibitemOpen
  \bibfield  {author} {\bibinfo {author} {\bibfnamefont {E.}~\bibnamefont {Keogh}}, \bibinfo {author} {\bibfnamefont {K.}~\bibnamefont {Chakrabarti}}, \bibinfo {author} {\bibfnamefont {M.}~\bibnamefont {Pazzani}},\ and\ \bibinfo {author} {\bibfnamefont {S.}~\bibnamefont {Mehrotra}},\ }\href {https://doi.org/10.1007/PL00011669} {\bibfield  {journal} {\bibinfo  {journal} {Knowl. Inf. Syst.}\ }\textbf {\bibinfo {volume} {3}},\ \bibinfo {pages} {263} (\bibinfo {year} {2001})}\BibitemShut {NoStop}%
\bibitem [{\citenamefont {Lin}\ \emph {et~al.}(2007)\citenamefont {Lin}, \citenamefont {Keogh}, \citenamefont {Wei},\ and\ \citenamefont {Lonardi}}]{lin_sax_2007}%
  \BibitemOpen
  \bibfield  {author} {\bibinfo {author} {\bibfnamefont {J.}~\bibnamefont {Lin}}, \bibinfo {author} {\bibfnamefont {E.}~\bibnamefont {Keogh}}, \bibinfo {author} {\bibfnamefont {L.}~\bibnamefont {Wei}},\ and\ \bibinfo {author} {\bibfnamefont {S.}~\bibnamefont {Lonardi}},\ }\href {https://doi.org/10.1007/s10618-007-0064-z} {\bibfield  {journal} {\bibinfo  {journal} {Data Mining and Knowledge Discovery}\ }\textbf {\bibinfo {volume} {15}},\ \bibinfo {pages} {107} (\bibinfo {year} {2007})}\BibitemShut {NoStop}%
\bibitem [{\citenamefont {Wang}\ \emph {et~al.}(2022)\citenamefont {Wang}, \citenamefont {Ma}, \citenamefont {Chen}, \citenamefont {Lai},\ and\ \citenamefont {Zhang}}]{wang_full_2022}%
  \BibitemOpen
  \bibfield  {author} {\bibinfo {author} {\bibfnamefont {H.}~\bibnamefont {Wang}}, \bibinfo {author} {\bibfnamefont {C.}~\bibnamefont {Ma}}, \bibinfo {author} {\bibfnamefont {H.-S.}\ \bibnamefont {Chen}}, \bibinfo {author} {\bibfnamefont {Y.-C.}\ \bibnamefont {Lai}},\ and\ \bibinfo {author} {\bibfnamefont {H.-F.}\ \bibnamefont {Zhang}},\ }\href {https://doi.org/10.1038/s41467-022-30706-9} {\bibfield  {journal} {\bibinfo  {journal} {Nature Communications}\ }\textbf {\bibinfo {volume} {13}},\ \bibinfo {pages} {3043} (\bibinfo {year} {2022})}\BibitemShut {NoStop}%
\bibitem [{\citenamefont {Panzeri}\ \emph {et~al.}(2007)\citenamefont {Panzeri}, \citenamefont {Senatore}, \citenamefont {Montemurro},\ and\ \citenamefont {Petersen}}]{panzeri2007correcting}%
  \BibitemOpen
  \bibfield  {author} {\bibinfo {author} {\bibfnamefont {S.}~\bibnamefont {Panzeri}}, \bibinfo {author} {\bibfnamefont {R.}~\bibnamefont {Senatore}}, \bibinfo {author} {\bibfnamefont {M.~A.}\ \bibnamefont {Montemurro}},\ and\ \bibinfo {author} {\bibfnamefont {R.~S.}\ \bibnamefont {Petersen}},\ }\href {https://doi.org/10.1152/jn.00559.2007} {\bibfield  {journal} {\bibinfo  {journal} {Journal of Neurophysiology}\ }\textbf {\bibinfo {volume} {98}},\ \bibinfo {pages} {1064} (\bibinfo {year} {2007})}\BibitemShut {NoStop}%
\bibitem [{\citenamefont {Williams}\ and\ \citenamefont {Beer}(2010)}]{williams2010nonnegative}%
  \BibitemOpen
  \bibfield  {author} {\bibinfo {author} {\bibfnamefont {P.~L.}\ \bibnamefont {Williams}}\ and\ \bibinfo {author} {\bibfnamefont {R.~D.}\ \bibnamefont {Beer}},\ }\href@noop {} {\bibinfo {title} {Nonnegative decomposition of multivariate information}} (\bibinfo {year} {2010}),\ \Eprint {https://arxiv.org/abs/1004.2515} {arXiv:1004.2515 [cs.IT]} \BibitemShut {NoStop}%
\bibitem [{\citenamefont {Rosas}\ \emph {et~al.}(2019)\citenamefont {Rosas}, \citenamefont {Mediano}, \citenamefont {Gastpar},\ and\ \citenamefont {Jensen}}]{rosas2019o}%
  \BibitemOpen
  \bibfield  {author} {\bibinfo {author} {\bibfnamefont {F.~E.}\ \bibnamefont {Rosas}}, \bibinfo {author} {\bibfnamefont {P.~A.~M.}\ \bibnamefont {Mediano}}, \bibinfo {author} {\bibfnamefont {M.}~\bibnamefont {Gastpar}},\ and\ \bibinfo {author} {\bibfnamefont {H.~J.}\ \bibnamefont {Jensen}},\ }\href {https://doi.org/10.1103/PhysRevE.100.032305} {\bibfield  {journal} {\bibinfo  {journal} {Physical Review E}\ }\textbf {\bibinfo {volume} {100}},\ \bibinfo {pages} {032305} (\bibinfo {year} {2019})}\BibitemShut {NoStop}%
\bibitem [{\citenamefont {Beck}\ and\ \citenamefont {Schögl}(1993)}]{Beck_Schögl_1993}%
  \BibitemOpen
  \bibfield  {author} {\bibinfo {author} {\bibfnamefont {C.}~\bibnamefont {Beck}}\ and\ \bibinfo {author} {\bibfnamefont {F.}~\bibnamefont {Schögl}},\ }\href@noop {} {\emph {\bibinfo {title} {Thermodynamics of Chaotic Systems: An Introduction}}},\ Cambridge Nonlinear Science Series\ (\bibinfo  {publisher} {Cambridge University Press},\ \bibinfo {year} {1993})\BibitemShut {NoStop}%
\bibitem [{\citenamefont {Daw}\ \emph {et~al.}(2003)\citenamefont {Daw}, \citenamefont {Finney},\ and\ \citenamefont {Tracy}}]{daw_review_2003}%
  \BibitemOpen
  \bibfield  {author} {\bibinfo {author} {\bibfnamefont {C.~S.}\ \bibnamefont {Daw}}, \bibinfo {author} {\bibfnamefont {C.~E.~A.}\ \bibnamefont {Finney}},\ and\ \bibinfo {author} {\bibfnamefont {E.~R.}\ \bibnamefont {Tracy}},\ }\href {https://doi.org/10.1063/1.1531823} {\bibfield  {journal} {\bibinfo  {journal} {Review of Scientific Instruments}\ }\textbf {\bibinfo {volume} {74}},\ \bibinfo {pages} {915} (\bibinfo {year} {2003})}\BibitemShut {NoStop}%
\bibitem [{\citenamefont {Sch{\"a}fer}\ and\ \citenamefont {H{\"o}gqvist}(2012)}]{Schafer2012SFA}%
  \BibitemOpen
  \bibfield  {author} {\bibinfo {author} {\bibfnamefont {P.}~\bibnamefont {Sch{\"a}fer}}\ and\ \bibinfo {author} {\bibfnamefont {M.}~\bibnamefont {H{\"o}gqvist}},\ }in\ \href {https://doi.org/10.1145/2168836.2168899} {\emph {\bibinfo {booktitle} {Proceedings of the 15th International Conference on Extending Database Technology}}}\ (\bibinfo {year} {2012})\ pp.\ \bibinfo {pages} {516--527}\BibitemShut {NoStop}%
\bibitem [{\citenamefont {Kedem}(1986)}]{Kedem1986ZeroCrossing}%
  \BibitemOpen
  \bibfield  {author} {\bibinfo {author} {\bibfnamefont {B.}~\bibnamefont {Kedem}},\ }\href {https://doi.org/10.1109/PROC.1986.13668} {\bibfield  {journal} {\bibinfo  {journal} {Proceedings of the {IEEE}}\ }\textbf {\bibinfo {volume} {74}},\ \bibinfo {pages} {1477} (\bibinfo {year} {1986})}\BibitemShut {NoStop}%
\bibitem [{\citenamefont {Kennel}\ and\ \citenamefont {Buhl}(2003)}]{Kennel2003SFNN}%
  \BibitemOpen
  \bibfield  {author} {\bibinfo {author} {\bibfnamefont {M.~B.}\ \bibnamefont {Kennel}}\ and\ \bibinfo {author} {\bibfnamefont {M.}~\bibnamefont {Buhl}},\ }\href {https://doi.org/10.1103/PhysRevLett.91.084102} {\bibfield  {journal} {\bibinfo  {journal} {Physical Review Letters}\ }\textbf {\bibinfo {volume} {91}},\ \bibinfo {pages} {084102} (\bibinfo {year} {2003})}\BibitemShut {NoStop}%
\bibitem [{\citenamefont {Hirata}\ and\ \citenamefont {Judd}(2004)}]{Hirata2004Shadowing}%
  \BibitemOpen
  \bibfield  {author} {\bibinfo {author} {\bibfnamefont {Y.}~\bibnamefont {Hirata}}\ and\ \bibinfo {author} {\bibfnamefont {K.}~\bibnamefont {Judd}},\ }\href {https://doi.org/10.1103/PhysRevE.69.066213} {\bibfield  {journal} {\bibinfo  {journal} {Physical Review E}\ }\textbf {\bibinfo {volume} {69}},\ \bibinfo {pages} {066213} (\bibinfo {year} {2004})}\BibitemShut {NoStop}%
\bibitem [{\citenamefont {Aminikhanghahi}\ and\ \citenamefont {Cook}(2017)}]{Aminikhanghahi2017Survey}%
  \BibitemOpen
  \bibfield  {author} {\bibinfo {author} {\bibfnamefont {S.}~\bibnamefont {Aminikhanghahi}}\ and\ \bibinfo {author} {\bibfnamefont {D.~J.}\ \bibnamefont {Cook}},\ }\href {https://doi.org/10.1007/s10115-016-0987-z} {\bibfield  {journal} {\bibinfo  {journal} {Knowledge and Information Systems}\ }\textbf {\bibinfo {volume} {51}},\ \bibinfo {pages} {339} (\bibinfo {year} {2017})}\BibitemShut {NoStop}%
\bibitem [{\citenamefont {Truong}\ \emph {et~al.}(2020)\citenamefont {Truong}, \citenamefont {Oudre},\ and\ \citenamefont {Vayatis}}]{Truong2020Selective}%
  \BibitemOpen
  \bibfield  {author} {\bibinfo {author} {\bibfnamefont {C.}~\bibnamefont {Truong}}, \bibinfo {author} {\bibfnamefont {L.}~\bibnamefont {Oudre}},\ and\ \bibinfo {author} {\bibfnamefont {N.}~\bibnamefont {Vayatis}},\ }\href {https://doi.org/10.1016/j.sigpro.2019.107299} {\bibfield  {journal} {\bibinfo  {journal} {Signal Processing}\ }\textbf {\bibinfo {volume} {167}},\ \bibinfo {pages} {107299} (\bibinfo {year} {2020})}\BibitemShut {NoStop}%
\bibitem [{\citenamefont {Kandel}\ \emph {et~al.}(1996)\citenamefont {Kandel}, \citenamefont {Matias}, \citenamefont {Unger},\ and\ \citenamefont {Winkler}}]{Kandel1996}%
  \BibitemOpen
  \bibfield  {author} {\bibinfo {author} {\bibfnamefont {D.}~\bibnamefont {Kandel}}, \bibinfo {author} {\bibfnamefont {Y.}~\bibnamefont {Matias}}, \bibinfo {author} {\bibfnamefont {R.}~\bibnamefont {Unger}},\ and\ \bibinfo {author} {\bibfnamefont {P.}~\bibnamefont {Winkler}},\ }\href {https://doi.org/10.1016/S0166-218X(97)81456-4} {\bibfield  {journal} {\bibinfo  {journal} {Discrete Applied Mathematics}\ }\textbf {\bibinfo {volume} {71}},\ \bibinfo {pages} {171} (\bibinfo {year} {1996})}\BibitemShut {NoStop}%
\bibitem [{\citenamefont {van~der Heyden}\ \emph {et~al.}(1998)\citenamefont {van~der Heyden}, \citenamefont {Diks}, \citenamefont {Hoekstra},\ and\ \citenamefont {De~Goede}}]{vanDerHeyden1998}%
  \BibitemOpen
  \bibfield  {author} {\bibinfo {author} {\bibfnamefont {M.~J.}\ \bibnamefont {van~der Heyden}}, \bibinfo {author} {\bibfnamefont {C.~G.~C.}\ \bibnamefont {Diks}}, \bibinfo {author} {\bibfnamefont {B.~P.~T.}\ \bibnamefont {Hoekstra}},\ and\ \bibinfo {author} {\bibfnamefont {J.}~\bibnamefont {De~Goede}},\ }\href {https://doi.org/10.1016/S0167-2789(97)00315-1} {\bibfield  {journal} {\bibinfo  {journal} {Physica D: Nonlinear Phenomena}\ }\textbf {\bibinfo {volume} {117}},\ \bibinfo {pages} {299} (\bibinfo {year} {1998})}\BibitemShut {NoStop}%
\bibitem [{\citenamefont {Das}\ and\ \citenamefont {Dai}(2007)}]{Das2007}%
  \BibitemOpen
  \bibfield  {author} {\bibinfo {author} {\bibfnamefont {M.~K.}\ \bibnamefont {Das}}\ and\ \bibinfo {author} {\bibfnamefont {H.-K.}\ \bibnamefont {Dai}},\ }\href {https://doi.org/10.1186/1471-2105-8-S7-S21} {\bibfield  {journal} {\bibinfo  {journal} {BMC Bioinformatics}\ }\textbf {\bibinfo {volume} {8}},\ \bibinfo {pages} {S21} (\bibinfo {year} {2007})}\BibitemShut {NoStop}%
\bibitem [{\citenamefont {Sinatra}\ \emph {et~al.}(2010)\citenamefont {Sinatra}, \citenamefont {Condorelli},\ and\ \citenamefont {Latora}}]{sinatra_motifs_2010}%
  \BibitemOpen
  \bibfield  {author} {\bibinfo {author} {\bibfnamefont {R.}~\bibnamefont {Sinatra}}, \bibinfo {author} {\bibfnamefont {D.}~\bibnamefont {Condorelli}},\ and\ \bibinfo {author} {\bibfnamefont {V.}~\bibnamefont {Latora}},\ }\href {https://doi.org/10.1103/PhysRevLett.105.178702} {\bibfield  {journal} {\bibinfo  {journal} {Phys. Rev. Lett.}\ }\textbf {\bibinfo {volume} {105}},\ \bibinfo {pages} {178702} (\bibinfo {year} {2010})}\BibitemShut {NoStop}%
\bibitem [{\citenamefont {Hashim}\ \emph {et~al.}(2019)\citenamefont {Hashim}, \citenamefont {Mabrouk},\ and\ \citenamefont {Al-Atabany}}]{Hashim2019}%
  \BibitemOpen
  \bibfield  {author} {\bibinfo {author} {\bibfnamefont {F.~A.}\ \bibnamefont {Hashim}}, \bibinfo {author} {\bibfnamefont {M.~S.}\ \bibnamefont {Mabrouk}},\ and\ \bibinfo {author} {\bibfnamefont {W.}~\bibnamefont {Al-Atabany}},\ }\href {https://www.ncbi.nlm.nih.gov/pmc/articles/PMC6490410/} {\bibfield  {journal} {\bibinfo  {journal} {Avicenna Journal of Medical Biotechnology}\ }\textbf {\bibinfo {volume} {11}},\ \bibinfo {pages} {130} (\bibinfo {year} {2019})},\ \bibinfo {note} {pMID: 31057715; PMCID: PMC6490410}\BibitemShut {NoStop}%
\bibitem [{\citenamefont {Lin}(1991)}]{lin_jensen_shannon_1991}%
  \BibitemOpen
  \bibfield  {author} {\bibinfo {author} {\bibfnamefont {J.}~\bibnamefont {Lin}},\ }\href {https://doi.org/10.1109/18.61115} {\bibfield  {journal} {\bibinfo  {journal} {IEEE Trans. Inf. Theory}\ }\textbf {\bibinfo {volume} {37}},\ \bibinfo {pages} {145} (\bibinfo {year} {1991})}\BibitemShut {NoStop}%
\bibitem [{\citenamefont {Abbott~et al.}(2019)}]{abbott_LIGO_prior_posterior}%
  \BibitemOpen
  \bibfield  {author} {\bibinfo {author} {\bibfnamefont {B.~P.}\ \bibnamefont {Abbott~et al.}} (\bibinfo {collaboration} {LIGO Scientific Collaboration and Virgo Collaboration}),\ }\href {https://doi.org/10.1103/PhysRevX.9.031040} {\bibfield  {journal} {\bibinfo  {journal} {Phys. Rev. X}\ }\textbf {\bibinfo {volume} {9}},\ \bibinfo {pages} {031040} (\bibinfo {year} {2019})}\BibitemShut {NoStop}%
\bibitem [{\citenamefont {Pratten}\ \emph {et~al.}(2020)\citenamefont {Pratten}, \citenamefont {Schmidt}, \citenamefont {Buscicchio},\ and\ \citenamefont {Thomas}}]{pratten_binaries_2020}%
  \BibitemOpen
  \bibfield  {author} {\bibinfo {author} {\bibfnamefont {G.}~\bibnamefont {Pratten}}, \bibinfo {author} {\bibfnamefont {P.}~\bibnamefont {Schmidt}}, \bibinfo {author} {\bibfnamefont {R.}~\bibnamefont {Buscicchio}},\ and\ \bibinfo {author} {\bibfnamefont {L.~M.}\ \bibnamefont {Thomas}},\ }\href {https://doi.org/10.1103/PhysRevResearch.2.043096} {\bibfield  {journal} {\bibinfo  {journal} {Phys. Rev. Res.}\ }\textbf {\bibinfo {volume} {2}},\ \bibinfo {pages} {043096} (\bibinfo {year} {2020})}\BibitemShut {NoStop}%
\bibitem [{\citenamefont {Manning}\ \emph {et~al.}(2008)\citenamefont {Manning}, \citenamefont {Raghavan},\ and\ \citenamefont {Sch{\"u}tze}}]{manning2008introduction}%
  \BibitemOpen
  \bibfield  {author} {\bibinfo {author} {\bibfnamefont {C.~D.}\ \bibnamefont {Manning}}, \bibinfo {author} {\bibfnamefont {P.}~\bibnamefont {Raghavan}},\ and\ \bibinfo {author} {\bibfnamefont {H.}~\bibnamefont {Sch{\"u}tze}},\ }\href {https://doi.org/10.1017/CBO9780511809071} {\emph {\bibinfo {title} {Introduction to Information Retrieval}}}\ (\bibinfo  {publisher} {Cambridge University Press},\ \bibinfo {address} {Cambridge, UK},\ \bibinfo {year} {2008})\BibitemShut {NoStop}%
\bibitem [{\citenamefont {Newman}(2005)}]{newman_zipf_2005}%
  \BibitemOpen
  \bibfield  {author} {\bibinfo {author} {\bibfnamefont {M.}~\bibnamefont {Newman}},\ }\href {https://doi.org/10.1080/00107510500052444} {\bibfield  {journal} {\bibinfo  {journal} {Contemporary Physics}\ }\textbf {\bibinfo {volume} {46}},\ \bibinfo {pages} {323} (\bibinfo {year} {2005})}\BibitemShut {NoStop}%
\bibitem [{\citenamefont {Propp}\ and\ \citenamefont {Wilson}(1998)}]{propp1998how}%
  \BibitemOpen
  \bibfield  {author} {\bibinfo {author} {\bibfnamefont {J.~G.}\ \bibnamefont {Propp}}\ and\ \bibinfo {author} {\bibfnamefont {D.~B.}\ \bibnamefont {Wilson}},\ }\href {https://doi.org/10.1006/jagm.1997.0917} {\bibfield  {journal} {\bibinfo  {journal} {Journal of Algorithms}\ }\textbf {\bibinfo {volume} {27}},\ \bibinfo {pages} {170} (\bibinfo {year} {1998})}\BibitemShut {NoStop}%
\bibitem [{\citenamefont {Jiang}\ \emph {et~al.}(2008)\citenamefont {Jiang}, \citenamefont {Anderson}, \citenamefont {Gillespie},\ and\ \citenamefont {Mayne}}]{jiang2008ushuffle}%
  \BibitemOpen
  \bibfield  {author} {\bibinfo {author} {\bibfnamefont {M.}~\bibnamefont {Jiang}}, \bibinfo {author} {\bibfnamefont {J.}~\bibnamefont {Anderson}}, \bibinfo {author} {\bibfnamefont {J.}~\bibnamefont {Gillespie}},\ and\ \bibinfo {author} {\bibfnamefont {M.}~\bibnamefont {Mayne}},\ }\href {https://doi.org/10.1186/1471-2105-9-192} {\bibfield  {journal} {\bibinfo  {journal} {BMC Bioinformatics}\ }\textbf {\bibinfo {volume} {9}},\ \bibinfo {pages} {192} (\bibinfo {year} {2008})}\BibitemShut {NoStop}%
\bibitem [{\citenamefont {Powers}\ and\ \citenamefont {Ailab}(2011)}]{powers_evaluation_2011}%
  \BibitemOpen
  \bibfield  {author} {\bibinfo {author} {\bibfnamefont {D.}~\bibnamefont {Powers}}\ and\ \bibinfo {author} {\bibnamefont {Ailab}},\ }\href {https://doi.org/10.9735/2229-3981} {\bibfield  {journal} {\bibinfo  {journal} {J. Mach. Learn. Technol}\ }\textbf {\bibinfo {volume} {2}},\ \bibinfo {pages} {2229} (\bibinfo {year} {2011})}\BibitemShut {NoStop}%
\bibitem [{\citenamefont {{Allen Institute for Brain Science}}(2019)}]{allen_whitepaper}%
  \BibitemOpen
  \bibfield  {author} {\bibinfo {author} {\bibnamefont {{Allen Institute for Brain Science}}},\ }\href@noop {} {\emph {\bibinfo {title} {{Neuropixels Visual Coding}}}},\ \bibinfo {type} {Tech. Rep.}\ (\bibinfo  {institution} {Allen Brain Observatory},\ \bibinfo {year} {2019})\BibitemShut {NoStop}%
\bibitem [{\citenamefont {{Yahoo Finance}}(2025)}]{yahoofinance}%
  \BibitemOpen
  \bibfield  {author} {\bibinfo {author} {\bibnamefont {{Yahoo Finance}}},\ }\href {https://finance.yahoo.com} {\bibinfo {title} {Historical stock data (1994--2024)}} (\bibinfo {year} {2025})\BibitemShut {NoStop}%
\bibitem [{\citenamefont {Aroussi}(2025)}]{yfinance}%
  \BibitemOpen
  \bibfield  {author} {\bibinfo {author} {\bibfnamefont {R.}~\bibnamefont {Aroussi}},\ }\href@noop {} {\bibinfo {title} {yfinance: Yahoo! finance market data downloader}},\ \bibinfo {howpublished} {\url{https://github.com/ranaroussi/yfinance}} (\bibinfo {year} {2025}),\ \bibinfo {note} {used to download stock price data from 1994 to 2024}\BibitemShut {NoStop}%
\bibitem [{\citenamefont {Bonanno}\ \emph {et~al.}(2001)\citenamefont {Bonanno}, \citenamefont {Lillo},\ and\ \citenamefont {Mantegna}}]{Bonannostocks2001}%
  \BibitemOpen
  \bibfield  {author} {\bibinfo {author} {\bibfnamefont {G.}~\bibnamefont {Bonanno}}, \bibinfo {author} {\bibfnamefont {F.}~\bibnamefont {Lillo}},\ and\ \bibinfo {author} {\bibfnamefont {R.}~\bibnamefont {Mantegna}},\ }\href {https://doi.org/10.1080/713665554} {\bibfield  {journal} {\bibinfo  {journal} {Quant. Finance}\ }\textbf {\bibinfo {volume} {1}},\ \bibinfo {pages} {96} (\bibinfo {year} {2001})}\BibitemShut {NoStop}%
\bibitem [{\citenamefont {Klimt}\ and\ \citenamefont {Yang}(2004)}]{klimt_enron_2004}%
  \BibitemOpen
  \bibfield  {author} {\bibinfo {author} {\bibfnamefont {B.}~\bibnamefont {Klimt}}\ and\ \bibinfo {author} {\bibfnamefont {Y.}~\bibnamefont {Yang}},\ }in\ \href@noop {} {\emph {\bibinfo {booktitle} {Machine Learning: ECML 2004}}},\ \bibinfo {editor} {edited by\ \bibinfo {editor} {\bibfnamefont {J.-F.}\ \bibnamefont {Boulicaut}}, \bibinfo {editor} {\bibfnamefont {F.}~\bibnamefont {Esposito}}, \bibinfo {editor} {\bibfnamefont {F.}~\bibnamefont {Giannotti}},\ and\ \bibinfo {editor} {\bibfnamefont {D.}~\bibnamefont {Pedreschi}}}\ (\bibinfo  {publisher} {Springer Berlin Heidelberg},\ \bibinfo {address} {Berlin, Heidelberg},\ \bibinfo {year} {2004})\ pp.\ \bibinfo {pages} {217--226}\BibitemShut {NoStop}%
\bibitem [{\citenamefont {Tang}\ \emph {et~al.}(2013)\citenamefont {Tang}, \citenamefont {Leontiadis}, \citenamefont {Scellato}, \citenamefont {Nicosia}, \citenamefont {Mascolo}, \citenamefont {Musolesi},\ and\ \citenamefont {Latora}}]{tang_enron_2013}%
  \BibitemOpen
  \bibfield  {author} {\bibinfo {author} {\bibfnamefont {J.}~\bibnamefont {Tang}}, \bibinfo {author} {\bibfnamefont {I.}~\bibnamefont {Leontiadis}}, \bibinfo {author} {\bibfnamefont {S.}~\bibnamefont {Scellato}}, \bibinfo {author} {\bibfnamefont {V.}~\bibnamefont {Nicosia}}, \bibinfo {author} {\bibfnamefont {C.}~\bibnamefont {Mascolo}}, \bibinfo {author} {\bibfnamefont {M.}~\bibnamefont {Musolesi}},\ and\ \bibinfo {author} {\bibfnamefont {V.}~\bibnamefont {Latora}},\ }\bibinfo {title} {Applications of temporal graph metrics to real-world networks},\ in\ \href {https://doi.org/10.1007/978-3-642-36461-7_7} {\emph {\bibinfo {booktitle} {Temporal Networks}}},\ \bibinfo {editor} {edited by\ \bibinfo {editor} {\bibfnamefont {P.}~\bibnamefont {Holme}}\ and\ \bibinfo {editor} {\bibfnamefont {J.}~\bibnamefont {Saram{\"a}ki}}}\ (\bibinfo  {publisher} {Springer Berlin Heidelberg},\ \bibinfo {address} {Berlin, Heidelberg},\ \bibinfo {year} {2013})\ pp.\ \bibinfo {pages} {135--159}\BibitemShut {NoStop}%
\bibitem [{\citenamefont {Theiler}\ \emph {et~al.}(1992)\citenamefont {Theiler}, \citenamefont {Eubank}, \citenamefont {Longtin}, \citenamefont {Galdrikian},\ and\ \citenamefont {{Doyne Farmer}}}]{theiler_surrogate_1992}%
  \BibitemOpen
  \bibfield  {author} {\bibinfo {author} {\bibfnamefont {J.}~\bibnamefont {Theiler}}, \bibinfo {author} {\bibfnamefont {S.}~\bibnamefont {Eubank}}, \bibinfo {author} {\bibfnamefont {A.}~\bibnamefont {Longtin}}, \bibinfo {author} {\bibfnamefont {B.}~\bibnamefont {Galdrikian}},\ and\ \bibinfo {author} {\bibfnamefont {J.}~\bibnamefont {{Doyne Farmer}}},\ }\href {https://doi.org/https://doi.org/10.1016/0167-2789(92)90102-S} {\bibfield  {journal} {\bibinfo  {journal} {Physica D: Nonlinear Phenomena}\ }\textbf {\bibinfo {volume} {58}},\ \bibinfo {pages} {77} (\bibinfo {year} {1992})}\BibitemShut {NoStop}%
\bibitem [{\citenamefont {Benjamini}\ and\ \citenamefont {Hochberg}(1995)}]{benjamini1995}%
  \BibitemOpen
  \bibfield  {author} {\bibinfo {author} {\bibfnamefont {Y.}~\bibnamefont {Benjamini}}\ and\ \bibinfo {author} {\bibfnamefont {Y.}~\bibnamefont {Hochberg}},\ }\href {https://doi.org/10.1111/j.2517-6161.1995.tb02031.x} {\bibfield  {journal} {\bibinfo  {journal} {Journal of the Royal Statistical Society: Series B (Methodological)}\ }\textbf {\bibinfo {volume} {57}},\ \bibinfo {pages} {289} (\bibinfo {year} {1995})}\BibitemShut {NoStop}%
\end{thebibliography}%

\clearpage

\onecolumngrid

\renewcommand{\theequation}{S\arabic{equation}}
\renewcommand{\thefigure}{S\arabic{figure}}
\renewcommand{\thetable}{S\arabic{table}}

\setcounter{equation}{0}
\setcounter{figure}{0}
\setcounter{table}{0}
\setcounter{section}{0}

\begin{center}
{\large\bf SUPPLEMENTAL MATERIAL}
\end{center}

\section{Computational Scaling and Optimality} \label{sec:SM_scaling}

We analyze the asymptotic runtime and memory requirements of the algorithm behind our symbolic Bayesian method (SYMBA).  
SYMBA is \emph{asymptotically optimal}, achieving the minimal possible scaling in both runtime and memory usage (up to constant factors) for sequences of length $L$ and all distinct tuples up to length $k_{\text{max}}$.

\subsection*{Algorithm running time}

\begin{enumerate}
    \item \textbf{Counting all overlapping tuples of lengths $1,\dots,k_{\text{max}}$.}\\
    For each tuple length $k$, the algorithm counts all $L - k + 1$ overlapping tuples of $k$ adjacent symbols in the sequence of length $L$.  
    The total number of tuples processed is
    \begin{equation*}
        \sum_{k=1}^{k_{\text{max}}} (L - k + 1) = k_{\text{max}}L - \frac{k_{\text{max}}(k_{\text{max}}-1)}{2}
    \end{equation*}
    For $k_{\text{max}} \ll L$, the quadratic term is negligible, and thus
    \begin{equation*}
        \sum_{k=1}^{k_{\text{max}}} (L - k + 1) \in \Theta(k_{\text{max}}L)
    \end{equation*}

    \item \textbf{Observed probabilities.}\\
    Normalizing the counts for all distinct observed tuples requires work proportional to the number of distinct tuples:
    \begin{equation*}
        \Theta\left(U^{(\leq k_{\text{max}})}\right)
    \end{equation*}
    where $U^{(\leq k_{\text{max}})} = \sum_{k=1}^{k_{\text{max}}}U^{(k)}$ denotes the total number of distinct tuples across lengths $1,\dots,k_{\text{max}}$.

    \item \textbf{Expected probabilities.}\\
    Each expected probability $p_{\text{exp}}$ involves $\Theta(1)$ hash lookups, for a total of
    \begin{equation*}
        \Theta\left(U^{(\leq k_{\text{max}})}\right)
    \end{equation*}

    It is worth noting that, when computing $p_{\text{exp}}$ for unordered tuples, we must account not only for the observed tuples $U^{(k)}$ but also for all their permutations, which may belong to the support $Q^{(k)}$ even if they are not observed in the specific sequence. However, for a fixed $k$, the number of possible permutations is constant and therefore does not alter the asymptotic scaling of the algorithm.

    \item \textbf{Beta prior and posterior updates.}\\
    Each tuple contributes constant work, giving
    \begin{equation*}
        \Theta\left(U^{(\leq k_{\text{max}})}\right)
    \end{equation*}

    \item \textbf{Jensen--Shannon distance (sampled at $P$ points).}\\
    Computing the distance for each tuple requires $\Theta(P)$ operations, leading to
    \begin{equation*}
        \Theta\left(PU^{(\leq k_{\text{max}})}\right)
    \end{equation*}
\end{enumerate}

\noindent
\textbf{Total runtime:}
\begin{equation}
    T\left(L,k_{\text{max}},U^{(\leq k_{\text{max}})},P\right) \in \Theta\left(k_{\text{max}}L + (P+1)U^{(\leq k_{\text{max}})}\right)
\end{equation}
For fixed $P$ and $k_{\text{max}}$, this simplifies to
\begin{equation}
    T\left(L,U^{(\leq k_{\text{max}})}\right) \in \Theta\left(L + U^{(\leq k_{\text{max}})}\right)
\end{equation}

\subsection*{Asymptotic optimality}

Any method that assigns scores to all $k$-tuples based on observed statistics must necessarily process each observed tuple to estimate frequencies and each distinct tuple at least once to assign a score; this fundamental requirement sets an unavoidable lower bound on the computational cost.
Accordingly, any algorithm that computes statistics for all overlapping $k$-tuples of adjacent symbols in a sequence of length $L$ must, at a minimum, inspect each of the $L - k + 1$ observed $k$-tuples once. Summing over $k = 1,\dots,k_{\text{max}}$ gives $k_{\text{max}}L - k_{\text{max}}(k_{\text{max}}-1)/2$ tuples.
In the common regime $k_{\text{max}} \ll L$, this establishes a lower bound of $\Omega\left(k_{\text{max}}L\right)$ time merely to read and count all tuples, which, for fixed $k_{\text{max}}$, reduces to $\Omega(L)$.
No algorithm can asymptotically improve upon this. Moreover, computing expected probabilities and significance scores requires accessing each distinct tuple at least once, an additional $\Omega\left(U^{(\leq k_{\text{max}})}\right)$ operations. Hence, SYMBA achieves the theoretical lower bound on the total amount of work required, making it asymptotically optimal in runtime.

\subsection*{Algorithm memory usage}

The algorithm stores:
\begin{itemize}
    \item the input sequence of length $L$, requiring $\Theta(L)$ memory, and
    \item counts and probabilities for all $U^{(\leq k_{\text{max}})}$ distinct observed tuples, requiring $\Theta\left(U^{(\leq k_{\text{max}})}\right)$ memory.
\end{itemize}
Thus, the total memory requirement is
\begin{equation}
    M\left(L,U^{(\leq k_{\text{max}})}\right) \in \Theta\left(L + U^{(\leq k_{\text{max}})}\right)
\end{equation}

\subsection*{Memory optimality}

This memory usage is asymptotically minimal: any algorithm computing tuple statistics must at least retain or access all $L$ input symbols and maintain per-tuple statistics for the $U^{(\leq k_{\text{max}})}$ distinct tuples observed.  
Reducing memory below $\Theta\left(L + U^{(\leq k_{\text{max}})}\right)$ would require recomputation, thereby increasing runtime beyond the theoretical lower bound.  
Consequently, SYMBA is asymptotically optimal in both runtime and memory.

It is worth noticing that we considered here the case in which one wants to assign a significant score to each observed tuple. However, one could instead be interested in assigning a significance score to a smaller or larger set of tuples, for example just to some specific subset of cardinality $E^{(k)}$ or to all the tuples on the support $Q^{(k)}$ or even to all possible combination of symbols, whose number is given by $N^k$ for ordered tuples of length $k$ created with an alphabet on size $N$. This is indeed possible, as the algorithm does not put any restriction on the set of tuples that can be evaluated. However, in this case the algorithm complexity and memory usage do not scale anymore with $U^{(k_{\max})}$, but instead with the number of tuples actually evaluated (e.g., with $E^{(\leq k_{\max})}$, $Q^{(\leq k_{\max})}$ or $N^{\leq k_{\max}}$ in the above examples). 
This does not affect the optimality of the scaling, as in order to assign a score to a tuple you indeed need to access it at least once and to save the score in memory.

\subsection*{Computational runtime comparison for finite sequence lengths}

\begin{figure}[htp!]
    \centering
    \includegraphics[width=0.7\linewidth]{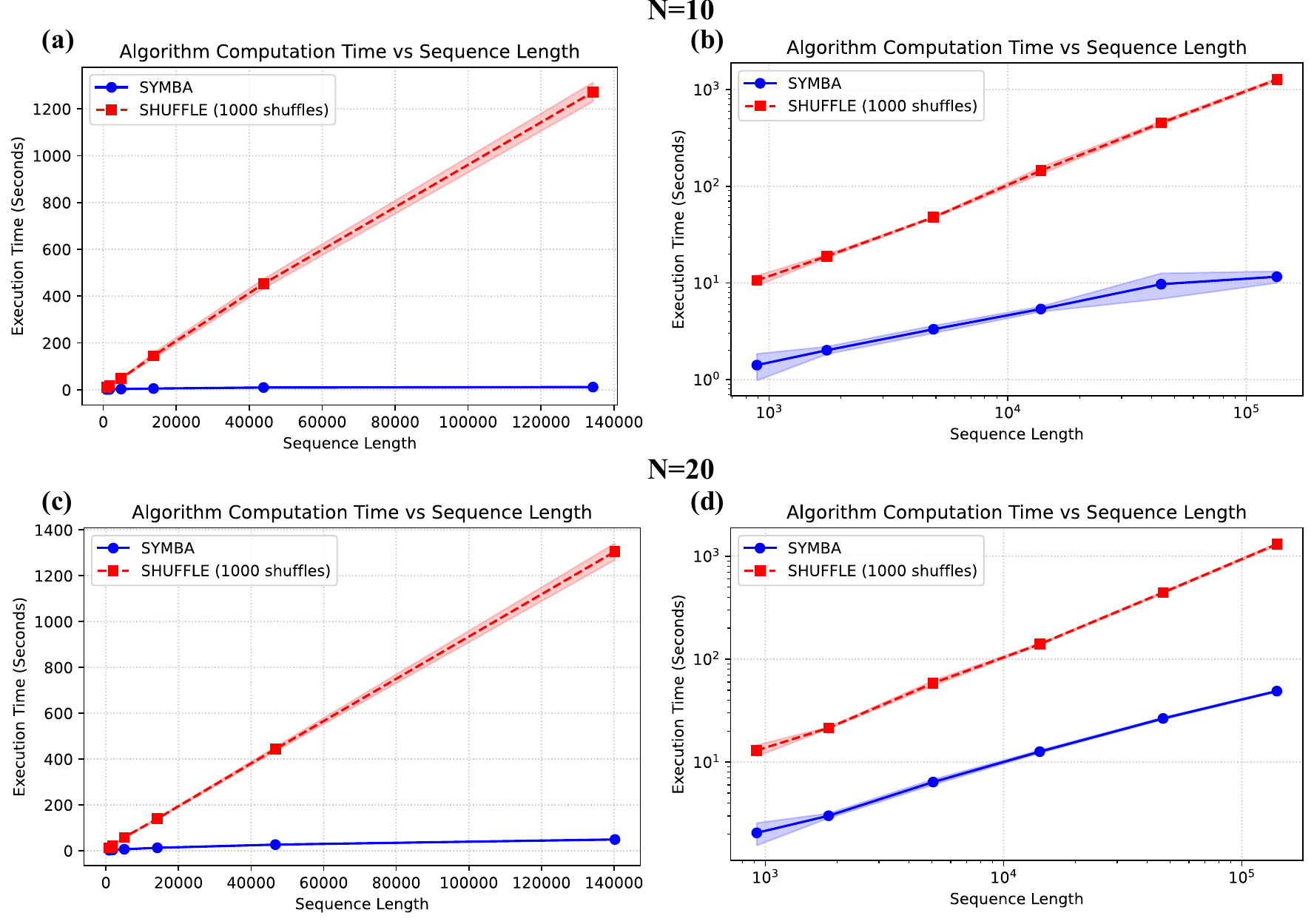}
    \caption{\textbf{Runtime comparison between SYMBA and the shuffle surrogate-data method.} Runtime of SYMBA and the shuffle surrogate-data method as a function of sequence length. Panels (a),(b) refer to synthetic sequences generated from an alphabet of $N=10$ symbols, while panels (c),(d) correspond to $N=20$. In the linear-scale plots (panels (a),(c)), the runtime of SYMBA appears nearly constant on the scale of the figure, as it is substantially faster than the shuffle method, whose computational cost depends on the number of reshuffles of the original sequence (here $B=1000$). To better compare the scaling of the two algorithms, panels (b),(d) show the same data on a log-log scale. The results demonstrate that the favorable asymptotic scaling of SYMBA translates into a significant computational advantage even for finite sequence lengths.}
    \label{fig:runtime}
\end{figure}

In addition to the analytical asymptotic scaling analysis, we show here a concrete runtime comparison between SYMBA and the surrogate-data method (shuffle, described in Sec.~\ref{sec:SM_shuffle}). As shown in Fig.~\ref{fig:runtime}, the runtime of SYMBA scales approximately linearly with the sequence length and is substantially faster than the surrogate approach, whose computational cost increases linearly with the sequence length for each reshuffling procedure and therefore also scales with the number of reshuffles performed.
To ensure a fair comparison, both methods were implemented entirely in Python. These results show that the favourable asymptotic scaling of SYMBA translates into a significant computational advantage also for finite sequence lengths.

\section{Algorithm for synthetic  data creation} \label{sec:SM_synthetic_creation}

We tested SYMBA on synthetic symbolic sequences containing known $k$-motifs with tunable noise. In our application we used $k$ up to $4$, although the algorithm can be applied to any $k$.
To generate these sequences, we used an alphabet of $N$ symbols (integers from $0$ to $N - 1$).

By randomly sampling symbols from the alphabet, we first created a dictionary containing $n_k$ $k$-motifs  (or \textit{$k$-words}) for $k = 2, \dots, k_{\max}$. 
The procedure differs slightly between the ordered and unordered cases, as described below.

\paragraph*{\bf Ordered tuples}
For ordered tuples, we first created the motifs and then discarded any duplicates, retaining only the unique ones.
We then generated a random sentence (a list of tuples) representing the signal by repeating each $k$-word in the dictionary exactly $m_k$ times, for $k = 2, \dots, k_{\max}$. 
Alternatively, the $k$-motifs in the signal could be sampled according to a specified probability distribution. 
We tested both approaches: using constant $m_k$ values (i.e., repeating each $k$-motif exactly $m_k$ times) and drawing instead each of the $m_kn_k$ motifs in the signal with uniform probability among the $n_k$ $k$-words in the dictionary. This second case is equivalent to draw the number of repetition of each $k$-motif from a multivariate distribution with expected value $m_k$ and variance $m_k(n_k - 1)/n_k$. Benchmarking our method for identifying significant tuples on sequences generated with the two sampling schemes yielded comparable predictive performance. 
All results presented in this Supplemental Material and in the main manuscript were obtained using synthetic sequences generated with constant $m_k$.
Finally, we randomized the order of the words in the sentence.

\paragraph*{\bf Unordered tuples}
For unordered tuples, since permutations of the same elements represent the same motif (e.g., $(a,b,c)$ and $(a,c,b)$ are distinct ordered tuples but the same unordered one), we first sorted the elements within each word (alphabetically or in increasing numerical order). 
After this reordering, we kept only unique words, thereby eliminating duplicates that differed only in element order.
As in the ordered case, we generated a random sentence by repeating each $k$-word in the dictionary $m_k$ times for $k = 2, \dots, k_{\max}$. 
Since the internal order of elements in each tuple does not matter in this case, we also applied a final shuffle to the elements within each word.

The resulting random sentence contains $\sum_{k=2}^{k_{\max}} m_k n_k$ motifs representing the signal. 
The total length of the sentence is
\begin{equation}
    L_{\text{s}} = \sum_{k=2}^{k_{\max}} k\, m_k n_k \label{eq:signal_length}
\end{equation}
We then added noise to the sentence by inserting symbols drawn from the alphabet (optionally including the special ``empty-space'' character) between adjacent words.
The level of noise is quantified by the noise-to-signal ratio $r_{\text{ns}}$, defined as the ratio between the total number of symbols in the noise, $L_{\text{n}}$, and the number of symbols in the signal, $L_{\text{s}}$:
\begin{equation}
    r_{\text{ns}} = \frac{L_{\text{n}}}{L_{\text{s}}}
\end{equation}
Therefore, after the addition of noise, we obtain a symbolic sequence of length
\begin{equation}
    L = L_{\text{n}} + L_{\text{s}} = \left(r_{\text{ns}} + 1\right) \sum_{k=2}^{k_{\max}} k\, m_k n_k,
\end{equation}
which scales linearly with the noise-to-signal ratio.
For a fixed $r_{\text{ns}}$, we generate $L_{\text{n}} = r_{\text{ns}} L_{\text{s}}$ noise symbols, which are randomly sampled from the alphabet and inserted between words to form the final noisy sequence.
The frequency distribution $f$ of the symbols in the noise is controlled by a tunable rank-frequency law of the form $f(r) \sim r^{-\gamma}$, where $r$ is the rank of a symbol according to its frequency~\cite{newman_zipf_2005}.
To produce noise with the desired rank-frequency distribution (characterized by exponent $\gamma$), we used the following algorithm:

\begin{enumerate}
    \item We assigned sequential ranks $r_i = 1, 2, \dots, N$ to the $N$ symbols in the alphabet and defined their frequencies as $f_i = r_i^{-\gamma}$. 
    This specifies the rank-frequency relation  corresponding to the chosen exponent $\gamma$.
    
    \item Noise symbols were then randomly drawn with probabilities proportional to their assigned frequencies $f_i$.
\end{enumerate}

\begin{figure}
    \centering
    \includegraphics[width=0.7\linewidth]{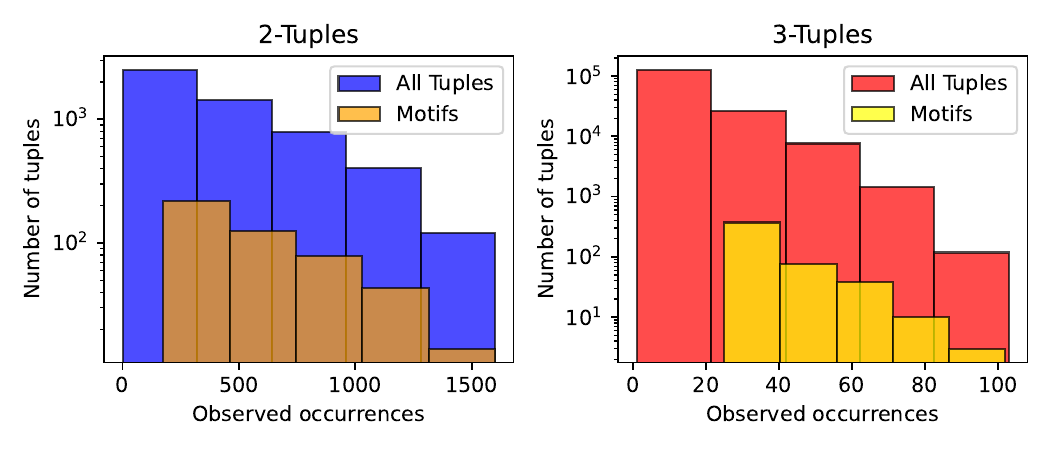}
    \caption{Distributions of the observed occurrences of unique tuples and motifs (i.e., significant tuples) in an artificial symbolic sequence created with our algorithm, for $2$-tuples (left) and $3$-tuples (right). These distributions refers to an artificial symbolic series created using an  alphabet of $100$ symbols (plus the empty-space) and the following parameters: $n_2 = 478$, $n_3 = 499$ (starting from 500 randomly created 2-motifs and 500 3-motifs, with duplicates subsequently removed), $m_2 = 175$, $m_3 = 25$, $r_{\text{ns}} = 10$, and $\gamma = -1$.}
    \label{fig:count_tuples_artificial}
\end{figure}

Fig.~\ref{fig:count_tuples_artificial} shows an example of the distribution of the number of occurrences of $2$- and $3$-tuples, along with the distribution of significant tuples (motifs) of the corresponding sizes, resulting from a realization of the algorithm for artificial symbolic sequences (unordered case). 

In this example, we generated a sequence starting from an alphabet of $100$ symbols (plus the empty-space symbol) and using the following parameters: $n_2 = 478$, $n_3 = 499$ (starting from $500$ randomly created $2$- and $500$ $3$-motifs, with duplicates subsequently removed), $m_2 = 175$, $m_3 = 25$, $r_{\text{ns}} = 10$, and $\gamma = -1$.

\section{Demonstration on synthetic datasets} \label{sec:SM_synthetic_demonstration}

Here, we illustrate how SYMBA operates using a synthetic dataset.

\begin{figure}[ht]
    \centering
    \includegraphics[width=0.6\linewidth]{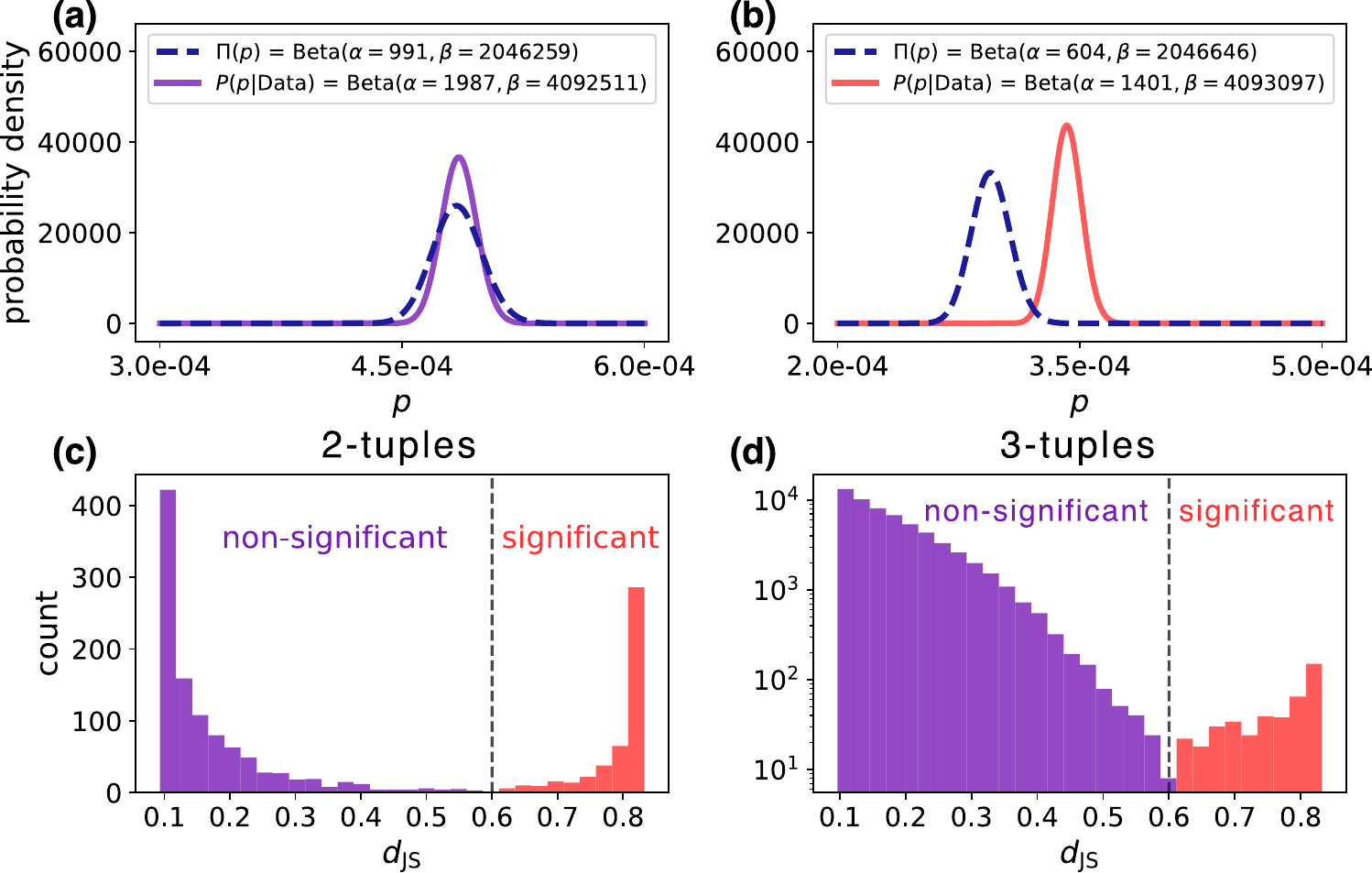}
    \caption{Testing the method on an artificial sequence with noise-to-signal ratio $r_{\text{ns}} = 10$ and $\gamma = -1$ (linear rank-frequency distribution). 
\textbf{(a–b)} Comparison between the marginal distribution of the prior $\Pi(p)$, and that of the posterior $P(p \mid \text{Data})$ for a non-significant tuple and a significant one (i.e., a motif), respectively. 
\textbf{(c–d)} Histogram of the number of $2$- and $3$-tuples as a function of the Jensen-Shannon distance $d_{\text{JS}}$. A logarithmic scale is used for the $y$-axis in panel (d) to improve readability, given the larger number of possible $3$-tuples compared to motifs.
}
    \label{fig:sign_andpdf}
\end{figure}

Figure~\ref{fig:sign_andpdf} shows the application of the method, as described in the main manuscript, to an artificial symbolic sequence generated according to the procedure outlined in Sec.~\ref{sec:SM_synthetic_creation} of the Supplemental Material (SM).
The result reported here are for the case of unordered tuples.
In particular, this sequence was generated using an alphabet of $100$ symbols (plus the special empty-space symbol), and with the following parameters: $n_2 = 478$, $n_3 = 499$ (starting from 500 randomly created 2-motifs and 500 3-motifs, with duplicates subsequently removed), $m_2 = 175$, $m_3 = 25$, $r_{\text{ns}} = 10$, and $\gamma = -1$.
The resulting distribution of the number of occurrences of $2$- and $3$-tuples, along with the distribution of significant tuples (motifs) of the corresponding sizes, is shown in Fig.~\ref{fig:count_tuples_artificial}.

In Fig.~\ref{fig:sign_andpdf}(a–b), we compare the marginal prior $\Pi(p)$ and posterior $P(p\mid\text{Data})$ distributions for a non-significant tuple and a significant one (i.e., a motif). The Jensen-Shannon distance is approximately $d_{\text{JS}} \simeq 0.1$ for the non-significant tuple and $d_{\text{JS}} \simeq 0.8$ for the motif.
In panels (c) and (d), we report the histograms of the number of $2$- and $3$-tuples, respectively, as a function of the Jensen-Shannon distance $d_{\text{JS}}$ between prior and posterior distributions (i.e., the BJS-score, as defined in the main manuscript).
In both cases, we observe that non-significant and significant tuples follow two distinct distributions: the former peaks around $0.1$, while the latter peaks just below $0.83 \approx \sqrt{\ln(2)}$, the maximum Jensen–Shannon distance for our logarithm base.
Here, we choose a hard significance threshold $\text{BJS}^{\text{thr}} = 0.6$, meaning that tuples with a BJS-score exceeding this threshold are classified as significant.

\begin{figure}[htp!]
    \centering
    \includegraphics[width=0.5\linewidth]{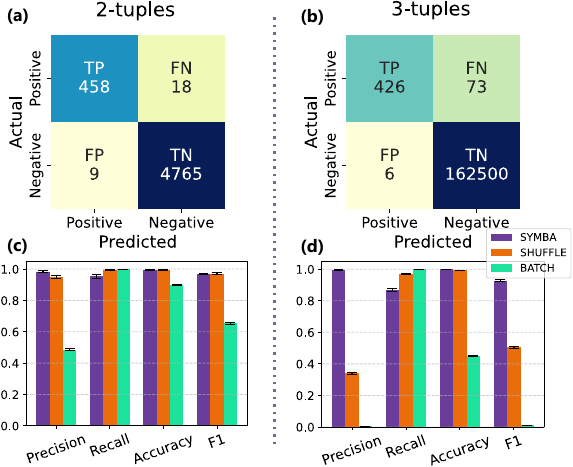}
    \caption{Confusion matrices for SYMBA in the classification of $2$-tuples \textbf{(a)} and $3$-tuples \textbf{(b)} in artificial data, using a significance threshold of $\text{BJS}^{\text{thr}} = 0.6$. \textbf{(c,d)} Bar plots showing the performance metrics of SYMBA with $\text{BJS}^{\text{thr}} = 0.6$ and those of shuffle and batch methods with a threshold of $z^{\text{thr}} = 3$, computed on the same data.
    While in our tests the shuffle and batch methods can achieve marginally better Recall than the SYMBA, we note that Recall alone is not a particularly meaningful indicator of classification performance. In fact, any method labelling all tuples as significant would achieve a Recall (true positive rate) of $1$. A more informative measure is the F1 score, which balances precision and recall.
    SYMBA consistently outperforms the two frequentist methods, particularly in the detection of higher-order motifs, showing higher Precision, Accuracy and F1 score.}
    \label{fig:confusion_metrics}
\end{figure}

In Fig.~\ref{fig:confusion_metrics}, we show the confusion matrices and performance metrics for the classification of $2$- and $3$-motifs, respectively, averaged over $10$ independent realizations of the artificial sequence. These $10$ artificial sequences contain on average $476 \pm 4$ and $499 \pm 1$ unique $2$- and $3$-motifs, respectively (we created $500$ of each, and then removed the duplicates).
Our method demonstrates excellent classification performance in this task, as shown by the confusion matrices and the evaluation metrics (Precision, Recall, Accuracy, and F1 score).
We recall, that Precision, Recall, Accuracy and F1 score are defined in terms of the confusion matrix entrances, as follow \cite{powers_evaluation_2011}: 
\begin{align}
\text{Precision} &= \frac{TP}{TP + FP} \\
\text{Recall} &= \frac{TP}{TP + FN} \\
\text{Accuracy} &= \frac{TP + TN}{TP + TN + FP + FN} \\
\text{F1} &= \frac{2 \cdot \text{Precision} \cdot \text{Recall}}{\text{Precision} + \text{Recall}} = \frac{2TP}{2TP + FP + FN}
\end{align}
where $TP$ is the number of true positive, $FP$ false positive, $TN$ true negative and $FN$ false negative.

For comparison, we report in Fig.~\ref{fig:confusion_metrics}(c-d) also the performance metrics of two alternative frequentist methods for measuring statistical significance, based on the z-score. For the two frequentist approaches, namely the shuffle and batch methods (described in Secs.~\ref{sec:SM_batch},~\ref{sec:SM_shuffle}) we used a standard significance threshold $z^{\text{thr}} = 3$ 
(i.e., considering as significant the tuples whose observed frequency falls outside $3$ standard deviations from the expected frequency).
We observe that SYMBA outperforms the frequentist methods in terms of Precision, Accuracy, and overall performance (measured by the F1 score).
This is especially true for the case of $3$-tuples, where the alternative frequentist approaches find many false positive, causing a dramatic degradation of performances.
For example, we observe that the the F1 score for the classification of $3$-motifs being  $\sim 0.9$ for SYMBA, while for the shuffle and batch methods is $\sim 0.5$ and $\sim 0$, respectively.

To guide the generation of the synthetic sequences and facilitate the interpretation of our tests, in Sec.~\ref{sec:SM_regimes} we identify, for the case of homogeneous symbol frequencies (e.g., $\gamma = 0$), two regimes of particular interest (namely, the data-sparsity and noise-dominated regimes) by characterizing them in terms of the parameters used to generate the symbolic sequence.
In Sec.~\ref{sec:SM_benchmarks_gamma0}, we then systematically benchmark the performance of SYMBA in identifying both ordered and unordered $k$-motifs on synthetic sequences generated using alphabets of different sizes and varying noise-to-signal ratios.
In Sec.~\ref{sec:SM_benchmarks_gammas}, we investigate instead the impact of noise heterogeneity on SYMBA’s performance. We show that, in all considered scenarios, SYMBA outperforms the frequentist batch (Sec.~\ref{sec:SM_batch}) and shuffle (Sec.~\ref{sec:SM_shuffle}) methods, not only in terms of computational scaling (see Sec.~\ref{sec:SM_scaling}), but also in terms of motif recognition, particularly in the relevant data-sparsity regime. Moreover, we find that the performance gap between SYMBA and standard frequentist approaches increases with the heterogeneity of the symbol frequency distribution.
These results demonstrate the robustness and generality of our findings, which hold independently of the specific threshold values and the rank-frequency distribution of symbols in the sequence.

\section{Relevant regimes as a function of the symbolic sequence parameters} \label{sec:SM_regimes}

To guide the generation of the synthetic datasets, facilitate the interpretation of the benchmarks in Secs.~\ref{sec:SM_benchmarks_gamma0},~\ref{sec:SM_benchmarks_gammas}, and gain a clearer understanding of the applicability of our method, we identify two regimes of particular interest and analytically characterize them in terms of the parameters used to generate the symbolic sequence.
These two regimes correspond to the cases of \textit{data-sparsity} and \textit{noise-dominated sequences}.
In the data-sparse regime, the symbolic sequence is too short for the sub-tuples of a motif to occur outside the motif itself, either in the noise or in other motifs, often enough compared with the number of repetitions of the motif.  Their counts then come mostly from the copies of the motif. A null model built from the sub-tuples therefore already predicts how often the motif occurs, and the motif does not stand out as significant.
As a result, motifs are hard to detect for any method that compares their counts with a null model built from their sub-tuples, including SYMBA and shuffling. Detection becomes easier as $r_{\text{ns}}$ grows. In the noise-dominated regime, instead, the counts of the motifs are lost in the fluctuations of the noise, and no method based on counts can identify them.
We characterize these two regimes in terms of the parameters used to generate the synthetic sequences, in particular the alphabet size, the number and repetition of motifs in the signal, and the motif length.
Since our goal is to provide general insight rather than a detailed analytical description, we rely on a set of simplifying assumptions.

We consider a sequence of length $L$ composed of symbols drawn from an alphabet $\mathcal{A}$ of size $|\mathcal{A}|=N$.
We recall that, by definition, $L = L_{\text{s}} + L_{\text{n}}$, where $L_{\text{s}}$ is the number of symbols in the signal (motifs) and $L_{\text{n}}$ is the number of symbols in the noise (i.e., uncorrelated symbols).
Assuming that each of the $n_k$ motifs is repeated in the signal the same number of times $m_k$, the length of the signal is given by Eq.~\eqref{eq:signal_length}:
\[
L_{\text{s}} = \sum_{k=2}^{k_{\max}} k\, m_k n_k
\]
Given the definition of the noise-to-signal ratio $r_{\text{ns}}$, the total length of the sequence can be written as
\begin{equation}
    L = (r_{\text{ns}} + 1) \sum_{k=2}^{k_{\max}} k\, m_k n_k
    \label{eq:length_explicit}
\end{equation}
and the number of noise symbols as $L_{\text{n}} = r_{\text{ns}} L_{\text{s}}$.

The occurrences of a tuple by chance arise from the noise, since the signal contains, by construction, the motifs themselves.
We assume that the noise symbols are independent and equiprobable, i.e., each appears with uniform probability $1/N$.
This corresponds to a rank--frequency distribution with exponent $\gamma = 0$ in our model for synthetic sequences.
Under this assumption, each of the $N^k$ possible ordered $k$-tuples is also independent and equiprobable, occurring with probability $1/N^k$.
Neglecting the tuples formed across motifs and noise, the noise can therefore be viewed as $L_{\text{n}} - k + 1$ independent trials, each corresponding to the realization of one $k$-tuple.
The joint distribution of the counts of all possible $k$-tuples is multinomial,
with $L_{\text{n}} - k + 1$ trials and $N^k$ equiprobable outcomes, each occurring with probability $1/N^k$.
The expected number of chance occurrences (successes) of each $k$-tuple is thus
\begin{equation}
    \mu_k = \frac{L_{\text{n}} - k + 1}{N^k} \label{eq:mu_gamma0}
\end{equation}
and the standard deviation is
\begin{equation}
    \text{std}_k = \sqrt{(L_{\text{n}}-k+1)\frac{N^k-1}{N^{2k}}} \label{eq:std_gamma0}
\end{equation}

Assuming $k \ll L_{\text{n}}$, as is generally the case in practical applications, we can replace $L_{\text{n}} - k + 1$ with $L_{\text{n}}$.
Moreover, we note that $\frac{N^k - 1}{N^{2k}} \simeq \frac{1}{N^k}$, even for small alphabet sizes $N$ and values of $k$ (for example, for $N = 4$ and $k = 2$ we have $\frac{N^k - 1}{N^{2k}} \approx 0.059$ and $\frac{1}{N^k} \approx 0.063$).
Therefore, Eqs.~\eqref{eq:mu_gamma0} and~\eqref{eq:std_gamma0} simplify to:
\begin{equation}
    \mu_k \simeq \frac{L_{\text{n}}}{N^k} = \frac{r_{\text{ns}} L_{\text{s}}}{N^k}
    \label{eq:mu_gamma0_brief}
\end{equation}
and the standard deviation becomes
\begin{equation}
    \text{std}_k \simeq \sqrt{\frac{L_{\text{n}}}{N^{k}}} = \sqrt{\mu_k}
    \label{eq:std_gamma0_brief}
\end{equation}
These expressions hold for tuples of any length; below we also use them for the sub-tuples of length $k-1$ and $k-2$ of a $k$-motif.
We also assume, for simplicity, that motifs are composed of $k$ distinct symbols. Motifs containing repeated symbols change the counts derived below only by factors of order one.

We can now characterize the two regimes of \textit{data-sparsity} and \textit{noise-dominated sequence}.

\subsection*{Data-sparsity regime}

Both SYMBA and the shuffle-based approach evaluate the number of occurrences of a $k$-tuple $\boldsymbol{s} = (s_1, \dots, s_k)$ against a null model built from its sub-tuples.
For $k \geq 3$, the expected number of occurrences of $\boldsymbol{s}$ is
\begin{equation}
    n_{\text{exp}}(\boldsymbol{s}) \simeq \frac{n(s_1, \dots, s_{k-1})\; n(s_2, \dots, s_k)}{n(s_2, \dots, s_{k-1})} ,
    \label{eq:nexp_subtuples}
\end{equation}
where $n(\cdot)$ denotes the observed number of occurrences of a tuple. This is the expected count $p_{\text{exp}}(L-k+1)$ of SYMBA, with $p_{\text{exp}}$ defined in Eq.~(1) of the main text, and it coincides with the mean count of $\boldsymbol{s}$ over sequences shuffled while preserving the counts of all $(k-1)$-tuples.
Consider an ordered $k$-motif repeated $m_k$ times. Each copy contains one occurrence of the prefix $(s_1, \dots, s_{k-1})$, of the suffix $(s_2, \dots, s_k)$ and of the inner $(k-2)$-sub-tuple $(s_2, \dots, s_{k-1})$.
Outside the copies, these sub-tuples occur only by chance, on average $\mu_{k-1}$ and $\mu_{k-2}$ times, respectively (Eq.~\eqref{eq:mu_gamma0_brief}).
The ratio between the observed and the expected number of occurrences of the motif is then
\begin{equation}
    \frac{m_k + \mu_k}{n_{\text{exp}}} \simeq \frac{(m_k + \mu_k)(m_k + \mu_{k-2})}{(m_k + \mu_{k-1})^2} \simeq 1 + \frac{\mu_{k-2}}{m_k} ,
    \label{eq:ratio_obs_exp}
\end{equation}
where in the last step we neglected $\mu_{k-1}$ and $\mu_k$, which are smaller than $\mu_{k-2}$ by a factor $N$ and $N^2$, respectively.
If the inner sub-tuple rarely occurs outside the motif, $\mu_{k-2} \ll m_k$, the null model expects almost all the observed copies, and the motif cannot be distinguished from a non-significant tuple, however many times it is repeated.
We therefore define the data-sparse regime as the condition in which the expected number of chance occurrences of the inner sub-tuple outside the copies of the motif is smaller than the number of repetitions of the motif:
\begin{equation}
    \mu_{k-2} \simeq \frac{r_{\text{ns}} \sum_{k'=2}^{k_{\max}} k'\, m_{k'} n_{k'}}{N^{k-2}} < m_k .
    \label{eq:data_sparse_simple}
\end{equation}
In this regime the data are sparse in the sense that the background (the noise) is too short to provide statistics on the sub-tuples of the motif that are independent of the motif itself.
Increasing $r_{\text{ns}}$ provides such occurrences and improves the detection.
Two remarks are in order.
First, for $k = 2$ the inner sub-tuple is empty, the expected number of occurrences is $n(s_1)\, n(s_2)/L$, and pairs have no data-sparse regime.
Second, if the inner sub-tuple is also part of other motifs, its occurrences in their copies add to $\mu_{k-2}$ and make detection easier; Eq.~\eqref{eq:data_sparse_simple} thus describes the least favourable case.
In the data-sparse regime the detection of motifs is limited for any method that assesses significance from the occurrences of the sub-tuples, including SYMBA and shuffling. Since there $\mu_k = \mu_{k-2}/N^2 < m_k/N^2 \ll m_k$, few non-significant $k$-tuples are expected to occur by chance as often as a motif, so that simply ranking the tuples by their number of occurrences can identify the motifs better than the methods based on a null model built from the sub-tuples. This criterion, however, ignores both the frequencies of the symbols and the fluctuations in the counts of the tuples. For heterogeneous frequencies, it fails when the expected number of chance occurrences of the $k$-tuples composed of the most frequent symbols becomes comparable with $m_k$ (Table~\ref{tab:counting_4tuples}); for equiprobable symbols, it fails, like any method based on counts, in the noise-dominated regime discussed below.

\subsection*{Noise-dominated regime}

We start by noting that $\mu_k$ represents the expected number of occurrences of a $k$-tuple under the assumption of equiprobable, uncorrelated symbols.
However, the motifs forming the signal are expected to appear more frequently than tuples composed of uncorrelated symbols.
In particular, each $k$-motif is repeated $m_k$ times within the signal.
Outside the portion of the signal where each $k$-motif is repeated $m_k$-times, each $k$-motif behaves like any other $k$-tuple, since it is simply one of the $N^k$ possible combinations.
Therefore, we can approximate the expected number of repetitions $\eta_k(\boldsymbol{s})$ as
\begin{align}
    \eta_k(\boldsymbol{s}) &\simeq \mu_k, && \text{for a non-significant $k$-tuple,} \label{eq:num_tuples_simple}
    \\
    \eta_k(\boldsymbol{s}) &\simeq \mu_k + m_k, && \text{for a $k$-motif.} \label{eq:num_motifs_simple}
\end{align}
Thus, for $\gamma = 0$, the expected number of occurrences of a $k$-motif, $m_k + \mu_k$, is always larger than that of a non-significant $k$-tuple, $\mu_k$.
However, this does not imply that a tuple can be correctly classified as a motif or as non-significant simply by counting its number of occurrences.
Indeed, the fluctuations in the occurrence counts of the $k$-tuples play a crucial role.
When the standard deviation is comparable to $m_k$, a substantial number of non-significant tuples may appear a number of times greater than $\mu_k$ and comparable to, or even larger than, $m_k + \mu_k$.
The condition that we use to define the \textit{noise-dominated regime} is therefore that the typical counts of the non-significant $k$-tuples and of the motifs overlap,
\begin{equation}
    \mu_k + \text{std}_k > m_k + \mu_k - \text{std}_k ,
\end{equation}
where $\text{std}_k \simeq \sqrt{\mu_k}$ is the standard deviation of the number of chance occurrences, Eq.~\eqref{eq:std_gamma0_brief}. This gives
\begin{equation}
    2\sqrt{\mu_k} = 2\sqrt{ \frac{r_{\text{ns}} \sum_{k'=2}^{k_{\max}} k'\, m_{k'} n_{k'}}{N^k} } > m_k .
    \label{eq:noise_dominate_simple}
\end{equation}

\subsection*{Noise-to-signal ratio in the two regimes}

We can use Eq.~\eqref{eq:data_sparse_simple} to derive the lower limit of $r_{\text{ns}}$ below which the system enters the data-sparse regime.
Similarly, Eq.~\eqref{eq:noise_dominate_simple} provides the upper limit of $r_{\text{ns}}$ above which the system enters the noise-dominated regime.
Taken together, these conditions delimit the range of $r_{\text{ns}}$ that lies outside the two regimes:
\begin{equation}
    r_{\text{ns}}^{\text{low}} = \frac{m_k N^{k-2}}{\sum_{k'=2}^{k_{\max}} k' m_{k'} n_{k'}}
    < r_{\text{ns}} <
    \frac{m_k^2 N^k}{4 \sum_{k'=2}^{k_{\max}} k' m_{k'} n_{k'}} = r_{\text{ns}}^{\text{up}} ,
    \label{eq:thresholds_noise}
\end{equation}
where the lower limit applies to $k \geq 3$ (for $k = 2$, $r_{\text{ns}}^{\text{low}} = 0$).
For the operationally simpler case in which the signal contains motifs of a single length $k$, these conditions reduce to
\begin{equation}
    r_{\text{ns}}^{\text{low}} = \frac{m_k N^{k-2}}{k\, m_k n_k}
    < r_{\text{ns}} <
    \frac{m_k^2 N^k}{4k\, m_k n_k}= r_{\text{ns}}^{\text{up}} \label{eq:thresholds_noise_onek}
\end{equation}

We note that as $r_{\text{ns}}$ increases while keeping all other parameters fixed (i.e., as the sequence length increases for a fixed signal length), the system moves, for $k \geq 3$, from the data-sparse regime to the noise-dominated regime, passing through an intermediate region where motif detection is expected to be easier: the sub-tuples of the motifs occur often enough outside the motifs, yet the noise does not dominate the signal.
Moreover, both thresholds grow with the alphabet size $N$, but at different rates: $r_{\text{ns}}^{\text{up}}$ scales as $N^k$, whereas $r_{\text{ns}}^{\text{low}}$ scales as $N^{k-2}$, i.e. it is absent for $k = 2$, grows linearly with $N$ for $k = 3$, and becomes large only for $k \geq 4$. Expressed in terms of the noise length, leaving the data-sparse regime requires roughly $L_{\text{n}} \gtrsim m_k N^{k-2}$, i.e., each inner sub-tuple must occur by chance about as many times as the motif is repeated. In particular, for large alphabets (e.g., symbolic sequences generated by systems with many units), the noise-dominated regime is reached only for very long sequences (i.e., for exceptionally high values of $r_{\text{ns}}$, given the signal parameters), whereas the data-sparse regime is relevant only for long motifs: for $N = 300$ and $m_k = 10$, leaving it requires $L_{\text{n}} \gtrsim 3 \times 10^3$ for $k = 3$, but $L_{\text{n}} \gtrsim 10^6$ for $k = 4$. This means that, in real-world sequences with many symbols obtained from the activity of systems with hundreds or thousands of units (for example neuropixel datasets), one can typically expect to be far from the noise-dominated regime, and outside the data-sparse regime for short motifs, while motifs of length $k \geq 4$ may lie in it.

We note that the identification of the two regimes was carried out under the assumption of ordered tuples.
For unordered tuples, the analysis of the noise-dominated regime can be readily extended by replacing $N^k$ (the number of distinct ordered $k$-tuples) with $\binom{N + k - 1}{k}$ (the number of distinct unordered $k$-tuples).
The data-sparse regime, instead, depends on how the copies of an unordered motif are distributed over the $k!$ possible orderings of its symbols.
We consider copies that appear with their symbols in random order, as in our synthetic sequences (for motifs with repeated symbols, $k!$ is replaced by the number of distinct orderings).
The null model for an unordered tuple sums Eq.~\eqref{eq:nexp_subtuples} over all its orderings.
If each ordering is repeated several times ($m_k \gg k!$), the copies of the motif contain each of its ordered sub-tuples in different positions: a given ordered $(k-1)$-sub-tuple appears in a copy with probability $2/k!$, and a given ordered $(k-2)$-sub-tuple with probability $6/k!$, but only in one of its three possible positions as the inner sub-tuple. In the absence of noise, summing Eq.~\eqref{eq:nexp_subtuples} over the $k!$ orderings then gives $n_{\text{exp}} \simeq k!\,(2 m_k/k!)^2/(6 m_k/k!) = 2 m_k/3$, for any $k \geq 3$.
The occurrences of the motif are thus not fully predictable from those of its sub-tuples: about one third of the copies remains in excess of the null expectation even without noise, and no data-sparse regime arises. For $k = 3$ the excess is even larger, since the inner sub-tuple is a single symbol, which typically also occurs in other motifs.
If, instead, the copies are few compared with the number of orderings ($m_k \lesssim k!$, as is typical for long motifs), most copies have an ordering not shared by any other copy. Each copy then behaves as an ordered motif repeated once, which the null model accounts for unless its inner sub-tuple also occurs by chance, i.e., unless $\mu_{k-2} \gtrsim 1$.
In this case the data-sparse regime is recovered, with the lower threshold of Eq.~\eqref{eq:thresholds_noise} obtained by replacing $m_k$ with 1.
For a signal containing motifs of a single length $k$, the thresholds for unordered tuples read
\begin{equation}
    r_{\text{ns}}^{\text{low}} =
    \frac{N^{k-2}}{k\, m_k n_k}
    < r_{\text{ns}} <
    \frac{m_k^2 \binom{N + k - 1}{k}}{4 k\, m_k n_k}
    = r_{\text{ns}}^{\text{up}} ,
    \label{eq:thresholds_noise_onek_unordered}
\end{equation}
where the lower threshold applies only for $m_k \lesssim k!$.
In the intermediate case, $m_k \sim k!$, the behaviour lies between these two limits.
Hence, for unordered motifs with $m_k \lesssim k!$ the system goes through the same three regimes as for ordered motifs, with a lower threshold smaller by a factor $m_k$, whereas for $m_k \gg k!$ only the noise-dominated regime is present.

Although these expressions rely on the assumption of equiprobable symbols (i.e., a noise rank--frequency distribution with $\gamma = 0$), they can still serve as a qualitative guide when generating artificial symbolic sequences with different values of $\gamma$. For $\gamma > 0$, sub-tuples made of frequent symbols cross the thresholds at lower values of $r_{\text{ns}}$, and those made of rare symbols at higher values.

Finally, we stress that the above derivation is an approximation, meant to provide only the order of magnitude of the two thresholds. In particular, Eq.~\eqref{eq:noise_dominate_simple} compares the counts of the motifs with those of a typical non-significant tuple. However, since the number of non-significant tuples is large, some of them deviate from their expected count by much more than one standard deviation, and can occur more often than the motifs already at lower values of $r_{\text{ns}}$. For example, for $k = 3$, with $N = 30$, $n_k = 20$ and $m_k = 10$, this lowers $r_{\text{ns}}^{\text{up}}$ by a factor of $2.5$--$4$, without changing its order of magnitude.

\section{Benchmarks on synthetic symbolic sequences} \label{sec:SM_benchmarks_gamma0}

\begin{figure}[htp!]
    \centering
    \includegraphics[width=1.\linewidth]{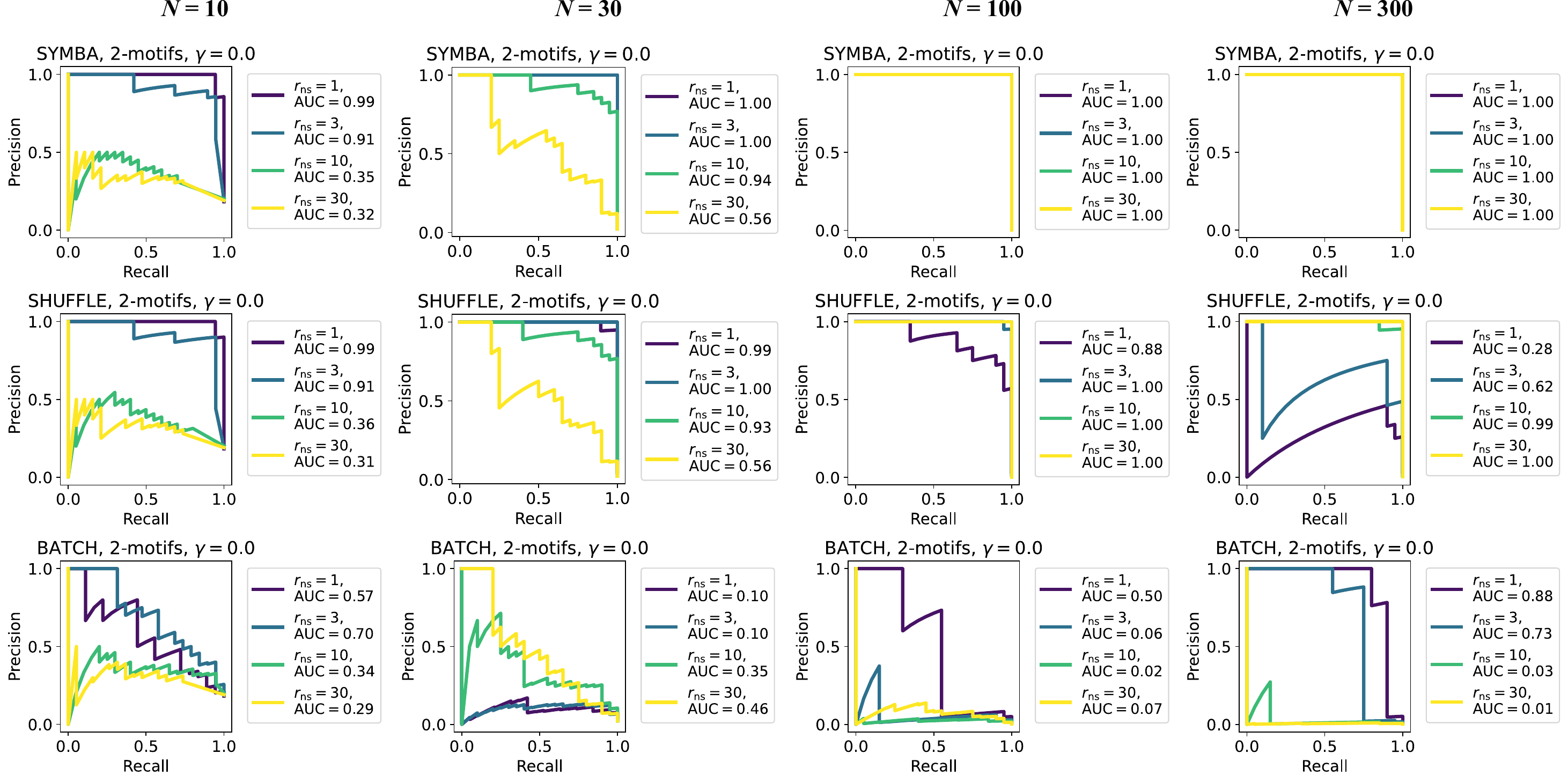}
    \caption{PR curves showing the performance of SYMBA (top panels), SHUFFLE (middle), and BATCH (bottom) in the identification of ordered $2$-motifs. 
    Each column corresponds to a different alphabet size $N$, increasing from left to right. In each panel, different curves correspond to different noise-to-signal ratios $r_{\text{ns}}$, and the corresponding Area Under the Curve (AUC) values are reported in the legend. The noise rank-frequency distribution follows $f(r) = r^{-\gamma}$ with $\gamma = 0$. The signal consists of $n_2 = 20$ distinct ordered $2$-tuples, each repeated $m_2 = 10$ times.}
\label{fig:pr_3methods_2_motifs_dir}
\end{figure}

\begin{figure}[htp!]
    \centering
    \includegraphics[width=1.\linewidth]{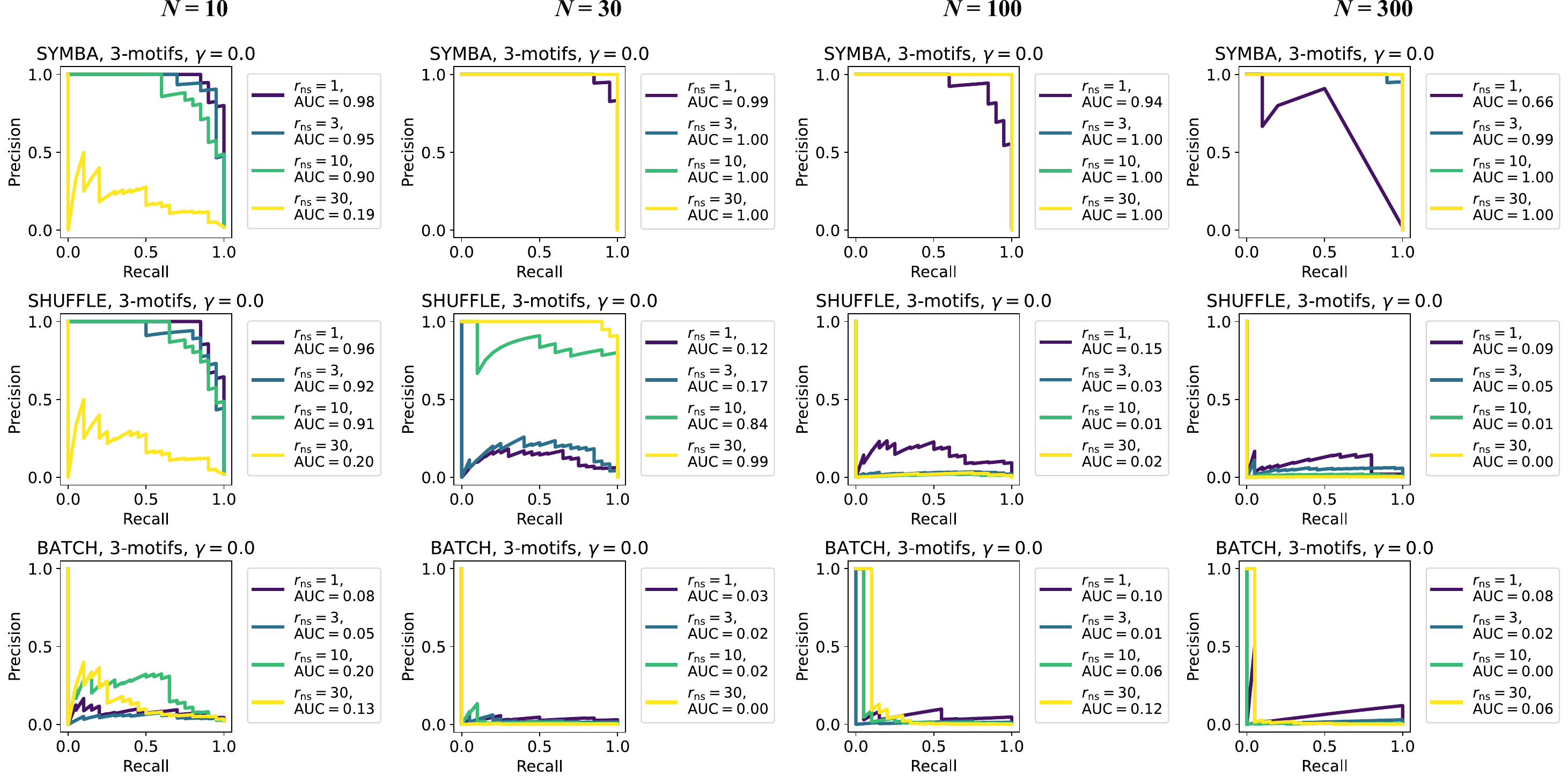}
    \caption{PR curves showing the performance of SYMBA (top panels), SHUFFLE (middle), and BATCH (bottom) in the identification of ordered $3$-motifs. 
    Each column corresponds to a different alphabet size $N$, increasing from left to right. In each panel, different curves correspond to different noise-to-signal ratios $r_{\text{ns}}$, and the corresponding Area Under the Curve (AUC) values are reported in the legend. The noise rank-frequency distribution follows $f(r) = r^{-\gamma}$ with $\gamma = 0$. The signal consists of $n_3 = 20$ distinct ordered $3$-tuples, each repeated $m_3 = 10$ times.}
    \label{fig:pr_3methods_3_motifs_dir}
\end{figure}

\begin{figure}[htp!]
    \centering
    \includegraphics[width=1.\linewidth]{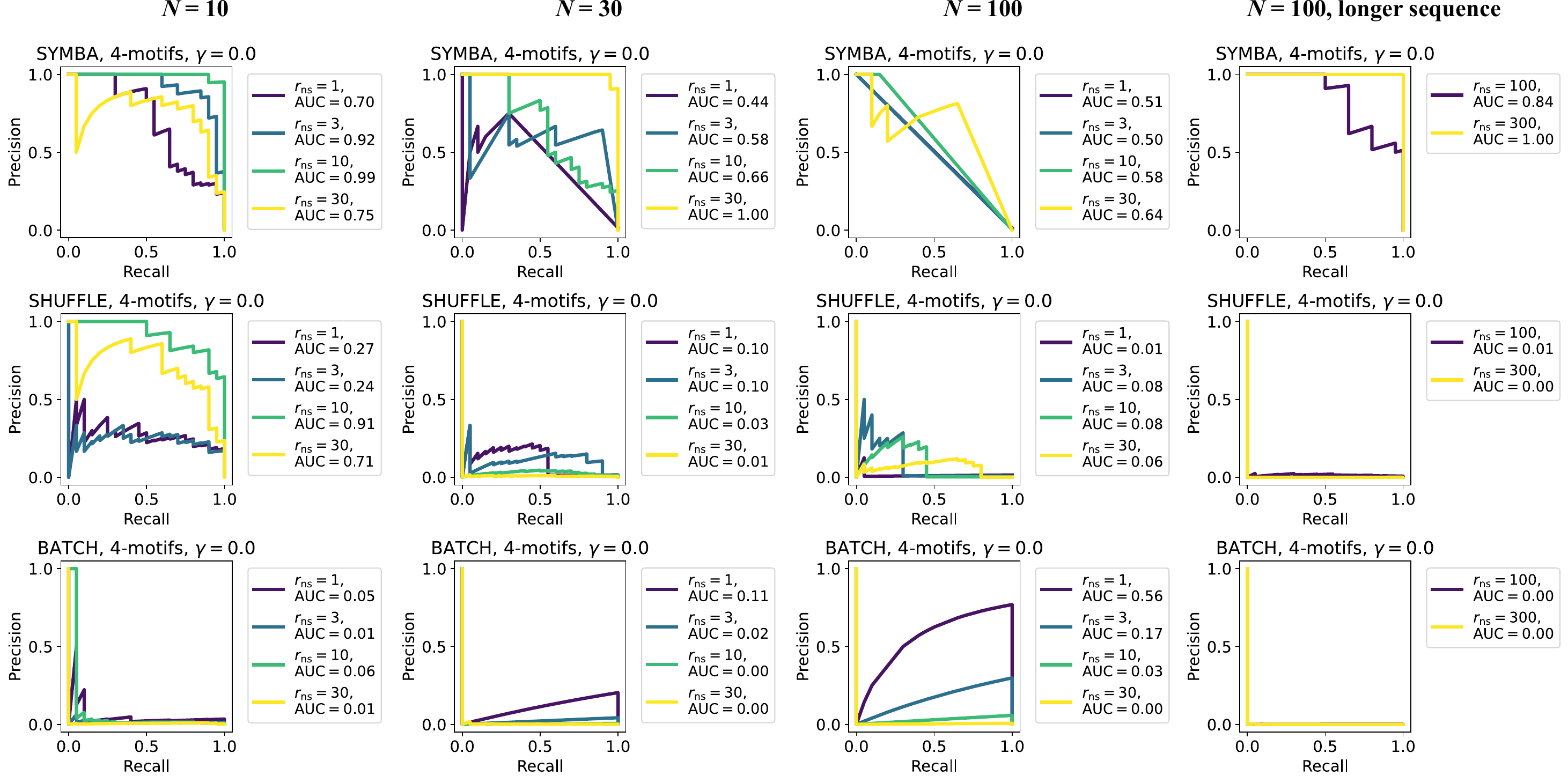}
    \caption{PR curves showing the performance of SYMBA (top panels), SHUFFLE (middle), and BATCH (bottom) in the identification of ordered $4$-motifs. 
    Each column corresponds to a different alphabet size $N$, increasing from left to right. In each panel, different curves correspond to different noise-to-signal ratios $r_{\text{ns}}$, and the corresponding Area Under the Curve (AUC) values are reported in the legend. Given the degradation in performance already observed for $N = 100$ due to the extremely sparse statistical regime, the rightmost column shows results for longer sequences (obtained by using higher values of $r_{\text{ns}}$), which mitigate data-sparsity and enable good classification performance for SYMBA. The noise rank-frequency distribution follows $f(r) = r^{-\gamma}$ with $\gamma = 0$. The signal consists of $n_4 = 20$ distinct ordered $4$-tuples, each repeated $m_4 = 10$ times.}
    \label{fig:pr_3methods_4_motifs_dir}
\end{figure}

\begin{figure}[htp!]
    \centering
    \includegraphics[width=1.\linewidth]{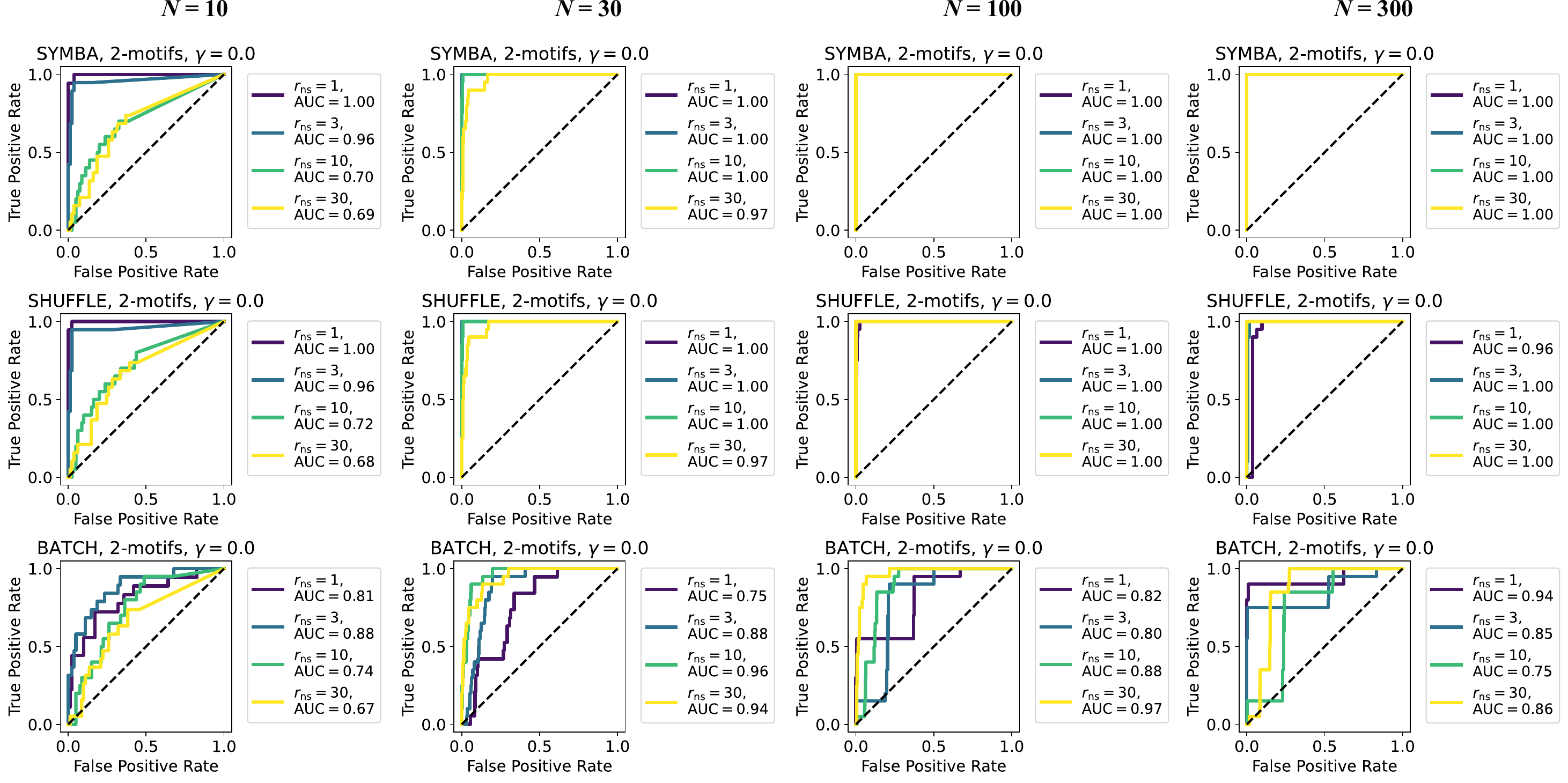}
    \caption{ROC curves showing the performance of SYMBA (top panels), SHUFFLE (middle), and BATCH (bottom) in the identification of ordered $2$-motifs. 
    Each column corresponds to a different alphabet size $N$, increasing from left to right. In each panel, different curves correspond to different noise-to-signal ratios $r_{\text{ns}}$, and the corresponding Area Under the Curve (AUC) values are reported in the legend. The noise rank-frequency distribution follows $f(r) = r^{-\gamma}$ with $\gamma = 0$. The signal consists of $n_2 = 20$ distinct ordered $2$-tuples, each repeated $m_2 = 10$ times.}
    \label{fig:roc_3methods_2_motifs_dir}
\end{figure}

\begin{figure}[htp!]
    \centering
    \includegraphics[width=1.\linewidth]{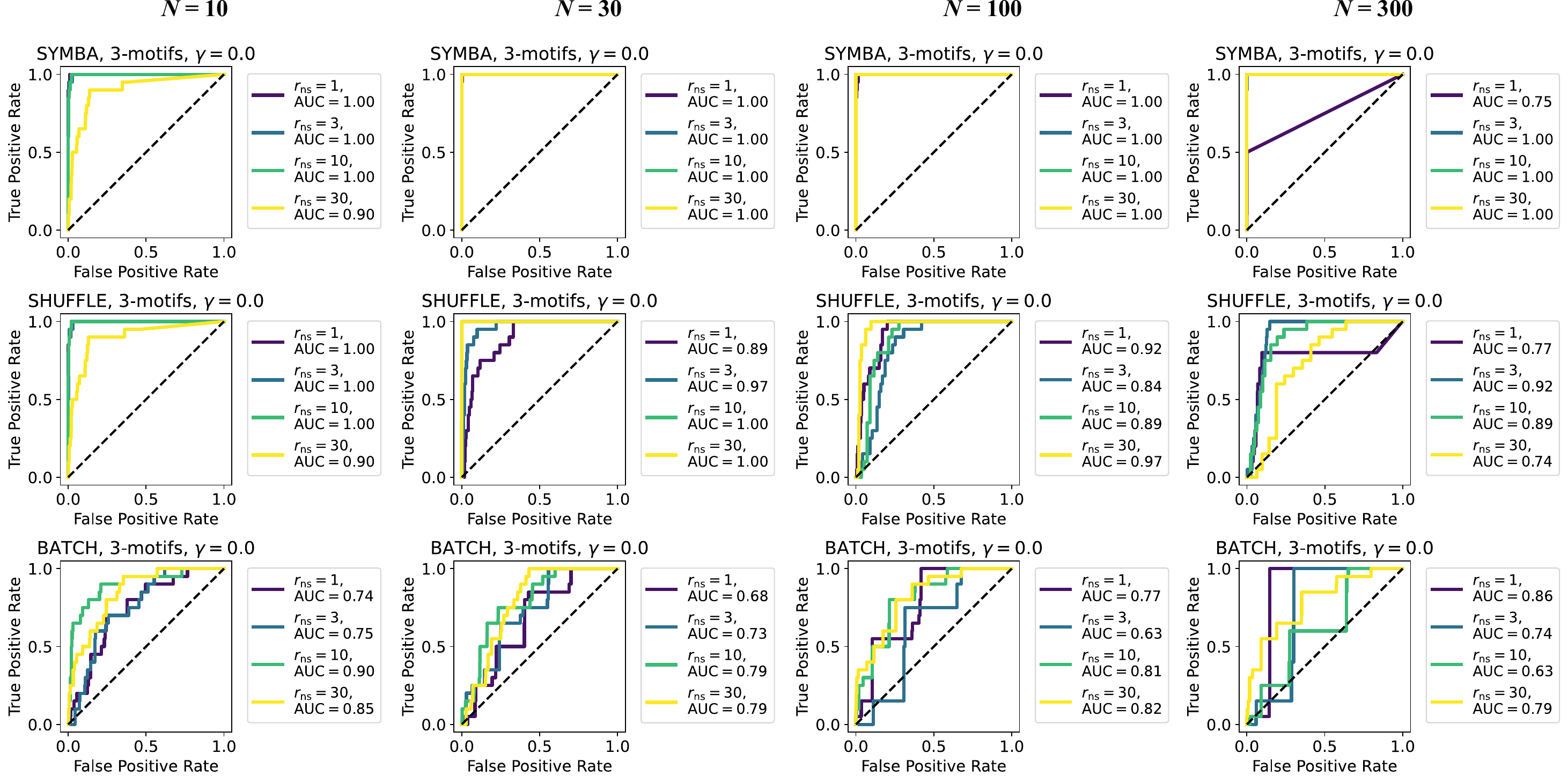}
    \caption{ROC curves showing the performance of SYMBA (top panels), SHUFFLE (middle), and BATCH (bottom) in the identification of ordered $3$-motifs. 
    Each column corresponds to a different alphabet size $N$, increasing from left to right. In each panel, different curves correspond to different noise-to-signal ratios $r_{\text{ns}}$, and the corresponding Area Under the Curve (AUC) values are reported in the legend. The noise rank-frequency distribution follows $f(r) = r^{-\gamma}$ with $\gamma = 0$. The signal consists of $n_3 = 20$ distinct ordered $3$-tuples, each repeated $m_3 = 10$ times.}
    \label{fig:roc_3methods_3_motifs_dir}
\end{figure}

\begin{figure}[htp!]
    \centering
    \includegraphics[width=1.\linewidth]{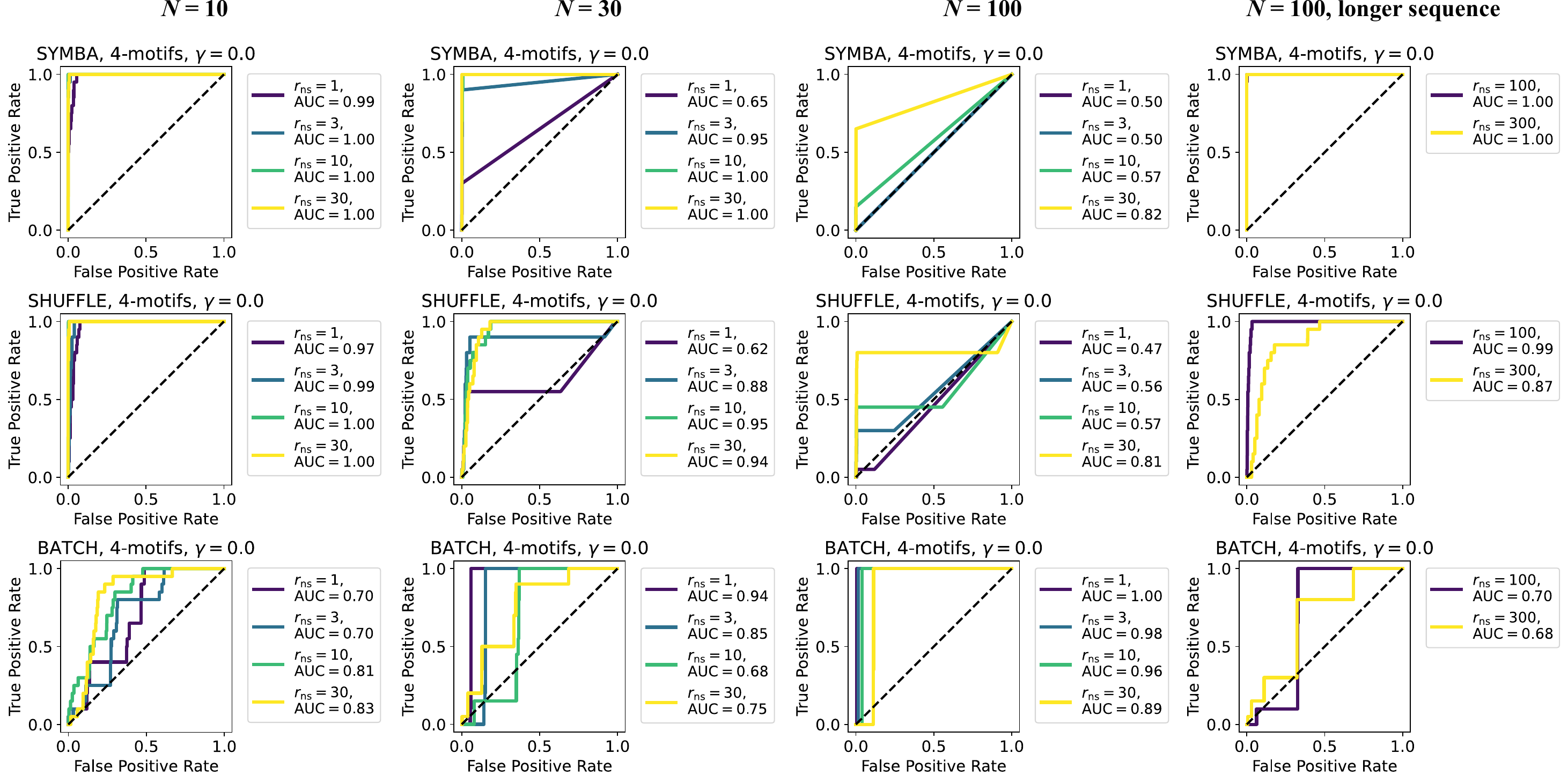}
    \caption{ROC curves showing the performance of SYMBA (top panels), SHUFFLE (middle), and BATCH (bottom) in the identification of ordered $4$-motifs. 
    Each column corresponds to a different alphabet size $N$, increasing from left to right. In each panel, different curves correspond to different noise-to-signal ratios $r_{\text{ns}}$, and the corresponding Area Under the Curve (AUC) values are reported in the legend. Given the degradation in performance already observed for $N = 100$ due to the extremely sparse statistical regime, the rightmost column shows results for longer sequences (obtained by using higher values of $r_{\text{ns}}$), which mitigate data-sparsity and enable good classification performance for SYMBA. The noise rank-frequency distribution follows $f(r) = r^{-\gamma}$ with $\gamma = 0$. The signal consists of $n_4 = 20$ distinct ordered $4$-tuples, each repeated $m_4 = 10$ times.}
    \label{fig:roc_3methods_4_motifs_dir}
\end{figure}

\begin{figure}[htp!]
    \centering
    \includegraphics[width=1.\linewidth]{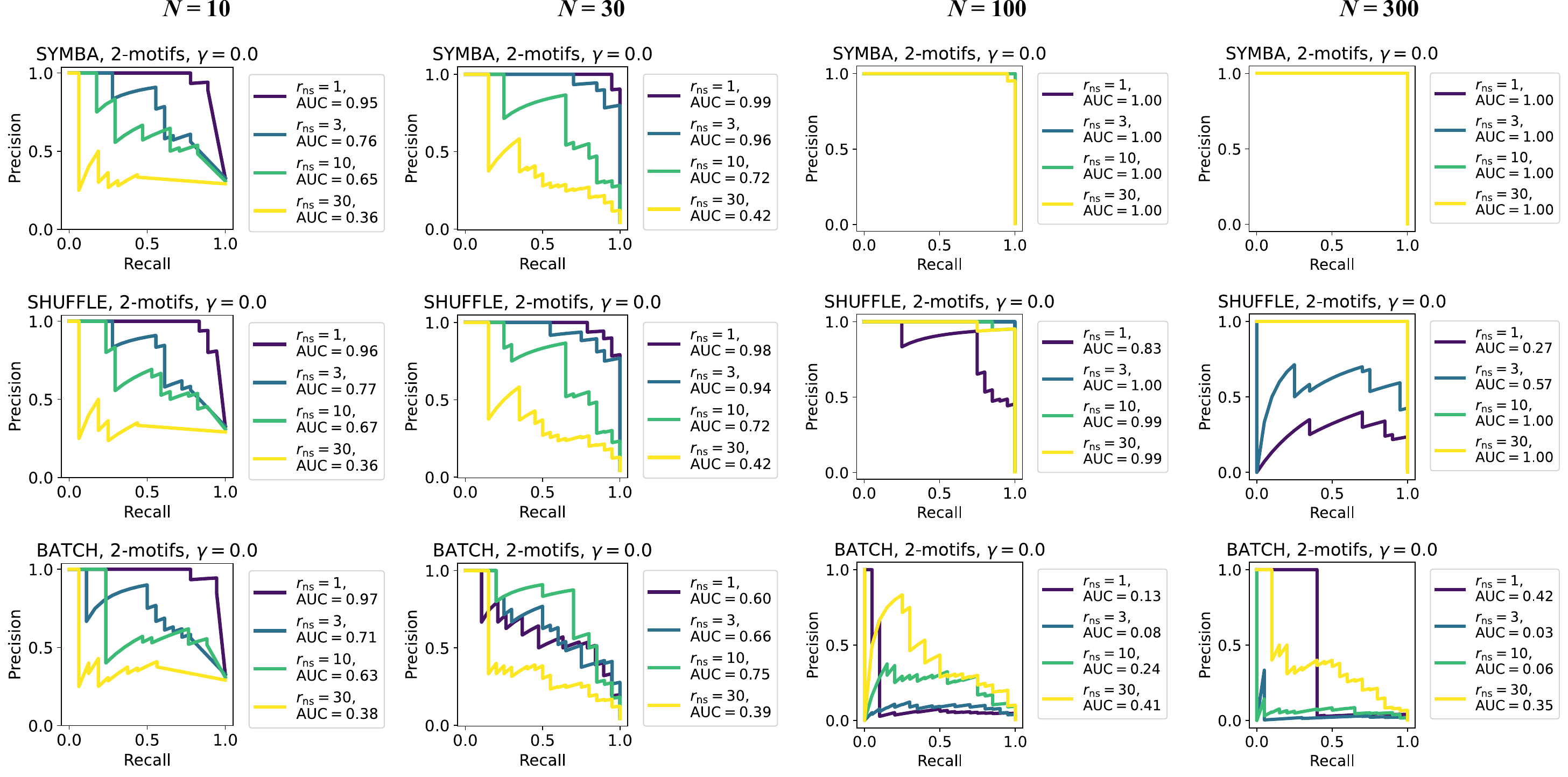}
    \caption{PR curves showing the performance of SYMBA (top panels), SHUFFLE (middle), and BATCH (bottom) in the identification of unordered $2$-motifs. 
    Each column corresponds to a different alphabet size $N$, increasing from left to right. In each panel, different curves correspond to different noise-to-signal ratios $r_{\text{ns}}$, and the corresponding Area Under the Curve (AUC) values are reported in the legend. The noise rank-frequency distribution follows $f(r) = r^{-\gamma}$ with $\gamma = 0$. The signal consists of $n_2 = 20$ distinct unordered $2$-tuples, each repeated $m_2 = 10$ times.}
    \label{fig:pr_3methods_2_motifs_und}
\end{figure}

\begin{figure}[htp!]
    \centering
    \includegraphics[width=1.\linewidth]{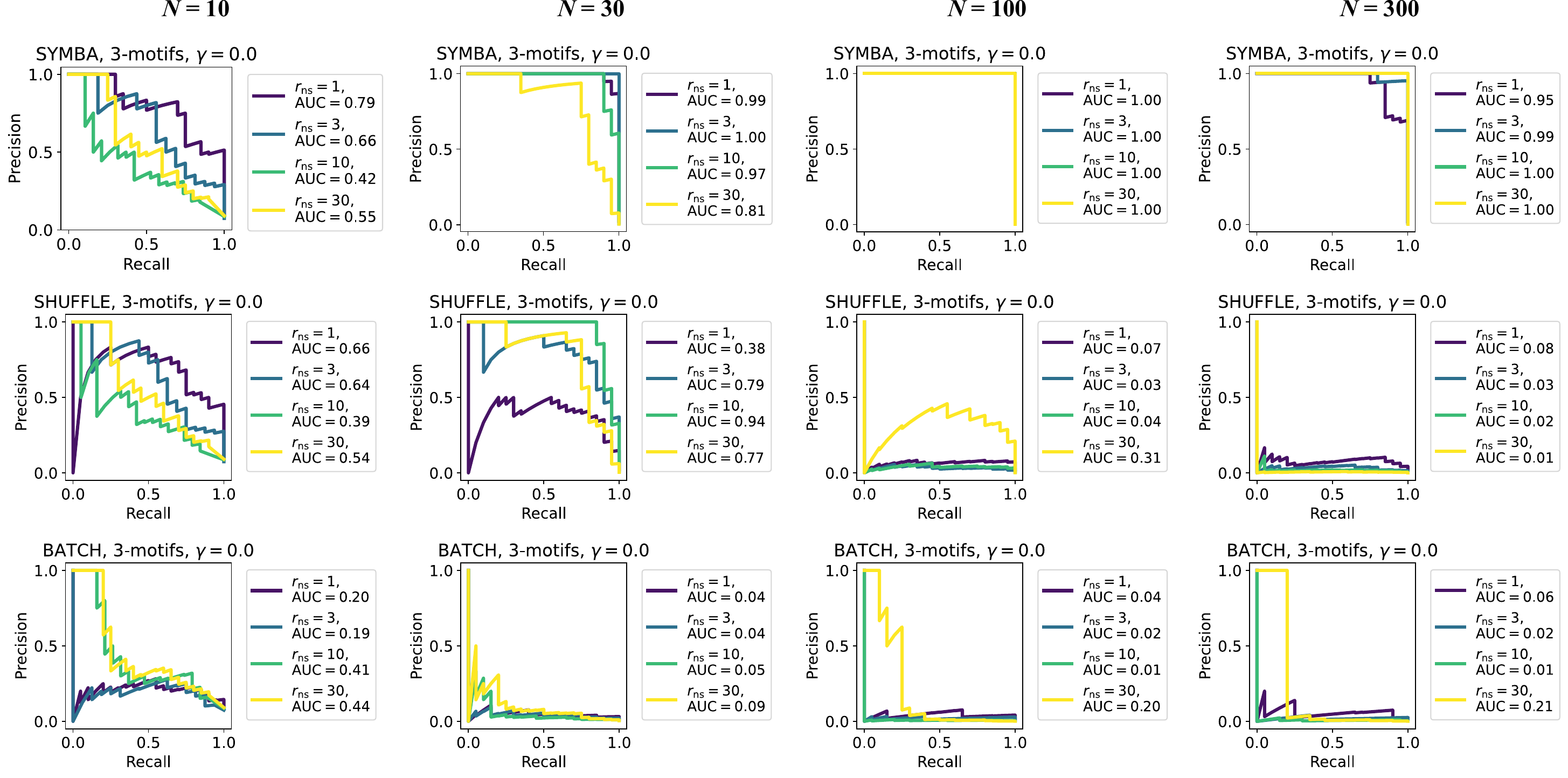}
    \caption{PR curves showing the performance of SYMBA (top panels), SHUFFLE (middle), and BATCH (bottom) in the identification of unordered $3$-motifs. 
    Each column corresponds to a different alphabet size $N$, increasing from left to right. In each panel, different curves correspond to different noise-to-signal ratios $r_{\text{ns}}$, and the corresponding Area Under the Curve (AUC) values are reported in the legend. The noise rank-frequency distribution follows $f(r) = r^{-\gamma}$ with $\gamma = 0$. The signal consists of $n_3 = 20$ distinct unordered $3$-tuples, each repeated $m_3 = 10$ times.}
    \label{fig:pr_3methods_3_motifs_und}
\end{figure}

\begin{figure}[htp!]
    \centering
    \includegraphics[width=1.\linewidth]{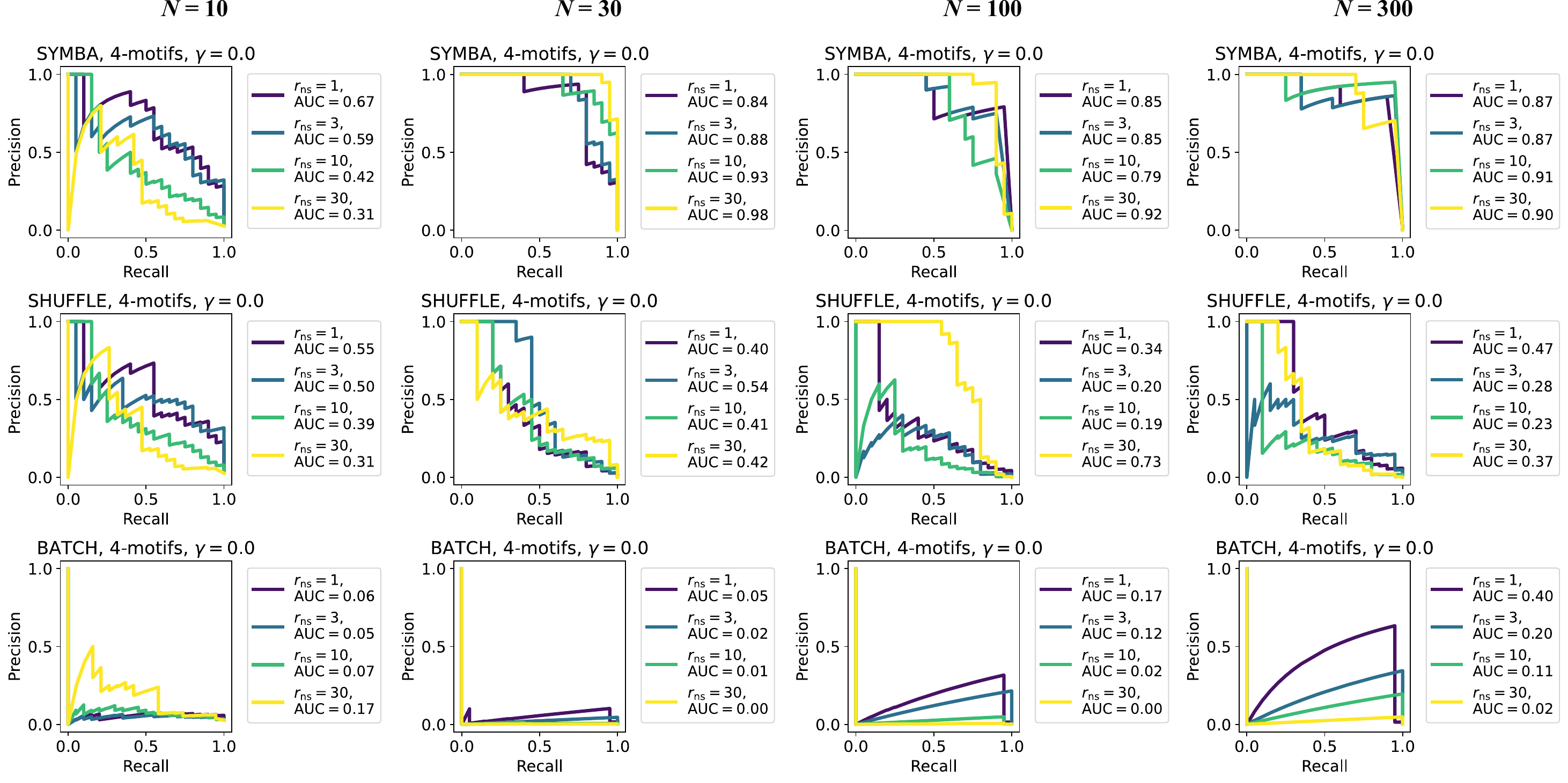}
    \caption{PR curves showing the performance of SYMBA (top panels), SHUFFLE (middle), and BATCH (bottom) in the identification of unordered $4$-motifs. 
    Each column corresponds to a different alphabet size $N$, increasing from left to right. In each panel, different curves correspond to different noise-to-signal ratios $r_{\text{ns}}$, and the corresponding Area Under the Curve (AUC) values are reported in the legend. The noise rank-frequency distribution follows $f(r) = r^{-\gamma}$ with $\gamma = 0$. The signal consists of $n_4 = 20$ distinct unordered $4$-tuples, each repeated $m_4 = 10$ times.}
    \label{fig:pr_3methods_4_motifs_und}
\end{figure}

\begin{figure}[htp!]
    \centering
    \includegraphics[width=1.\linewidth]{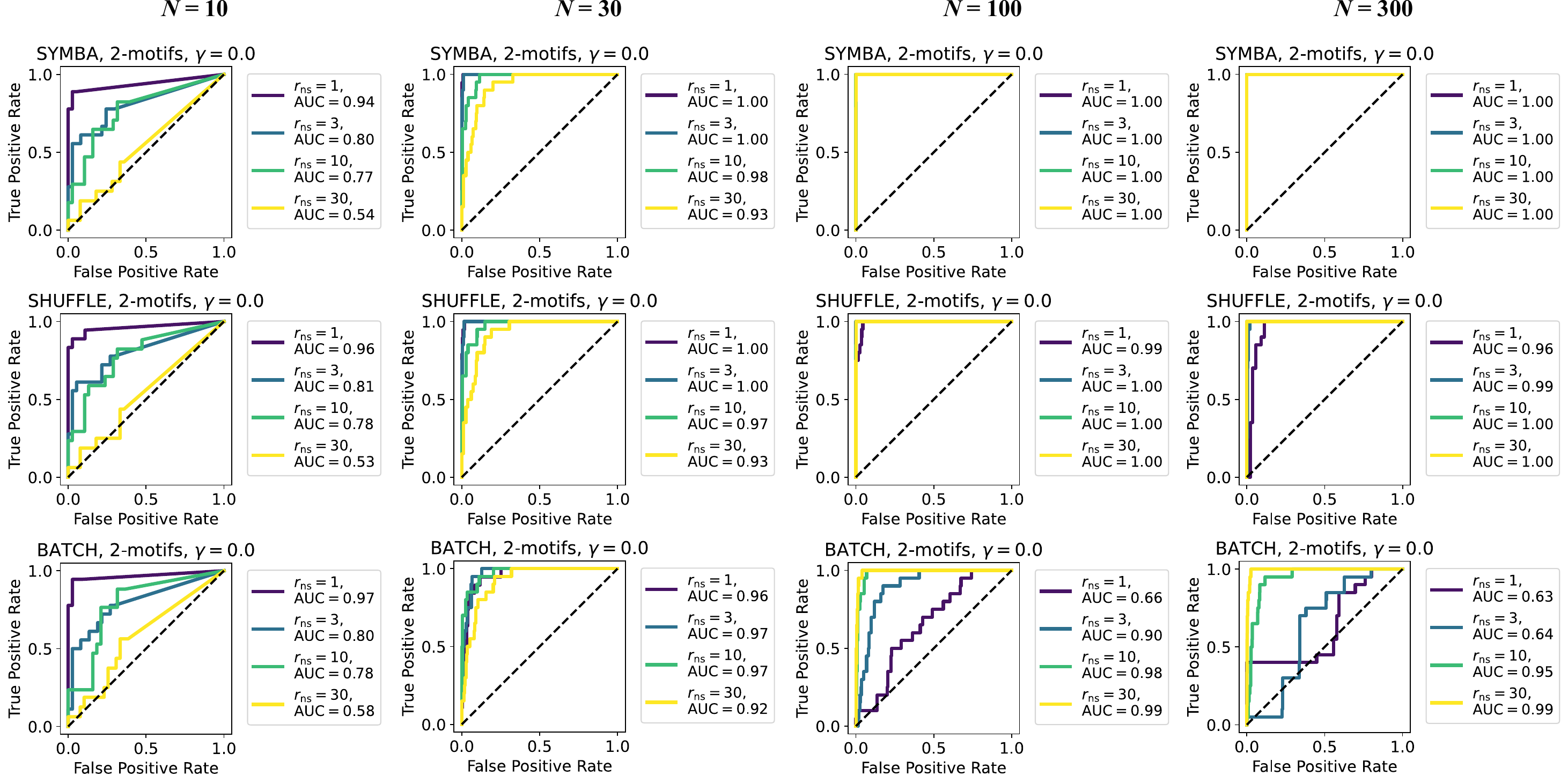}
    \caption{ROC curves showing the performance of SYMBA (top panels), SHUFFLE (middle), and BATCH (bottom) in the identification of unordered $2$-motifs. 
    Each column corresponds to a different alphabet size $N$, increasing from left to right. In each panel, different curves correspond to different noise-to-signal ratios $r_{\text{ns}}$, and the corresponding Area Under the Curve (AUC) values are reported in the legend. The noise rank-frequency distribution follows $f(r) = r^{-\gamma}$ with $\gamma = 0$. The signal consists of $n_2 = 20$ distinct unordered $2$-tuples, each repeated $m_2 = 10$ times.}
    \label{fig:roc_3methods_2_motifs_und}
\end{figure}

\begin{figure}[htp!]
    \centering
    \includegraphics[width=1.\linewidth]{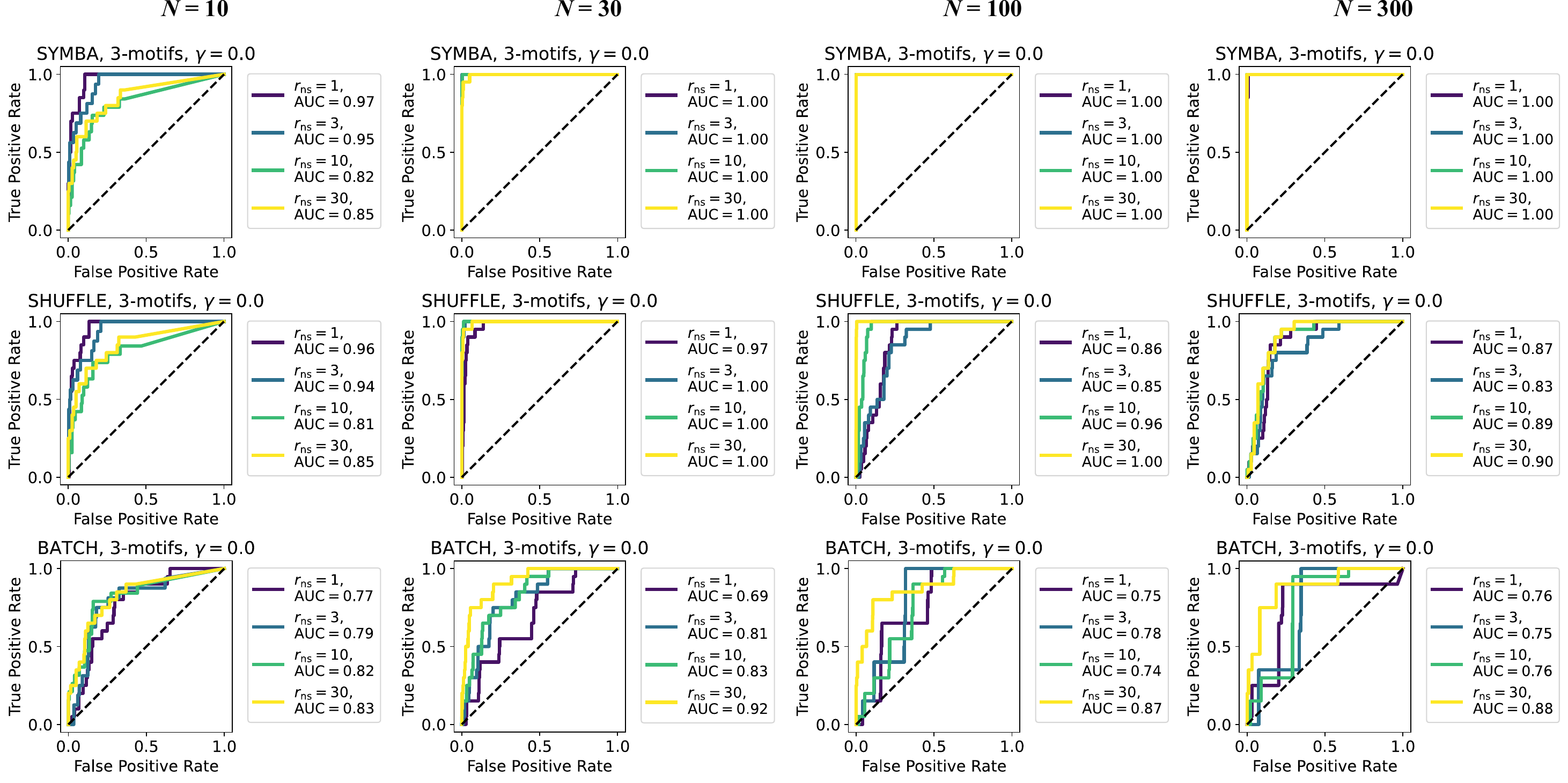}
    \caption{ROC curves showing the performance of SYMBA (top panels), SHUFFLE (middle), and BATCH (bottom) in the identification of unordered $3$-motifs. 
    Each column corresponds to a different alphabet size $N$, increasing from left to right. In each panel, different curves correspond to different noise-to-signal ratios $r_{\text{ns}}$, and the corresponding Area Under the Curve (AUC) values are reported in the legend. The noise rank-frequency distribution follows $f(r) = r^{-\gamma}$ with $\gamma = 0$. The signal consists of $n_3 = 20$ distinct unordered $3$-tuples, each repeated $m_3 = 10$ times.}
    \label{fig:roc_3methods_3_motifs_und}
\end{figure}

\begin{figure}[htp!]
    \centering
    \includegraphics[width=1.\linewidth]{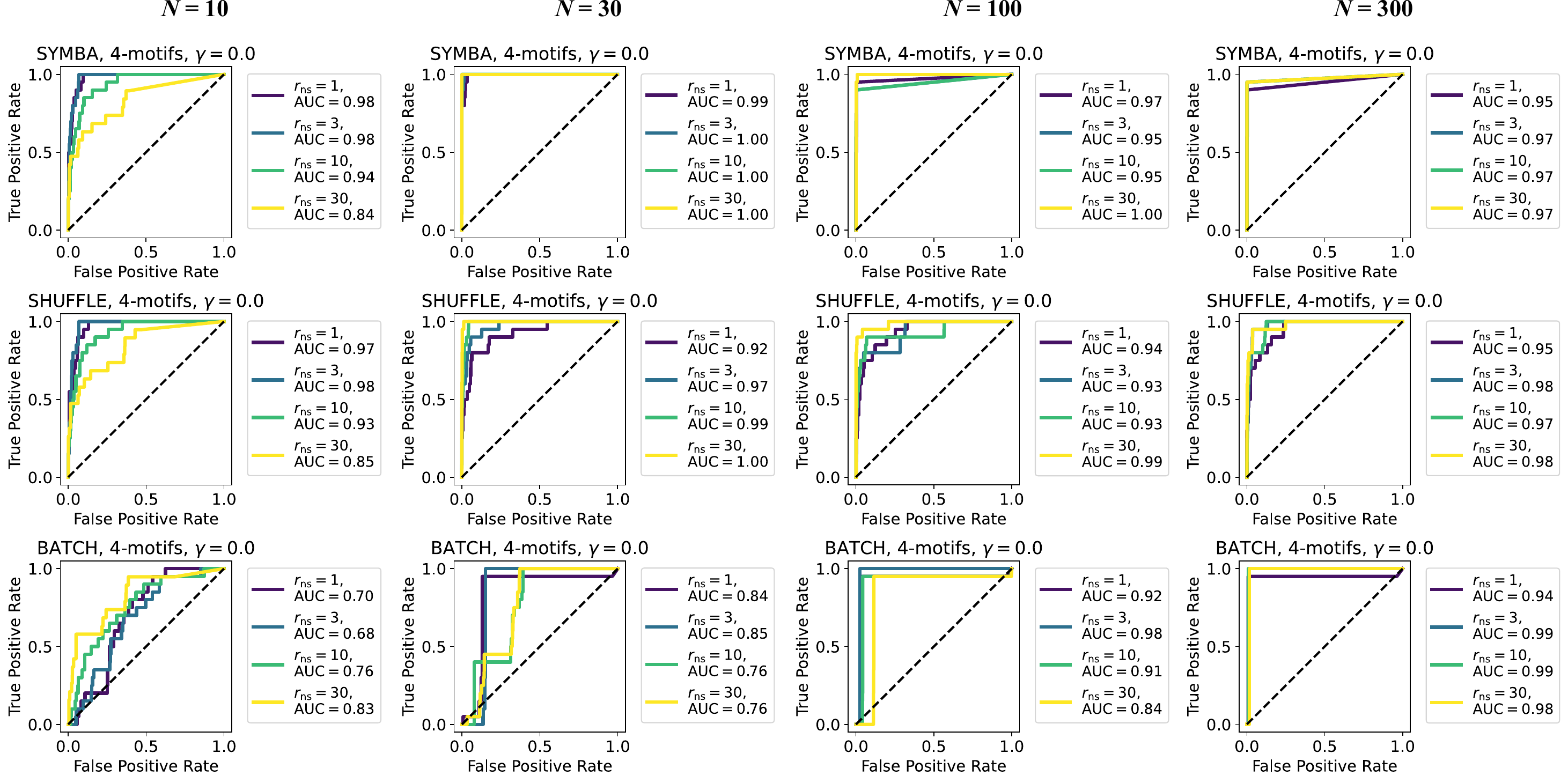}
    \caption{ROC curves showing the performance of SYMBA (top panels), SHUFFLE (middle), and BATCH (bottom) in the identification of unordered $4$-motifs. 
    Each column corresponds to a different alphabet size $N$, increasing from left to right. In each panel, different curves correspond to different noise-to-signal ratios $r_{\text{ns}}$, and the corresponding Area Under the Curve (AUC) values are reported in the legend. The noise rank-frequency distribution follows $f(r) = r^{-\gamma}$ with $\gamma = 0$. The signal consists of $n_4 = 20$ distinct unordered $4$-tuples, each repeated $m_4 = 10$ times.}
    \label{fig:roc_3methods_4_motifs_und}
\end{figure}

We benchmarked SYMBA against the classification performance of the shuffle (Sec.~\ref{sec:SM_shuffle}) and batch (Sec.~\ref{sec:SM_batch}) methods, showing that SYMBA outperforms both frequentist approaches, not only in computational efficiency (as demonstrated by the optimality analysis), but also in motif-recognition accuracy.
Figures~\ref{fig:pr_3methods_2_motifs_dir}–\ref{fig:roc_3methods_4_motifs_dir} report the precision–recall (PR) and receiver operating characteristic (ROC) curves for the recognition of ordered $2$-, $3$-, and $4$-motifs. Figures~\ref{fig:pr_3methods_2_motifs_und}–\ref{fig:roc_3methods_4_motifs_und} report the corresponding results for unordered motifs. In both cases, results are shown for different values of the alphabet size $N$ and the noise-to-signal ratio $r_{\text{ns}}$, with $\gamma = 0$ for the noise distribution. For all synthetic sequences, we used $n_k = 20$ distinct $k$-tuples, each repeated $m_k = 20$ times.
In Tables~\ref{tab:thresholds_noise_ordered} and \ref{tab:thresholds_noise_unordered}, we report the values of $r_{\text{ns}}^{\text{low}}$ and $r_{\text{ns}}^{\text{up}}$, which are respectively the thresholds for the data-sparse and noise-dominated regimes, for ordered (Eq.~\eqref{eq:thresholds_noise_onek}) and unordered (Eq.~\eqref{eq:thresholds_noise_onek_unordered}) tuples.
We observe that the performance of SYMBA is consistently at least as good as that of shuffle method, and the advantage becomes increasingly pronounced as data-sparsity increases. Batch method is consistently the method with the worst performance. This is expected, since dividing the original sequence into multiple batches exacerbates the data-sparsity problem in each subsequence.
In the noise-dominated regime (i.e., for values of $r_{\text{ns}} > r_{\text{ns}}^{\text{up}}$), we observe, as expected, a degradation of performance for all methods, with SYMBA and shuffle method showing similar classification accuracy. In this regime, the main advantage of SYMBA lies in its computational efficiency, being orders of magnitude faster than the surrogate-based method (depending on the number of reshuffles).

\begin{table}[htp!]
\centering
\begin{tabular}{|c|c|c|c|c|c|c|c|c|}
\hline
\multirow{2}{*}{\textbf{Ordered}} 
  & \multicolumn{2}{c|}{$N=10$} 
  & \multicolumn{2}{c|}{$N=30$} 
  & \multicolumn{2}{c|}{$N=100$} 
  & \multicolumn{2}{c|}{$N=300$} \\
\cline{2-9}
  & $r_{\text{ns}}^{\text{low}}$ & $r_{\text{ns}}^{\text{up}}$
  & $r_{\text{ns}}^{\text{low}}$ & $r_{\text{ns}}^{\text{up}}$
  & $r_{\text{ns}}^{\text{low}}$ & $r_{\text{ns}}^{\text{up}}$ 
  & $r_{\text{ns}}^{\text{low}}$ & $r_{\text{ns}}^{\text{up}}$ \\
\hline
$2$-tuples & 0.25 & 6.25 & 2.25 & 56.25 & 25 & 625 & 225 & 5625 \\
$3$-tuples & 1.67 & 41.67 & 45 & 1125 & 1666.67 & 41666.67 & 45000 & 1125000 \\
$4$-tuples & 12.5 & 312.5 & 1012.5 & 25312.5 & 125000 & 3125000 & 10125000 & 253125000 \\
\hline
\end{tabular}
\caption{Noise-to-signal ratio thresholds for the data-sparse ($r_{\text{ns}}^{\text{low}}$) and noise-dominated ($r_{\text{ns}}^{\text{up}}$) regimes, for the case of ordered tuples Eq.~\eqref{eq:thresholds_noise_onek}. For all values of $k$ and alphabet size $N$, we used $n_k = 20$ and $m_k = 10$.} \label{tab:thresholds_noise_ordered}
\end{table}

\begin{table}[htp!]
\centering
\begin{tabular}{|c|c|c|c|c|c|c|c|c|}
\hline
\multirow{2}{*}{\textbf{Unordered}} 
  & \multicolumn{2}{c|}{$N=10$} 
  & \multicolumn{2}{c|}{$N=30$} 
  & \multicolumn{2}{c|}{$N=100$} 
  & \multicolumn{2}{c|}{$N=300$} \\
\cline{2-9}
  & $r_{\text{ns}}^{\text{low}}$ & $r_{\text{ns}}^{\text{up}}$
  & $r_{\text{ns}}^{\text{low}}$ & $r_{\text{ns}}^{\text{up}}$
  & $r_{\text{ns}}^{\text{low}}$ & $r_{\text{ns}}^{\text{up}}$ 
  & $r_{\text{ns}}^{\text{low}}$ & $r_{\text{ns}}^{\text{up}}$ \\
\hline
$2$-tuples & 0.14 & 3.44 & 1.16 & 29.06 & 12.63 & 315.63 & 112.88 & 2821.88 \\
$3$-tuples & 0.37 & 9.17 & 8.27 & 206.67 & 286.17 & 7154.17 & 7575.17 & 189379.17 \\
$4$-tuples & 0.89 & 22.35 & 51.15 & 1278.75 & 5526.59 & 138164.85 & 430364.16 & 10759103.91 \\
\hline
\end{tabular}
\caption{Noise-to-signal ratio thresholds for the data-sparse ($r_{\text{ns}}^{\text{low}}$) and noise-dominated ($r_{\text{ns}}^{\text{up}}$) regimes, for the case of unordered tuples Eq.~\eqref{eq:thresholds_noise_onek_unordered}. For all values of $k$ and alphabet size $N$, we used $n_k = 20$ and $m_k = 10$.}
\label{tab:thresholds_noise_unordered}
\end{table}


\section{Impact of noise distribution} \label{sec:SM_benchmarks_gammas}

We tested the impact of the symbol distribution and noise level on the method's performance by using artificial symbolic sequences with varying values of $\gamma$, the exponent of the rank-frequency distribution $f(r)\sim r^{-\gamma}$. To isolate the effect of $\gamma$, all other parameters in the generation of the artificial sequences were kept fixed. In particular, we employed an alphabet of $N=30$ symbols and we generated $n_3=20$ $3$-motifs, each motif was repeated $m_3=10$ times.
For each value of $\gamma$, we generated symbolic sequences with various noise levels. Specifically, we used noise-to-signal ratios $r_{\text{ns}} \in \{1, 3, 10, 30\}$. 

\begin{figure}[htp!]
    \centering
    \includegraphics[width=1.\linewidth]{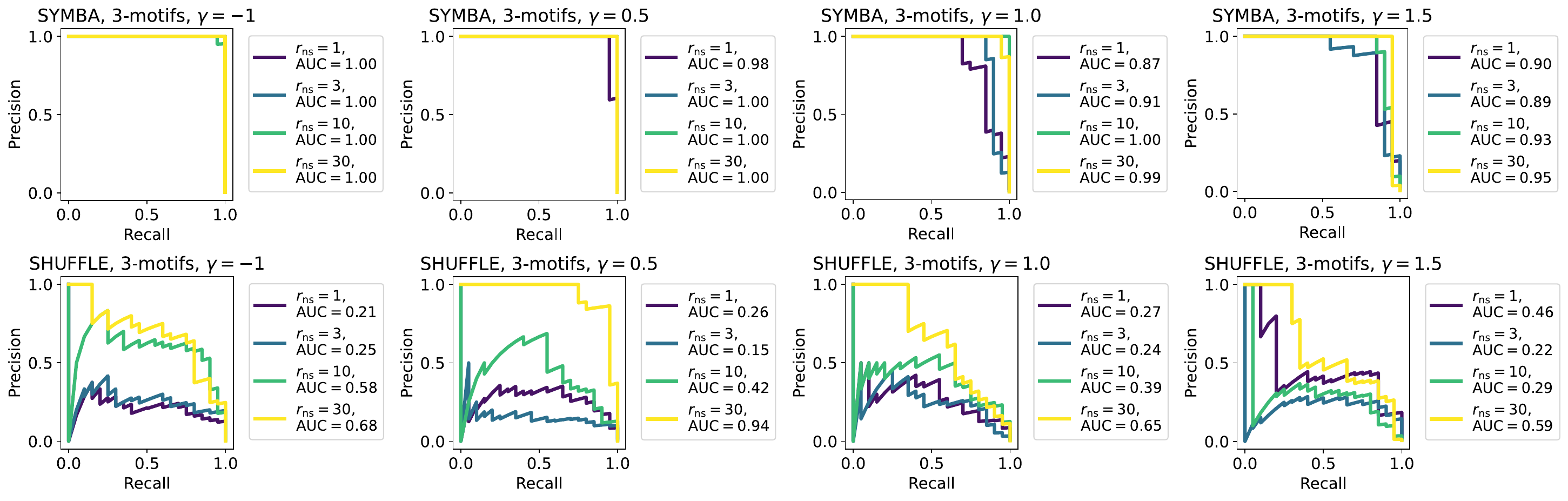}
    \caption{PR curves showing the performance of SYMBA (top panels) and shuffle (bottom) in the identification of ordered $3$-motifs. 
    Each column corresponds to a different exponent $\gamma$ of noise rank-frequency distribution follows $f(r) = r^{-\gamma}$, increasing from left to right. In each panel, different curves correspond to different noise-to-signal ratios $r_{\text{ns}}$, and the corresponding Area Under the Curve (AUC) values are reported in the legend. The signal consists of $n_3 = 20$ distinct ordered $3$-tuples, each repeated $m_3 = 10$ times.}
    \label{fig:pr_symba_vs_shuffle_3_motifs_dir}
\end{figure}
\begin{figure}[htp!]
    \centering
    \includegraphics[width=1.\linewidth]{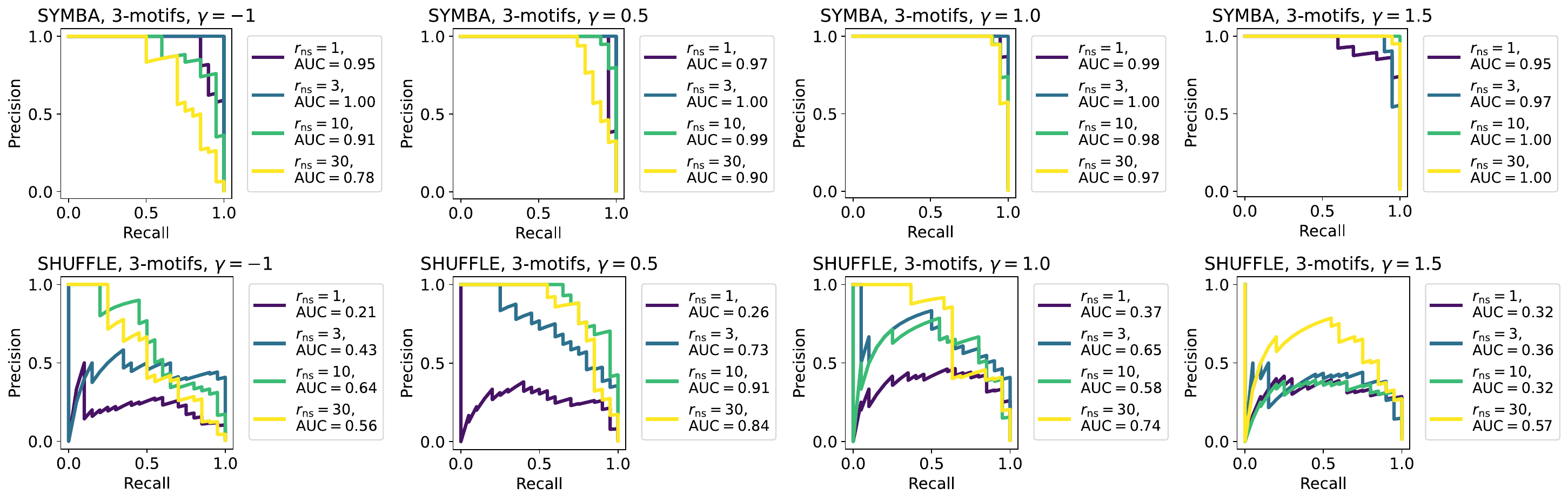}
    \caption{PR curves showing the performance of SYMBA (top panels) and shuffle (bottom) in the identification of unordered $3$-motifs. 
    Each column corresponds to a different exponent $\gamma$ of noise rank-frequency distribution follows $f(r) = r^{-\gamma}$, increasing from left to right. In each panel, different curves correspond to different noise-to-signal ratios $r_{\text{ns}}$, and the corresponding Area Under the Curve (AUC) values are reported in the legend. The signal consists of $n_3 = 20$ distinct unordered $3$-tuples, each repeated $m_3 = 10$ times.}
    \label{fig:pr_symba_vs_shuffle_3_motifs_und}
\end{figure}

In Fig.~\ref{fig:pr_symba_vs_shuffle_3_motifs_dir} and Fig.~\ref{fig:pr_symba_vs_shuffle_3_motifs_und}, we report the Precision-Recall (PR) curves for ordered and unordered $3$-tuples respectively, and we compare the classification performance of SYMBA (top panels) to those of the shuffle method (bottom panels). 
From the curves and the corresponding areas under the curve (AUC) \cite{powers_evaluation_2011}, we observe that SYMBA consistently outperforms the shuffle method across all tested values of $\gamma$. Notably, for $\gamma>0$, this performance advantage becomes increasingly pronounced as $\gamma$ increases, corresponding to progressively more heterogeneous noise frequency distributions.

\subsection*{Comparison between p-value and z-score significance assessment for shuffle method}

For the surrogate-based significance assessment (shuffle method) in the benchmarks reported so far, we used an empirical $z$-score. Here we investigate whether an empirical $p$-value could provide a competitive alternative, particularly in regimes where the available statistics are limited.

\begin{figure}[htp!]
    \centering
    \includegraphics[width=0.9\linewidth]{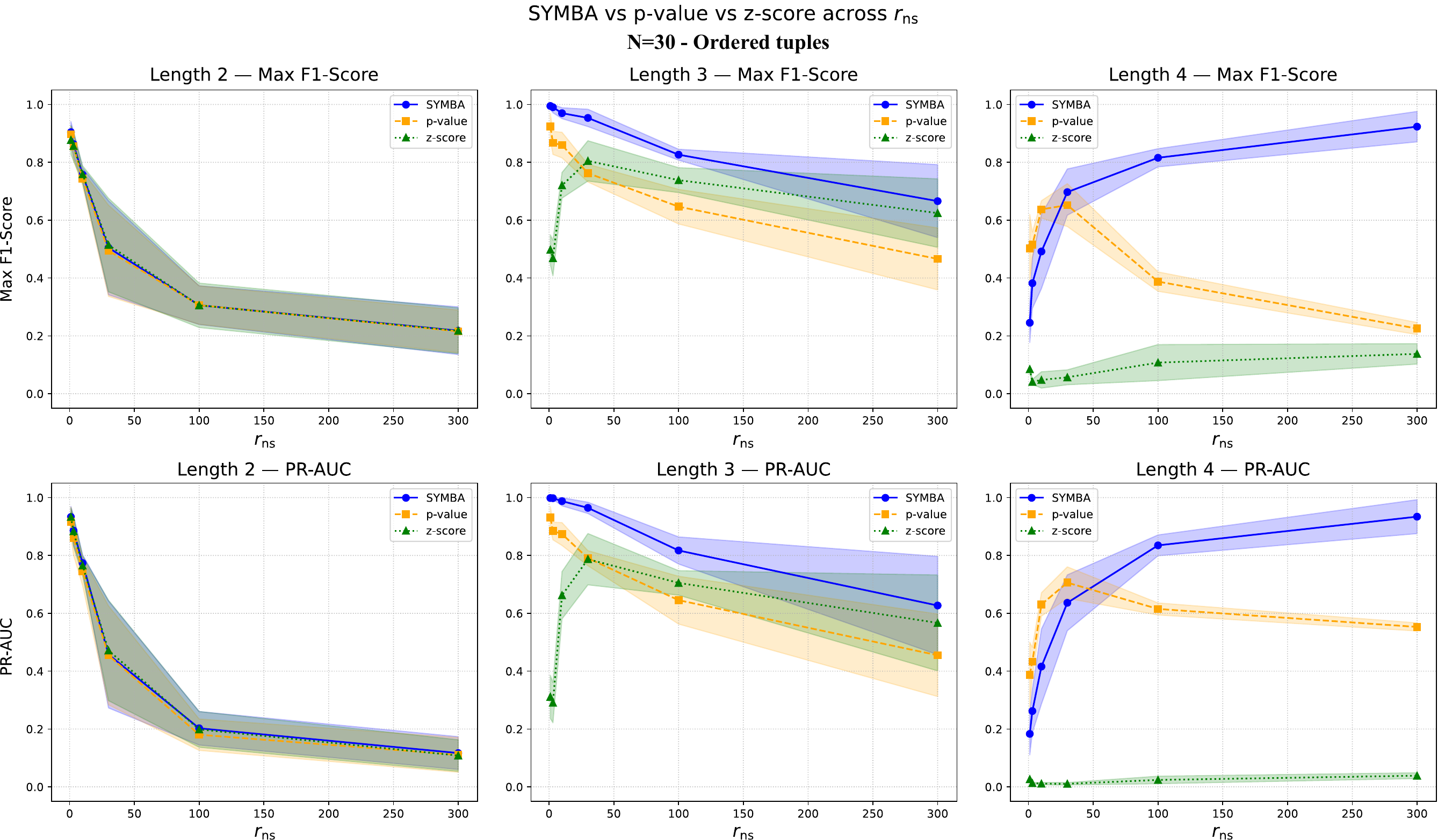}
    \caption{\textbf{Performance comparison of surrogate-based significance measures and SYMBA.} Maximum $F_1$ score (top) and Area Under the Precision--Recall Curve (AUC, bottom) as a function of the noise-to-signal ratio $r_{\mathrm{ns}}$ for synthetic sequences containing ordered motifs sampled from an alphabet of $N=30$ symbols. Results are averaged over $5$ independent realizations. SYMBA is compared with significance assessment based on the empirical $p$-value and the $z$-score computed from $B=1000$ surrogate sequences. The relative performance of the $p$-value and the $z$-score depends on the statistical regime, while SYMBA provides comparable or superior classification performance in all considered cases outside the data-sparsity regime. For longer tuples, the performance of the surrogate-based methods is limited by the minimum attainable $p$-value, $1/(B+1)$, and by the Gaussian approximation underlying the $z$-score (see text).}
    \label{fig:pvalue_zscore_symba_ordered}
\end{figure}

\begin{figure}[htp!]
    \centering
    \includegraphics[width=0.9\linewidth]{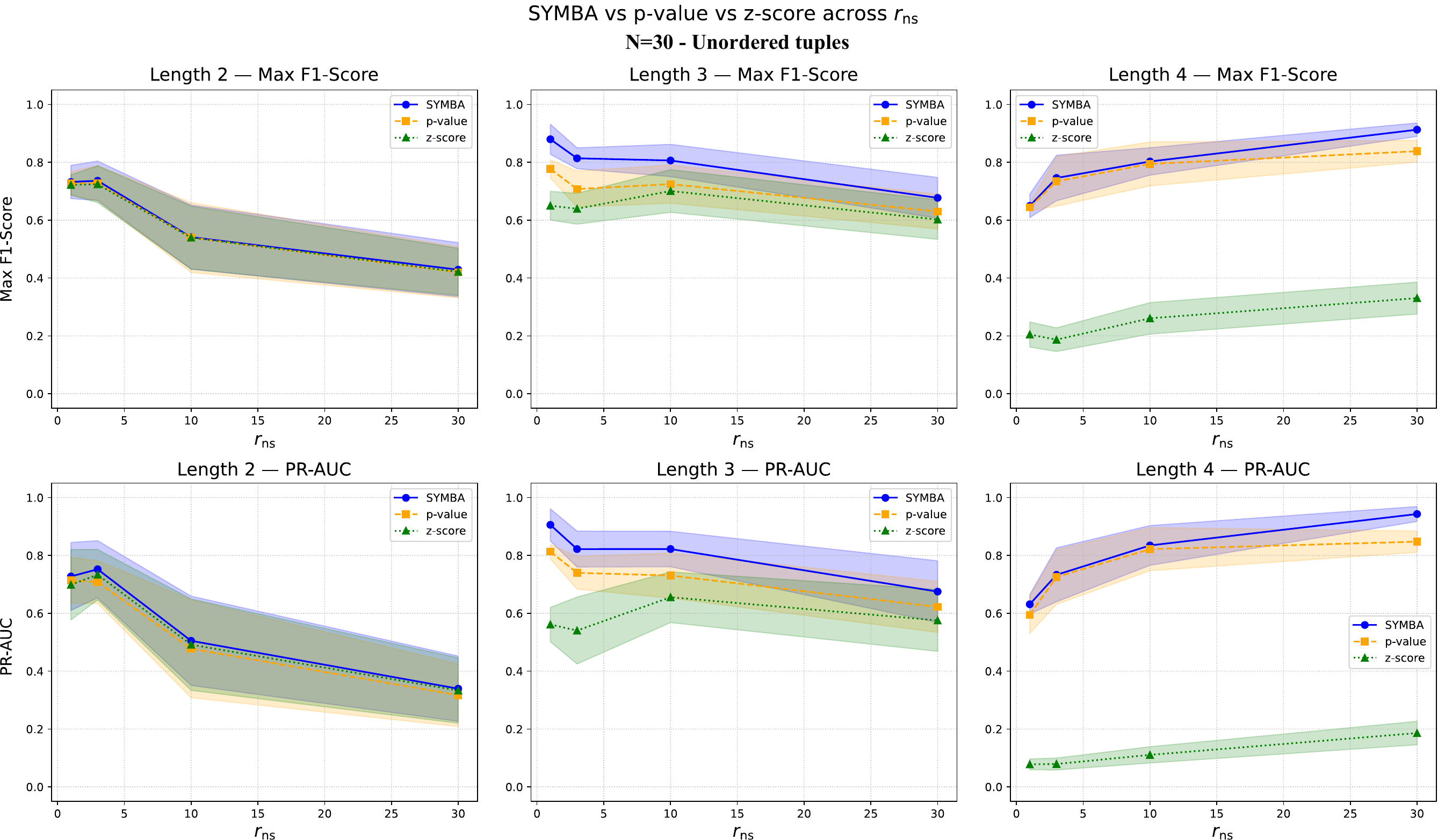}
    \caption{\textbf{Performance comparison of surrogate-based significance measures and SYMBA.} Maximum $F_1$ score (top) and Area Under the Precision--Recall Curve (AUC, bottom) as a function of the noise-to-signal ratio $r_{\mathrm{ns}}$ for synthetic sequences containing unordered motifs sampled from an alphabet of $N=30$ symbols. Results are averaged over $5$ independent realizations. SYMBA is compared with significance assessment based on the empirical $p$-value and the $z$-score computed from $B=1000$ surrogate sequences. For $2$-tuples, the three methods show similar performance. For longer tuples, SYMBA achieves the best classification performance, while the surrogate-based methods exhibit a progressive degradation, due to the minimum attainable $p$-value, $1/(B+1)$, and to the Gaussian approximation underlying the $z$-score (see text).}
\label{fig:pvalue_zscore_symba_unordered}
\end{figure}

We compared the performance of motif significance assessment based on the empirical $p$-value and on the $z$-score computed from surrogate sequences. Figures~\ref{fig:pvalue_zscore_symba_ordered} and \ref{fig:pvalue_zscore_symba_unordered} report the corresponding classification performance for ordered and unordered tuples, respectively, together with the performance of SYMBA.
For these tests, we generated synthetic sequences from an alphabet of $N=30$ symbols, containing $20$ motifs of length $2$, $20$ motifs of length $3$, and $20$ motifs of length $4$, each repeated $10$ times. Background noise was then added with variable noise-to-signal ratio $r_{\text{ns}}$, using a symbol frequency distribution following a Zipf law with exponent $\gamma=1$ (see Sec.~\ref{sec:SM_synthetic_creation}).

The results show that the relative performance of the $p$-value and the $z$-score depends on the statistical regime, which is determined by the amount of available data, the tuple length $k$, and the alphabet size.
As discussed in Sec.~\ref{sec:SM_regimes}, for a fixed signal, increasing the sequence length (equivalently the background noise-to-signal ratio $r_{\mathrm{ns}}$) leads to three qualitatively different regimes: a data-sparsity regime ($r_{\mathrm{ns}} < r_{\mathrm{ns}}^{\mathrm{low}}$), where the occurrences of the motifs are largely accounted for by those of their sub-tuples; an intermediate regime, where the available statistics are sufficient to reliably detect the motifs; and a noise-dominated regime ($r_{\mathrm{ns}} > r_{\mathrm{ns}}^{\mathrm{up}}$), where the signal is overwhelmed by the background noise. The values of the thresholds depend on both the tuple length and the alphabet size.
Although the analytical estimates derived in Sec.~\ref{sec:SM_regimes} were obtained for $\gamma=0$, we expect the same qualitative picture to hold also for the Zipf-distributed symbols ($\gamma=1$) considered in this test, although the values of the thresholds can be shifted.
The figures show that the relative performance of the $p$-value and the $z$-score changes across these regimes.

For 2-tuples, increasing $r_{\mathrm{ns}}$ rapidly drives the system into the noise-dominated regime, and all methods exhibit a similar degradation in performance.
For 3-tuples, SYMBA consistently achieves the highest classification performance, while the relative ranking of the surrogate-based $p$-value and $z$-score depends on whether ordered or unordered tuples are considered.
In particular, for ordered tuples the $p$-value performs better at low $r_{\mathrm{ns}}$, where most $3$-tuples are expected to occur by chance far less than once, whereas the $z$-score becomes competitive as $r_{\mathrm{ns}}$ increases.
For unordered tuples, instead, the two measures rapidly converge to similar performance, with the $p$-value retaining a small advantage.
For 4-tuples, the differences become more pronounced. For ordered tuples, at low $r_{\mathrm{ns}}$ the $p$-value performs better than SYMBA. In this data-sparsity regime the performance of all methods that assess significance from the occurrences of the sub-tuples is limited, and simply ranking the tuples by their number of occurrences can perform better than both (see Table~\ref{tab:counting_4tuples}).
Both null models predict the same expected number of occurrences of a motif, $p_{\text{exp}}(L-k+1)$, with $p_{\text{exp}}$ given by Eq.~(1) of the main text, to which the average over the surrogate sequences converges as $B$ increases. The advantage of the $p$-value in this regime comes from the width of the null distribution. When the inner sub-tuple occurs mostly within the copies of the motif, the null distribution of the shuffle method is narrow, whereas the prior of SYMBA, because of the smoothing, has a width of at least about one occurrence, so that SYMBA requires a larger excess of occurrences to assign a high score. The same property keeps the number of false positives of SYMBA much smaller than that of the shuffle $p$-value at the conventional threshold, as discussed below.
As $r_{\mathrm{ns}}$ increases, SYMBA continues to improve, approaching nearly perfect classification, and outperforms the $p$-value, whose performance deteriorates, while the $z$-score performs poorly at all values of $r_{\mathrm{ns}}$.
The two surrogate-based measures are limited for different reasons. The $z$-score assumes a Gaussian null distribution, which is inadequate for the many tuples whose expected number of occurrences is much smaller than one: a tuple observed by chance only once then receives a very large $z$-score. Indeed, for an alphabet of $N=30$ symbols there are $30^4=810{,}000$ possible ordered 4-tuples, so that most of them are expected to occur far less than once. The empirical $p$-value, instead, cannot be smaller than $1/(B+1)$, with $B=1000$ surrogate sequences, which is the value assigned to a tuple that never occurs as often as observed in any of the surrogates. As $r_{\mathrm{ns}}$ increases, many motifs and many tuples observed by chance reach this minimum value. Since they all receive the same $p$-value, the motifs can no longer be distinguished from these tuples.
This limitation naturally becomes more severe as the tuple length and the alphabet size increase, unless the number of surrogate sequences is correspondingly increased, at the expense of a proportional increase in computational cost.

To clarify the behaviour of the methods in the data-sparsity regime, Table~\ref{tab:counting_4tuples} reports, for ordered $4$-tuples, the maximum $F_1$ score and the PR-AUC of SYMBA, of the shuffle method ($p$-value and $z$-score), and of a simple classifier that ranks the tuples by their number of occurrences, over $20$ independent realizations with the same parameters as in Fig.~\ref{fig:pvalue_zscore_symba_ordered}. At the lowest values of $r_{\mathrm{ns}}$ counting performs best, but its performance drops sharply for $r_{\mathrm{ns}} \geq 10$. SYMBA is the only method whose performance improves steadily with $r_{\mathrm{ns}}$: it becomes comparable to the $p$-value at $r_{\mathrm{ns}}=30$ and performs best for larger values, while the performance of the $p$-value deteriorates. The transition is spread over a wide range of $r_{\mathrm{ns}}$ because, for Zipf-distributed symbols, the threshold differs from motif to motif: it depends on the frequencies $f_b$ and $f_c$ in the background noise of the symbols of the inner sub-tuple, $r_{\mathrm{ns}}^{\mathrm{low}} \simeq m_k/(L_{\text{s}}\,f_b\,f_c)$, which ranges from about $2$ to $50$ for $80\%$ of the motifs.

\begin{table}[htp!]
\centering
\begin{tabular}{|c|c|c|c|c|c|c|}
\hline
\textbf{Ordered $4$-tuples} & $r_{\text{ns}}=1$ & $r_{\text{ns}}=3$ & $r_{\text{ns}}=10$ & $r_{\text{ns}}=30$ & $r_{\text{ns}}=100$ & $r_{\text{ns}}=300$ \\
\hline
SYMBA & 0.32 / 0.25 & 0.39 / 0.28 & 0.48 / 0.39 & 0.67 / 0.62 & 0.83 / 0.84 & 0.90 / 0.90 \\
Shuffle, $p$-value & 0.50 / 0.38 & 0.58 / 0.49 & 0.65 / 0.62 & 0.59 / 0.67 & 0.37 / 0.61 & 0.21 / 0.53 \\
Shuffle, $z$-score & 0.09 / 0.03 & 0.05 / 0.02 & 0.04 / 0.01 & 0.07 / 0.02 & 0.09 / 0.02 & 0.11 / 0.03 \\
Counting & 1.00 / 1.00 & 0.95 / 0.49 & 0.46 / 0.15 & 0.10 / 0.03 & 0.02 / 0.00 & 0.00 / 0.00 \\
\hline
\end{tabular}
\caption{Maximum $F_1$ score / PR-AUC for the classification of ordered $4$-motifs, averaged over $20$ independent realizations with the same parameters as in Fig.~\ref{fig:pvalue_zscore_symba_ordered} ($N=30$, $\gamma=1$, $20$ motifs of each length $2$, $3$ and $4$, each repeated $10$ times, $B=1000$). Counting ranks the tuples by their number of occurrences; its results were obtained from $20$ further realizations with the same parameters.} \label{tab:counting_4tuples}
\end{table}

Finally, we note that the maximum $F_1$ score and the PR-AUC assess the ranking of the tuples, but not the choice of a significance threshold, which is needed in applications where the ground truth is unknown. In this respect, SYMBA is conservative by construction: the smoothing of the prior requires an excess of several occurrences before a tuple obtains a high score, so that in the data-sparsity regime SYMBA tends to miss motifs rather than report false positives. With the threshold of $0.6$ used in the main text, SYMBA reported at most $2$ false positives per sequence, for any tuple length and value of $r_{\mathrm{ns}}$, over $20$ independent realizations with the same parameters as in Fig.~\ref{fig:pvalue_zscore_symba_ordered}. The shuffle $p$-value at the conventional threshold $p<0.05$ reported instead between $75$ ($r_{\mathrm{ns}}=1$) and about $11000$ ($r_{\mathrm{ns}}=300$) false $4$-tuples per sequence. Corrections for multiple testing, such as the Bonferroni correction or the Benjamini--Hochberg procedure for controlling the false discovery rate~\cite{benjamini1995}, remove these false positives but, with $B=1000$ surrogate sequences, no $3$- or $4$-tuple reaches the corrected thresholds, since the minimum $p$-value, $1/(B+1)$, is too large compared with these thresholds when thousands of tuples are tested. A much larger number of surrogates would be required, with a proportional increase in computational cost.

Overall, these results indicate that neither the empirical $p$-value nor the $z$-score is uniformly superior across all statistical regimes. Their relative performance depends on the available statistics and on the tuple length under consideration. In particular, the $p$-value performs better than, or comparably to, the $z$-score in most of the considered cases, while the $z$-score is superior for ordered $3$-tuples at large $r_{\mathrm{ns}}$. In contrast, SYMBA provides comparable or superior classification performance in all the considered scenarios outside the data-sparsity regime, without requiring the generation of large ensembles of surrogate sequences.

\section{SYMBA robustness to observational noise}

\begin{figure}[htp!]
    \centering
    \includegraphics[width=1.\linewidth]{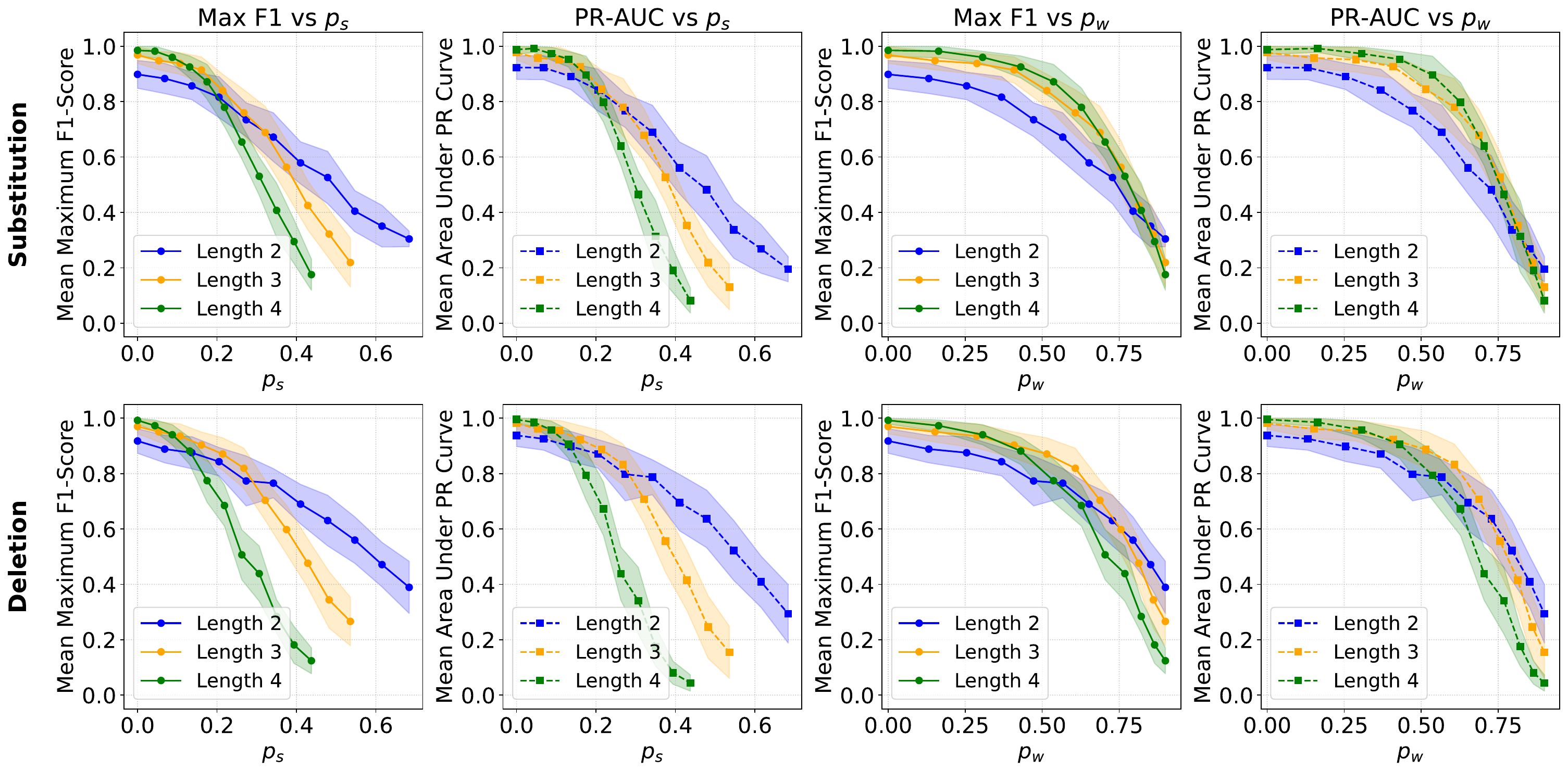}
    \caption{
    \textbf{SYMBA classification performance under symbol substitution and deletion noise.} Maximum $F_1$ score and Area Under the Curve (AUC) as a function of the symbol error probability $p_s$ and the word damage probability $p_w = 1 - (1-p_s)^k$. The quantity $p_w$ is the probability that at least one symbol of a planted word (motif) of length $k$ is altered. Curves show the performance degradation for motif lengths $k=2$, $3$, and $4$. Synthetic sequences were generated by first creating $20$ random words (motifs) of length $k$ (sampled from an alphabet of $N=20$ symbols), each repeated $30$ times. Each symbol was then independently substituted with a random symbol from the same alphabet (substitution) or deleted (deletion) with probability $p_s$. Finally, background noise was added to obtain a noise-to-signal ratio $r_{ns}=10$.
    }
    \label{fig:word_error_sub_remove}
\end{figure}
We performed additional robustness tests with respect to observational noise. In particular, Fig.~\ref{fig:word_error_sub_remove} reports the performance degradation of SYMBA under symbol substitution and symbol deletion. In both cases, each symbol in every planted word is independently substituted, or removed, with probability $p_s$.
The results show that the quantity that primarily governs the performance is the word damage probability, $p_w = 1 - (1-p_s)^k$, namely the probability that at least one symbol of a planted word of length $k$ is altered. Since $p_w$ captures the dependence on the word length, plotting the performance as a function of $p_w$ reveals qualitatively similar behaviour for the two types of perturbation: the performance degrades gradually and drops significantly only for large values of $p_w$.

While symbol substitution and deletion alter the sequence composition directly, observational limitations can also affect the exact timing of recorded events. Specifically, instrument noise and finite temporal resolution (e.g., discrete sampling or binning) introduce uncertainty in event timestamps. Because SYMBA relies on temporal adjacency, the impact of this timing uncertainty, often modeled as temporal jitter, depends on its magnitude relative to the threshold $\Delta t$ used to define symbol adjacency and to the distribution of inter-event intervals in the sequence. Three qualitatively different regimes can be distinguished:
\begin{enumerate}
    \item %
    If the timing uncertainty only increases the inter-event interval beyond the adjacency threshold $\Delta t$, thereby introducing an erroneous separation between otherwise related symbols, this effect can be mitigated by adopting a larger value of $\Delta t$, which corresponds to probing functional relations over a broader temporal scale.
    \item %
    If the timing uncertainty is sufficient to alter the relative ordering of nearby events, the corresponding motifs are no longer reliably represented as ordered tuples. In this case, the analysis can instead be performed using unordered tuples, which do not rely on the precise ordering of events within a motif.
    \item %
    If the timing uncertainty is so large that the symbols of a motif are no longer adjacent in the symbolic sequence because they become separated by other events, the adjacency condition is violated and the motif may no longer be detected. This leads to a loss of motif detectability qualitatively similar to that produced by symbol deletion noise.
\end{enumerate}

\section{Benchmarking with higher-order Markov processes}

\begin{figure}[!htbp]
    \centering
    \includegraphics[width=0.9\linewidth]{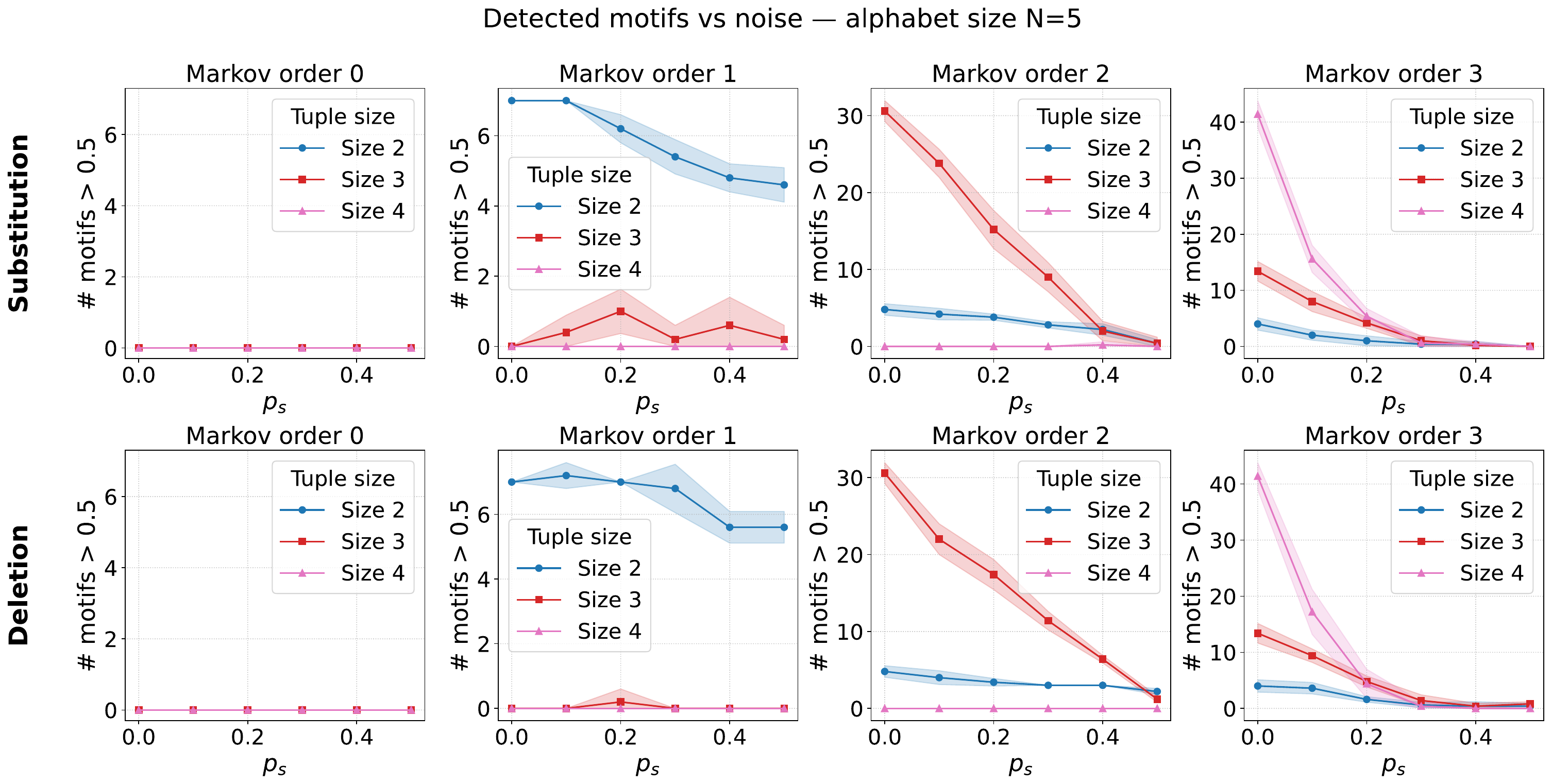}
    \caption{ \textbf{SYMBA identification of significant motifs in higher-order Markov sequences under symbol substitution and deletion noise.} Number of significant motifs, defined as tuples with a BJS score exceeding the threshold $\mathrm{BJS}^{\mathrm{thr}}=0.50$, as a function of the symbol substitution (top panels) and deletion (bottom panels) probability $p_s$. The alphabet size is fixed at $N=5$. Panels from left to right correspond to sequences generated by Markov processes of order $0$--$3$, respectively. In each panel, curves report the number of significant tuples of length $k=2$, $3$, and $4$. As expected, SYMBA correctly identifies tuples as significant when the Markov memory of the generating process exceeds the memory assumed by the corresponding null model, while increasing symbol noise progressively destroys these underlying correlations.}
    \label{fig:markov_noise_n5}
\end{figure}
\begin{figure}[!htbp]
    \centering
    \includegraphics[width=0.9\linewidth]{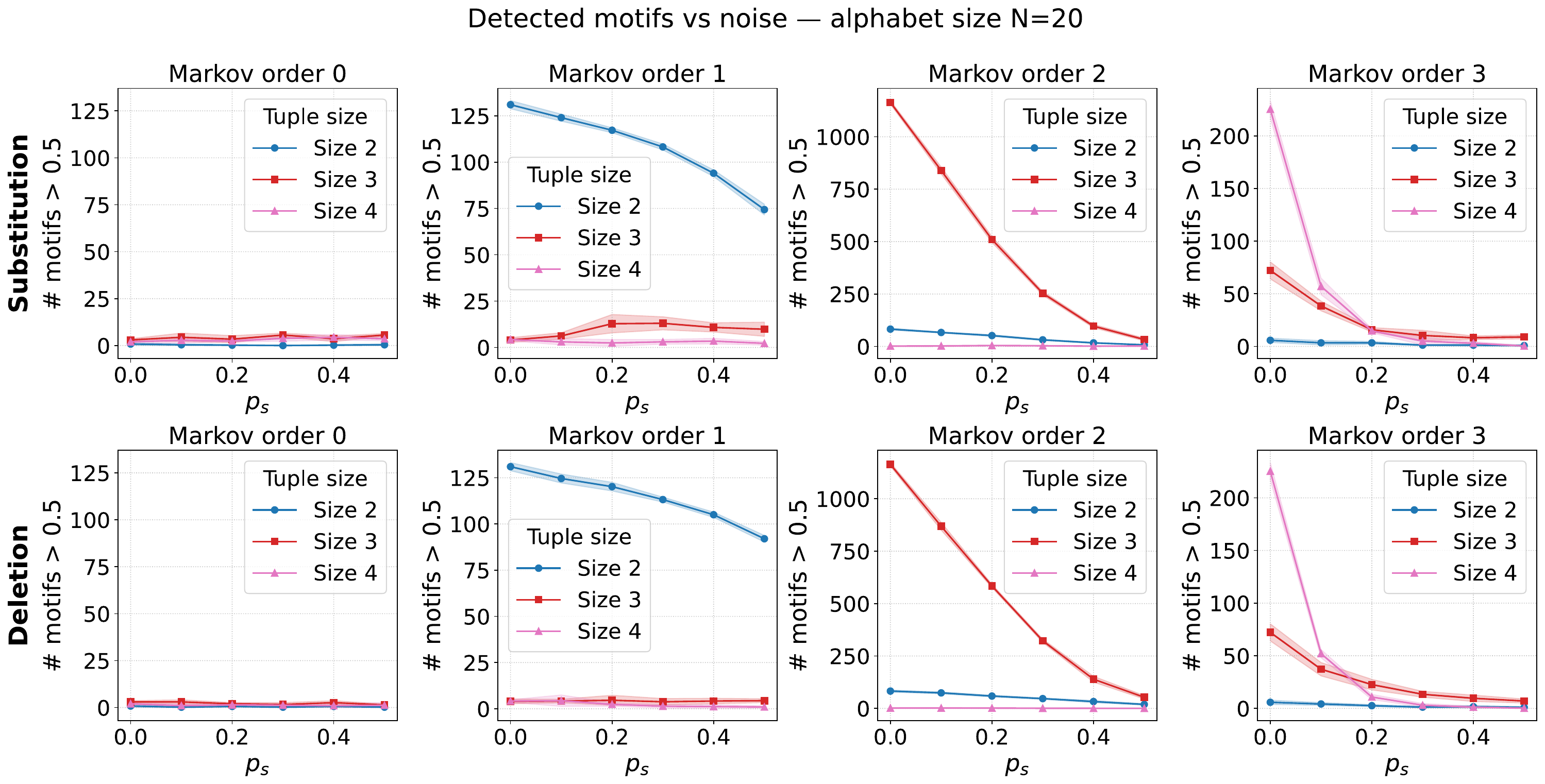}
    \caption{ \textbf{SYMBA identification of significant motifs in higher-order Markov sequences under symbol substitution and deletion noise.} Number of significant motifs, defined as tuples with a BJS score exceeding the threshold $\mathrm{BJS}^{\mathrm{thr}}=0.50$, as a function of the symbol substitution (top panels) and deletion (bottom panels) probability $p_s$. The alphabet size is fixed at $N=20$. Panels from left to right correspond to sequences generated by Markov processes of order $0$--$3$, respectively. In each panel, curves report the number of significant tuples of length $k=2$, $3$, and $4$. As expected, SYMBA correctly identifies tuples as significant when the Markov memory of the generating process exceeds the memory assumed by the corresponding null model, while increasing symbol noise progressively destroys these underlying correlations.}
    \label{fig:markov_noise_n20}
\end{figure}

To benchmark SYMBA under the null hypothesis associated with $p_{exp}$ (Eq.~1 of the main text), we generated synthetic symbolic sequences from Markov processes of order $k-2$, for $k\in\{2,3,4,5\}$.
In the SYMBA formulation, $p_{exp}$ corresponds to the expected probability under the assumption that the sequence is generated by a stationary Markov process of order $k-2$, such that correlations at length $k$ are explained by correlations at length $k-1$ and shorter. Therefore, for sequences generated by a Markov process of order $k-2$, we expect SYMBA not to identify $k$-tuples as statistically significant. Conversely, for shorter tuples of length $k'<k$, the generating process has a longer memory than that assumed by the corresponding null hypothesis ($k'-2$), and these tuples are therefore expected to be identified as significant.
We generated the transition probabilities by sampling each conditional distribution from a Dirichlet distribution with parameter $\alpha=1$. We considered alphabets of size $N=5$ and $N=20$, using sequences of length $5000$ and $100000$, respectively. 
We verified numerically that, in all generated sequences, the mutual information between symbols at distance $\ell$ drops, within at most $\ell = 7$ symbols, to the value obtained for the same sequence randomly shuffled, so that correlations decay on a scale much shorter than the sequence length.
We adopted a BJS significance threshold of $0.5$, lower than the value $0.6$ used elsewhere in this work, so as to test the method in a less conservative setting.
Figures~\ref{fig:markov_noise_n5},\ref{fig:markov_noise_n20} show the number of significant tuples identified by SYMBA for different tuple lengths for alphabets of size $N=5$ and $N=20$, respectively. We further assessed the robustness to observational noise by introducing symbol substitution and symbol deletion with probability $p_s$.
As expected, SYMBA correctly does not identify tuples of length $k$ as statistically significant, with only a small number of false positives due to finite-sample fluctuations, while identifying tuples of length $k'<k$ as significant. Adopting a larger BJS significance threshold further reduces these false positives.
As the noise level $p_s$ increases, the correlations are progressively destroyed, the sequences become increasingly similar to a memoryless random process, and the number of significant tuples identified by SYMBA correspondingly decreases towards zero.
Overall, these results provide empirical support that the Bayesian inference adopted by SYMBA remains effective when the data are generated according to the Markov null hypothesis underlying $p_{exp}$, while also demonstrating the robustness of the method across different data-generating mechanisms and observational noise levels.

\section{\texorpdfstring{Choosing the threshold $\text{BJS}^{\text{thr}}(k)$}{Choosing the threshold BJSthr(l)}} \label{sec:SM_bjs_threshold}

The choice of a significance hard threshold $\text{BJS}^{\text{thr}}(k)$ is inherently arbitrary.
However, we outline below some rules of thumb and methodologies that can serve as useful guides for selecting a meaningful threshold.

We found in our tests on synthetic sequences that in general the optimal threshold for the BJS-score, corresponding to the maximum value of F1 score, depends on the amount of data available,  with sparser datasets showing smaller $\text{BJS}^{\text{thr}}(k)$, correctly indicating a lower significance of the finding due to the scarce statistic. A good way to estimate the data sparsity of a real-world dataset is looking how many of all $k$-tuples on the support of $p_{\exp}$ are actually observed in the sequence (i.e., looking at $\sum_{\boldsymbol{s}\in \mathcal{U}^{(k)}} p_{\text{exp}}(\boldsymbol{s})$, which is equal to $1$ only if all the possible $k$-tuples $\boldsymbol{s}$ are observed in the sequence). For values of $\sum_{\boldsymbol{s}\in \mathcal{U}^{(k)}} p_{\text{exp}}(\boldsymbol{s}) > 0.8$, as those observed in our real-world datasets, we found on synthetic sequences $\text{BJS}^{\text{thr}}(k) \simeq [0.5, 0.7]$ for both $k=2$ ($2$-tuples) and $k=3$ ($3$-tuples), independently from the specific noise distribution.
Therefore, in the applications to real-world datasets, we will adopt thresholds in this range, in particular we show results corresponding to $\text{BJS}^{\text{thr}}=0.6$ and $\text{BJS}^{\text{thr}}=0.7$ for both $k=2$ and $k=3$.

The threshold can also be estimated on the sequence under study, by using shuffled sequences that satisfy the null hypothesis of SYMBA. For $k$-tuples, a sequence shuffled while preserving the counts of all $(k-1)$-tuples, as in the shuffle method (see Sec.~\ref{sec:SM_shuffle}), has the same expected probabilities $p_{\text{exp}}(\boldsymbol{s})$, Eq.~(1) of the main text, as the original sequence, so that any $k$-tuple with a high BJS-score in a shuffled sequence is a false positive. Two choices are then possible. A conservative threshold is the $95$th percentile, over the shuffled sequences, of the maximum BJS-score of the $k$-tuples in each shuffled sequence, so that the probability of reporting even one false positive is about $5\%$. A less conservative threshold is the smallest value for which the mean number of $k$-tuples above the threshold in the shuffled sequences is smaller than $5\%$ of the number of $k$-tuples above the threshold in the original sequence, so that the false positives are expected to be less than $5\%$ of the detected tuples (false discovery rate~\cite{benjamini1995}). The second choice detects more motifs, at the cost of a few false positives.
In our tests on synthetic sequences, about ten shuffled sequences were sufficient to estimate both thresholds, whereas the empirical $p$-value of each tuple requires a number of shuffled sequences that grows with the number of tested tuples when corrections for multiple testing are applied (see the discussion following Table~\ref{tab:counting_4tuples}). Both thresholds increase with the length of the sequence, consistently with the dependence on the amount of data described above.

\section{z-score computation from batches on a single symbolic sequence (BATCH)} \label{sec:SM_batch}

In the batch method, a z-score is assigned to each $k$-tuple $\boldsymbol{s}$, defined as follows:
\begin{equation}
    \text{z-score}(\boldsymbol{s})=\frac{p_{\text{obs}}(\boldsymbol{s}) - p_{\text{exp}}(\boldsymbol{s})}{\text{std}_{\text{exp}}(\boldsymbol{s})}
\end{equation}
where $p_{\text{exp}}(\boldsymbol{s})$ is the expected probability for observing  $\boldsymbol{s}$ taking into account the effect of lower-order correlations coming from tuples of length smaller than $k$~\cite{sinatra_motifs_2010}, defined in the main manuscript. For the ordered case we have:
\begin{align}
    p_{\text{exp}}(\boldsymbol{s}=s_1, s_2) &= p_{\text{obs}}(s_1)p_{\text{obs}}(s_2) \label{eq:exp_length2_main}
\end{align}
and: 
\begin{align}
    p_{\text{exp}}(\boldsymbol{s}=s_1, \dots, s_k) &= \frac{p_{\text{obs}}(s_1,\dots, s_{k-1})p_{\text{obs}}(s_2, \dots, s_k)}{p_{\text{obs}}(s_2, \dots, s_{k-1})} \label{eq:exp_lengthk_main}
\end{align}
for $k>2$. 
We recall from the manuscript, that the observed probability of $\boldsymbol{s}$ is computed as:
\begin{align}
    p_{\text{obs}}(\boldsymbol{s})
    = \frac{n_{\text{obs}}(\boldsymbol{s})}{L-k+1}
\end{align}
where 
$n_{\text{obs}}(\boldsymbol{s})$ represents the number of occurrences of $\boldsymbol{s}$ in the sequence of length $S$ and $L-k+1$ is the total number of $k$-tuples.
For the application to the case of unordered tuples, all we have to do is to sum the results for the ordered case over $S_p(\boldsymbol{s})$, the set of unique permutations $\boldsymbol{\sigma}$ of the ordered tuple $\boldsymbol{s}=s_1, \dots, s_k$, obtaining:
\begin{equation}
    p_{\text{exp}}^{\text{und}}(\boldsymbol{s}) = \sum_{\boldsymbol{\sigma} \in S_p(\boldsymbol{s})}p_{\text{exp}}(\boldsymbol{\sigma})
\end{equation}
and 
\begin{equation}
    n_{\text{obs}}^{\text{und}}(\boldsymbol{s}) = \sum_{\boldsymbol{\sigma} \in S_p(\boldsymbol{s})}n_{\text{obs}}(\boldsymbol{\sigma})
\end{equation}
In order to estimate the standard deviation $\text{std}_{\text{exp}}(\boldsymbol{s})$, we partition the original sequence in $m$ successive non overlapping subsequences (namely, batches) of equal length. In each batch $i$ we compute $ p_{\text{exp}}^{(i)}(\boldsymbol{s})$ and then we compute the standard deviation of $\boldsymbol{s}$ as:
\begin{equation}
    \text{std}_{\text{exp}}(\boldsymbol{s}) \approx \sqrt{\frac{\langle  p_{\text{exp}}^{(i)}(\boldsymbol{s}) ^ 2\rangle - \langle  p_{\text{exp}}^{(i)}(\boldsymbol{s})\rangle ^ 2}{m-1}}
\end{equation}
where $\langle \cdot \rangle$ is the average over the $m$ batches.
For our results, we partitioned the original sequence in $m=20$ batches.

\section{Shuffle-based surrogate-data method (SHUFFLE) }\label{sec:SM_shuffle}

We compare our method with a shuffle-based procedure to assess the statistical significance of observed $k$-tuples in a symbolic sequence while preserving exactly the counts of all $(k-1)$-tuples. This provides an empirical null model that maintains the local sequential structure up to order $k-1$.

Let $\mathcal{A}$ denote the alphabet and let the observed sequence be $\mathcal{S}=(s_1,s_2,\ldots,s_L)$, where $s_i\in\mathcal{A}$, of length $L$. For any integer $k\geq2$ and any $k$-tuple $\boldsymbol{s}\in\mathcal{A}^k$, let $n_{\mathrm{obs}}(\boldsymbol{s})$ denote the number of occurrences of $\boldsymbol{s}$ in $\mathcal{S}$, computed using a sliding window so that there are $L-k+1$ windows of length $k$. Our goal is to determine whether each observed $k$-tuple occurs significantly more or less frequently than expected under random sequences preserving exactly the observed $(k-1)$-tuple counts.

\paragraph*{De Bruijn multigraph construction.}
For $k=2$, preserving 1-tuple counts trivially reduces to a standard random permutation of the sequence. For $k \geq 3$, to generate constrained random sequences, we construct the directed De Bruijn multigraph $G$ of order $k-1$. Specifically:

\begin{itemize}
    \item each distinct observed $(k-2)$-tuple defines a node of $G$;
    \item each occurrence of a $(k-1)$-tuple $(s_i,s_{i+1},\ldots,s_{i+k-2})$ defines a directed edge from node $(s_i,\ldots,s_{i+k-3})$ to node $(s_{i+1},\ldots,s_{i+k-2})$;
    \item repeated occurrences of the same $(k-1)$-tuple generate parallel edges, so that $G$ is, in general, a directed multigraph.
\end{itemize}

Every Eulerian trail of $G$ traverses each edge exactly once and therefore reconstructs a symbolic sequence whose multiset of $(k-1)$-tuples is identical to that of the original sequence. Randomized surrogate sequences are thus obtained by generating Eulerian trails on $G$ and reconstructing the corresponding symbolic sequence through the standard overlap between consecutive nodes.

\paragraph*{Uniform sampling of surrogate sequences.}
Simply randomizing the outgoing edges during Hierholzer's algorithm does not, in general, produce a uniform distribution over all valid Eulerian trails~\cite{Kandel1996}. To avoid this bias, we adopt the uniform sampling framework implemented in \emph{uShuffle}~\cite{jiang2008ushuffle}, following the construction of Kandel \emph{et al.}~\cite{Kandel1996}.

The algorithm first generates a random spanning arborescence rooted at the sink node
$t=(s_{L-k+3},\ldots,s_L)$,
using the loop-erased random walk algorithm of Propp and Wilson~\cite{propp1998how}. Starting from each node not yet belonging to the tree, a random walk is performed by selecting uniformly among its outgoing edges until a node already contained in the tree is reached. Loops are erased as they are formed, and the resulting loop-erased path is incorporated into the arborescence. Repeating this procedure for all nodes generates a uniformly distributed spanning arborescence rooted at $t$.
The arborescence defines a unique reserved outgoing edge for every non-sink node. These reserved edges are constrained to be traversed last, while the remaining outgoing edges are randomly permuted. Hierholzer's algorithm is then applied to this edge ordering to construct an Eulerian trail from the source
$u=(s_1,\ldots,s_{k-2})$
to the sink $t$.

According to Kandel \emph{et al.}~\cite{Kandel1996}, the combination of a uniformly sampled spanning arborescence and the corresponding last-edge reservation yields a uniform distribution over all valid Eulerian trails, and therefore over the constrained surrogate sequences preserving exactly the observed $(k-1)$-tuple counts. The empirical null distribution of each $k$-tuple is finally estimated from $B$ independently generated surrogate sequences. 

In particular, for each $k$-tuple $\boldsymbol{s} \in \mathcal{A}^k$, we record its count in the $i$-th randomized sequence as $c_i(\boldsymbol{s})$, with $i = 1, \dots, B$. The empirical null distribution associated with $\boldsymbol{s}$ is therefore
$\{ c_1(\boldsymbol{s}), \dots, c_B(\boldsymbol{s}) \}$. 
From this distribution, we compute the empirical mean and standard deviation of each tuple as
\begin{align}
    \mu(\boldsymbol{s}) &= \frac{1}{B}\sum_{i=1}^B c_i(\boldsymbol{s}), \\
    \mathrm{std}(\boldsymbol{s}) &= \sqrt{\frac{1}{B-1}\sum_{i=1}^B \big[c_i(\boldsymbol{s}) - \mu(\boldsymbol{s})\big]^2}.
\end{align}
The corresponding empirical expected probability for the shuffle-based method is then
\begin{equation}
    p_{\text{exp}}^{\text{shuffle}}(\boldsymbol{s}) = \frac{\mu(\boldsymbol{s})}{L - k + 1},
\end{equation}
where $k$ is the tuple length and $L - k + 1$ is the number of $k$-tuples observed in a sequence of length $L$.
Finally, statistical significance can be assessed using empirical measures such as the standard $z$-score or the empirical $p$-value. The $z$-score is defined as
\begin{equation}
    z(\boldsymbol{s}) = \frac{n_{\mathrm{obs}}(\boldsymbol{s}) - \mu(\boldsymbol{s})}{\mathrm{std}(\boldsymbol{s})}.
\end{equation}
Alternatively, the empirical one-sided $p$-value is defined as the fraction of surrogate sequences in which the simulated motif count equals or exceeds the observed count $n_{\mathrm{obs}}(\boldsymbol{s})$.

We note that all the quantities defined above depend on $B$, the number of reshuffles. We therefore compare the empirical expected probability $p_{\text{exp}}^{\text{shuffle}}$ for increasing values of $B$ with the analytical expectation computed using our symbolic Bayesian approach (SYMBA), denoted as $p_{\text{exp}}^{\text{SYMBA}}$.

\begin{figure}[!htbp]
    \centering
     \includegraphics[width=1\linewidth]{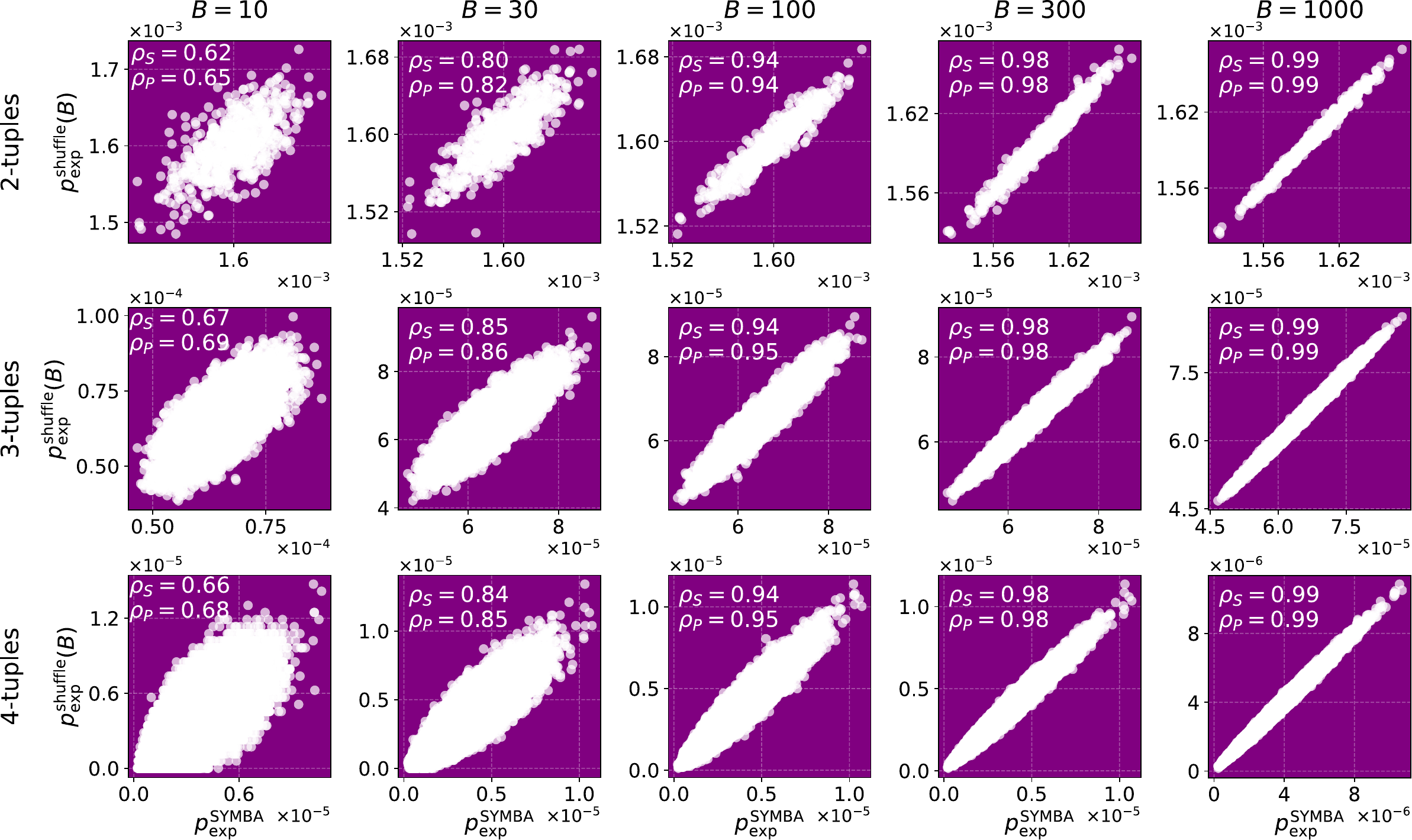}
    \caption{Scatter plots comparing the analytical expected probability computed with our symbolic Bayesian method (SYMBA), $p_{\text{exp}}^{\text{SYMBA}}$, and the empirical expected probability estimated from surrogate data obtained by reshuffling the sequence while preserving the counts of $(k-1)$-tuples, $p_{\text{exp}}^{\text{shuffle}}(\boldsymbol{s})=\frac{1}{B}\frac{\sum_{i=1}^B c_i(\boldsymbol{s})}{L-k+1}=\frac{\mu(\boldsymbol{s})}{L-k+1}$. The plots are shown as a function of the number of reshuffles $B$ and for different tuple lengths $k$. Note that only $p_{\text{exp}}^{\text{shuffle}}$ depends on $B$.   Each point corresponds to a distinct tuple. Columns correspond to increasing values of $B$ from left to right ($B = 10, 30, 100, 300, 1000$), and rows correspond to increasing tuple lengths ($k = 2, 3, 4$). For all values of $k$, as the number of reshuffles increases, $p_{\text{exp}}^{\text{shuffle}}$ converges toward the analytical value $p_{\text{exp}}^{\text{SYMBA}}$. This convergence is also reflected in the increasing Spearman and Pearson correlation coefficients ($\rho_S$ and $\rho_P$) reported in each panel. For this plot we used the parameters: $N=25$, $\gamma=0$, $n_2 = 50$, $n_3 = 50$, $m_2=10$, $m_3=5$ and $r_{\text{ns}}=100$, resulting in a sequence of length $L=176750$.}
    \label{fig:scatter_symba_vs_shuffle}
\end{figure}

Fig.~\ref{fig:scatter_symba_vs_shuffle} shows scatter plots of $p_{\text{exp}}^{\text{shuffle}}(B)$ versus $p_{\text{exp}}^{\text{SYMBA}}$ for $B = (10, 30, 100, 300, 1000)$ and for $2$-, $3$-, and $4$-tuples (each point corresponding to a distinct tuple $\boldsymbol{s}$). We observe that $p_{\text{exp}}^{\text{shuffle}}(B)$ converges to $p_{\text{exp}}^{\text{SYMBA}}$ as $B$ increases, as also indicated by the Spearman and Pearson correlation coefficients ($\rho_S$ and $\rho_P$) reported in each plot, both approaching $1$ for all $k$ as $B$ grows.

\begin{figure}[!htbp]
    \centering
    \includegraphics[width=0.8\linewidth]{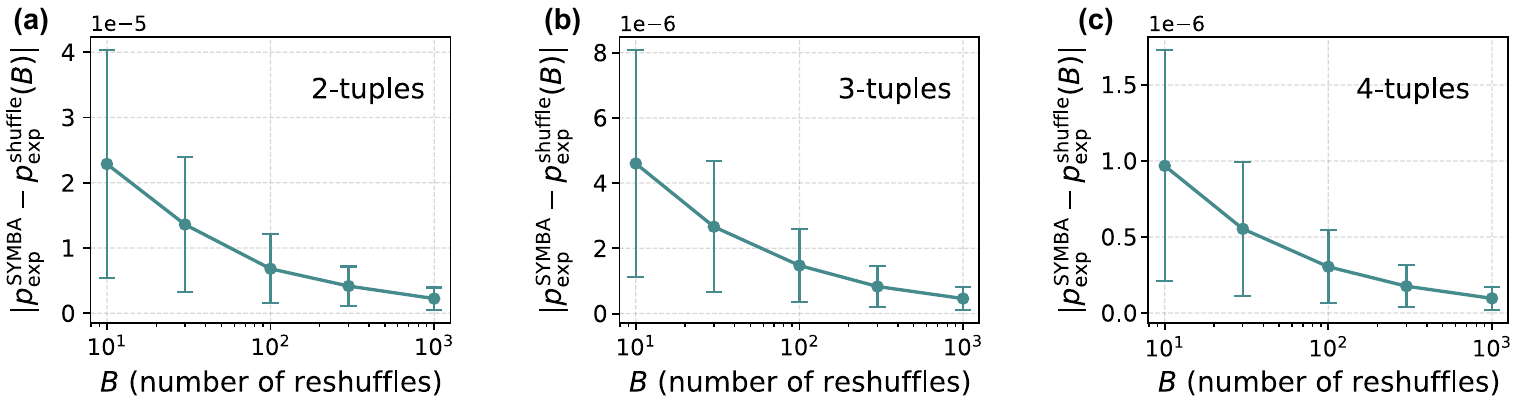}
    \caption{Absolute difference between $p_{\text{exp}}^{\text{SYMBA}}$ and $p_{\text{exp}}^{\text{shuffle}}$ as a function of $B$. Each panel corresponds to a different tuple length (from left to right, $k = 2, 3, 4$). Each point represents the mean absolute difference across all tuples, and the bars indicate the corresponding standard deviation. For all tuple lengths, we observe that $p_{\text{exp}}^{\text{shuffle}}$ (which depends on $B$) converges toward $p_{\text{exp}}^{\text{SYMBA}}$ as $B$ increases. For this plot we used the parameters: $N=25$, $\gamma=0$, $n_2 = 50$, $n_3 = 50$, $m_2=10$, $m_3=5$ and $r_{\text{ns}}=100$, resulting in a sequence of length $L=176750$.}
    \label{fig:diff_pexp_symba_shuffle}
\end{figure}

This convergence is further illustrated in Fig.~\ref{fig:diff_pexp_symba_shuffle}, which reports the absolute difference $|p_{\text{exp}}^{\text{shuffle}}(B) - p_{\text{exp}}^{\text{SYMBA}}|$ averaged over all tuples of length $k$, together with its standard deviation. Both the mean difference and its variability decrease monotonically with $B$, vanishing for large $B$.

Tuples are then ranked by their \(z\)-values.
This shuffle-based null model thus provides a fully empirical estimate of
$k$-tuple significance  under the constraint of preserving all $(k-1)$-tuple counts.

\section{Real-world data} \label{sec:real_world_data}

In our applications to real-world datasets, we consider unordered tuples. There are two main reasons for this choice. First, for these specific datasets we are interested in the co-activation of units rather than in their specific activation order, which we do not expect to be particularly meaningful.
 Second, from a practical perspective, the number of possible unordered tuples for $N$ symbols is $\binom{N + k - 1}{k}$, while the number of ordered tuples is $N^k$. For fixed $k$ and large $N$, $\binom{N+k-1}{k} \sim N^k/k!$, yielding a factorial reduction in the number of configurations.
This reduction improves the statistical reliability of the analysis, particularly for systems with a large number of units $N$ (see Sec.~\ref{sec:SM_regimes}).
For each dataset we selected a time bin $\Delta t$ commensurate with the resolution of the corresponding time series. For the brain data, we used $\Delta t = 0.0015$\,s, which corresponds to the neuronal refractory period—the typical time required for a neuron to become re-excitable after firing a spike. This choice was made to prevent the repetition of the same neuron within a single tuple, thereby avoiding self-loops or autocorrelations in neuronal activity.
We verified that the results remain consistent across a wide range of $\Delta t$ values. In particular, for the datasets we analyzed, using $\Delta t = 0.015$\,s and $\Delta t = 0.0015$\,s, the overlap between significant hyperedges with a BJS-score $\geq 0.6$ is approximately $97\%$ for $2$-motifs and $90\%$ for $3$-motifs (relative to the smaller set of hyperedges), while the Jaccard similarity is around $94\%$ for $2$-motifs and $79\%$ for $3$-motifs. 
Further increasing $\Delta t$ leaves the results essentially unchanged, as for $\Delta t = 0.015$\,s the symbolic sequence already contains few to no space characters. Conversely, decreasing $\Delta t$ significantly below $0.0015$s begins to break the symbolic sequence, as the number of space characters becomes very large, resulting in many isolated spikes. Nonetheless, even at $\Delta t = 0.0005$\,s, when compared to $\Delta t = 0.0015$\,s, the overlap relative to the smaller set of hyperedges remains high: approximately $98\%$ for $2$-motifs and $78\%$ for $3$-motifs. The corresponding Jaccard similarity in this case is about $72\%$ for $2$-motifs and $48\%$ for $3$-motifs.

For the stock prices and email exchange time series, we instead used $\Delta t = 1$\,day, corresponding to the time resolution of the stock price data. 
Given the temporal resolution of these two datasets, we verified that the choice of $\Delta t$ plays only a marginal role: for essentially any reasonable value of the time window, the resulting symbolic sequences contain very few space characters.
Although $\Delta t$ does not appear to have a significant impact for these specific datasets, this may not hold true for other types of data. In general, one should assess the influence of different $\Delta t$ values on the identification of motifs.

\begin{table}
\begin{tabular}{ |p{4.5cm}||p{1cm}|p{1cm}|p{1cm}|p{1cm}| }
\hline
 \bf{Data set}  & $N$  & $E_2$ & $E_3$ & $\text{BJS}^{\text{thr}}$ \\
 \hline
 \hline
  Brain 1 activity (micro-scale)         &  346   & 3458   &   599 & 0.7\\
  Brain 1 activity (micro-scale)       &  346   & 4523   &   1441 & 0.6\\
  Brain 1 activity (macro-scale)    &  15   & 38   &   115 & 0.7\\
  Brain 1 activity (macro-scale)     &  15   & 42   &   130 & 0.6\\
  Brain 2 activity (micro-scale)       &  361   & 3331   &   313 & 0.7\\ 
  Brain 2 activity (micro-scale)        &  361   & 4519   &   875 & 0.6\\ 
  Brain 2 activity (macro-scale)       &  18   & 56   &   146 & 0.7\\
  Brain 2 activity (macro-scale)      &  18   & 61   &   187 & 0.6\\
  Brain 3 activity (micro-scale)     &  392   & 3428   &   230 & 0.7\\ 
  Brain 3 activity (micro-scale)      &  392   & 4690   &   851 & 0.6\\ 
  Brain 3 activity (macro-scale)      &  18   & 57   &   166 & 0.7\\
  Brain 3 activity (macro-scale)       &  18   & 60   &   201 & 0.6\\
  Stock price changes (unsigned)        &   24   & 16   &   4 & 0.6\\
  Stock price changes (signed)        &   48   & 110   &   4 & 0.6\\
  Email exchanges    &  143   & 131   &   5 & 0.6\\
  \hline
\end{tabular} 
\caption{Basic properties of the unweighted and undirected hypergraphs obtained from multivariate time-series, respectively from single-cell neuronal activity \cite{allen_whitepaper}, financial stocks \cite{yahoofinance, yfinance}, and email exchanges \cite{klimt_enron_2004}. 
Values of $\text{BJS}^{\text{thr}} = 0.6, 0.7$ have been adopted to threshold the weighted hypergraphs. A larger threshold means that fewer hyperedges on average satisfy the condition $\text{BJS}^{\text{thr}} \leq \text{BJS}(\boldsymbol{s})$, resulting in more sparse hypergraphs (i.e., with less hyperedges). For the stocks and the emails network we show the results ony for $\text{BJS}^{\text{thr}} = 0.6$, as adopting a threshold of $0.7$ results in no higher-order hyperedges. 
}  \label{tab:datasets_SM}
\end{table}

We note that, in real-world data with limited temporal resolution, multiple events from different time series may occur at exactly the same time. We treat these co-occurrences by randomizing the order of the corresponding symbols in the symbolic sequence. If simultaneous events occur frequently, we recommend analyzing the sequence as unordered, since the relative order of adjacent symbols is then not meaningful.

\paragraph*{\bf Brain data.}
We analyzed publicly available time series of neural activity with single-neuron resolution, provided by the Allen Institute. These datasets consist of multiple recording sessions performed on mice engaged in a visual task. For each mouse, approximately $500$ individual neurons were recorded during each session using Neuropixels technology. After filtering out neurons based on quality metrics—specifically retaining only those with an inter-spike interval (ISI) violation fraction $\text{isi} \leq 0.1$ and a signal-to-noise ratio $\text{snr} \geq 1.5$—we typically retained approximately $300$–$400$ neurons per session \cite{allen_whitepaper}.
We applied our method to both the individual-neuron time series (and corresponding symbolic sequences), and an aggregated version in which neurons were grouped by their functional brain areas, as provided in the dataset metadata.
Table~\ref{tab:datasets_SM} reports the number of nodes $N$, the number of unordered $2$-motifs $E_2$, and the number of unordered $3$-motifs $E_3$ in the unweighted hypergraphs obtained using significance thresholds $\text{BJS}^{\text{thr}} = 0.6$ and $\text{BJS}^{\text{thr}} = 0.7$.
We present results for three different sessions of the ``Brain Observatory'' type \cite{allen_whitepaper}, each corresponding to a different mouse. For each mouse, we analyzed all data available from the session, resulting in symbolic sequences of approximately $3 \times 10^7$ symbols.
In particular, ``Brain 1'' refers to session 773418906, ``Brain 2'' to session 791319847, and ``Brain 3'' to session 797828357 of the ``Brain Observatory 1.1'' dataset.

{\bf Stock market.}
We considered the historical time series publicly available through Yahoo finance, represented by the closure prices of $24$ stocks from NASDAQ Stock Market and New York Stock Exchange over $30$ years, from 1994 to 2024 \cite{yahoofinance, yfinance}. In particular, we selected $3$ stocks from $8$ different categories: 
\begin{itemize}
    \item technology:
    \begin{itemize}
        \item  Apple Inc. (AAPL)
        \item  Microsoft Corporation (MSFT)
        \item International Business Machines Corporation (IBM)
    \end{itemize}
    \item chemicals:
    \begin{itemize}
        \item Air Products and Chemicals, Inc. (APD)
        \item PPG Industries, Inc. (PPG)
        \item DuPont de Nemours, Inc. (DD)
    \end{itemize}
    \item  banking:
    \begin{itemize}
        \item JPMorgan Chase \& Co. (JPM)
        \item Bank of America Corporation (BAC)
        \item Citigroup Inc. (C)
    \end{itemize}
    \item  entertainment:
    \begin{itemize}
        \item The Walt Disney Company (DIS)
        \item  Comcast Corporation (CMCSA)
        \item Sony Group Corporation (SONY)
    \end{itemize}

    \item food and beverage:
    \begin{itemize}
        \item the Coca-Cola Company (KO)
        \item PepsiCo, Inc. (PEP)
        \item General Mills, Inc. (GIS)
    \end{itemize}
    
    \item pharmaceutical:
    \begin{itemize}
        \item Pfizer Inc. (PFE)
        \item Merck \& Co., Inc. (MRK)
        \item Johnson \& Johnson (JNJ)
    \end{itemize}
    
    \item energy:
    \begin{itemize}
        \item Exxon Mobil Corporation (XOM)
        \item Chevron Corporation (CVX)
        \item ConocoPhillips (COP)
    \end{itemize}
    
    \item consumer goods:
    \begin{itemize}
        \item The Procter \& Gamble Company (PG)
        \item Colgate-Palmolive Company (CL)
        \item Kimberly-Clark Corporation (KMB)
    \end{itemize}
    
\end{itemize}

We first converted each time series of closing prices into a binary time series as follows: if the change in closing price of a stock exceeded $2\%$ compared to the previous closing price, we assigned a value of $1$ in the corresponding binary time series (representing a \emph{spike}); otherwise, we assigned a $0$.
We then applied the procedure described in the manuscript to map the binary multivariate time series of the $24$ stocks into a unique symbolic sequence of $25$ symbols (one for each stock, plus a special empty-space symbol). Given the daily timescale of this dataset, we inserted the empty symbol in the sequence whenever the time gap between two successive symbols exceeded $24$ hours.
Finally, we extracted the unordered $2$- and $3$-motifs from the symbolic sequence, as described in the manuscript.
For this dataset, we repeated the entire procedure several times and observed minor fluctuations (on the order of $\pm 1$) in the number of motifs identified. This variability arises from the limited temporal resolution of the dataset, which causes many spikes (i.e., entries with value $1$ in the binary time series) to occur simultaneously. In such cases, to construct the symbolic time series, we must randomize the order of events that occur at the same time. This step introduces a degree of stochasticity in the mapping process, which in turn affects the final set of motifs.
In all realizations of the procedure, we found that approximately $85\%$ of the pairwise motifs involved stocks belonging to the same category—an intuitively plausible result that supports the validity of our methodology.
Regarding the $3$-motifs, our method identified as the most significant motifs the triplet (BAC, C, JPM), corresponding to the three banking stocks, and (COP, CVX, XOM), corresponding to the three energy stocks. Both motifs had a BJS-score of approximately $0.75 \pm 0.02$.

We also considered the case of signed variations in stock prices by assigning each stock two distinct symbols: one representing a positive variation and the other a negative variation.
Interestingly, in the case of signed changes, all the $2$-motifs involved stock price changes of the same sign ($54$ among symbols representing positive variations and $56$ among those representing negative variations). 
In Table~\ref{tab:datasets_SM} we report the results relative to $\text{BJS}^{\text{thr}} = 0.6$.

{\bf Email exchange.}
We constructed binary time series from the Enron dataset of internal emails exchanged by the company's employees \cite{klimt_enron_2004}. 
Specifically, we set $x_i(t) = 1$ at time $t$ if employee $i$ sent an email at that time, and considered only email exchanges within the company—that is, from one Enron employee to other Enron employees.
A centrality analysis performed on the resulting unweighted hypergraph identified the most central nodes as key figures within the company.
In particular, we measured betweenness, closeness, and eigenvector centrality for the network nodes \cite{newman_networks_2010, latora_complex_2017}, consistently finding key figures in the company (e.g., vice presidents and chief operating officers) among the top 10 nodes across all centrality measures.
In Table~\ref{tab:datasets_SM}, we report the results corresponding to the threshold $\text{BJS}^{\text{thr}} = 0.6$.

\subsection*{Comparison with surrogate-based significance measures}

To further validate our approach on empirical data, we performed a direct comparison between SYMBA and a well-established surrogate-data method based on sequence shuffling, using both empirical $p$-values and $z$-scores to assess tuple significance.

We compared the significance assigned by the different methods on the stock market and Enron email datasets, as well as on the Neuropixels recordings. For the latter, to make the surrogate-data approach computationally feasible, we coarse-grained the neuronal activity by replacing each neuron with its corresponding brain area, and additionally considered a representative subsample of $100\,000$ symbols. Applying the surrogate-data approach to the full Neuropixels recordings at single-neuron resolution is computationally prohibitive: those sequences contain approximately $300$--$500$ symbols and have lengths of $2.7\times10^{7}$--$4.0\times10^{7}$ symbols, requiring an impractically large number of surrogate sequences to obtain reliable empirical $p$-values and $z$-scores.

\begin{table}[htp!]
\centering
\begin{tabular}{llcc}
\hline
Dataset & Tuple length & SYMBA vs $p$-value & SYMBA vs $z$-score \\
\hline
Stock market & 2 & 0.996 & 0.998 \\
             & 3 & 0.992 & 0.984 \\
\hline
Enron email & 2 & 0.967 & 0.915 \\
             & 3 & 0.935 & 0.840 \\
\hline
Neuropixels (brain areas) & 2 & 0.996 & 0.992 \\
                          & 3 & 0.985 & 0.957 \\
\hline
\end{tabular}
\caption{Spearman rank correlation between the significance scores assigned by SYMBA and the surrogate-based significance measures (empirical $p$-values and empirical $z$-scores). Correlations are computed over all observed tuples of the corresponding length. All reported correlations are highly significant ($p<10^{-50}$; in several cases the numerical value underflows to zero).}
\label{tab:realdata_correlations}
\end{table}

The results, reported in Table~\ref{tab:realdata_correlations}, show an excellent agreement between SYMBA and the surrogate-based significance measures on all the considered real datasets, with Spearman correlation coefficients close to one. These results indicate that, whenever surrogate-based significance estimation is both computationally feasible and statistically reliable, SYMBA yields significance rankings highly consistent with established surrogate-data methods, while providing orders-of-magnitude faster computation.

\clearpage

\end{document}